\documentclass[prd,twocolumn,showpacs,nofootinbib]{revtex4}

\usepackage{amsfonts}
\usepackage{amsmath}
\usepackage{amssymb}
\usepackage{aas_macros}
\usepackage{upgreek}
\usepackage{bm}
\usepackage{dsfont}
\usepackage{dcolumn}
\usepackage{epsfig}
\usepackage{graphicx}
\usepackage{graphics}
\usepackage{placeins}
\usepackage{subfigure}

\usepackage{color} 
\definecolor{dark_red}{rgb}{0.7, 0.1, 0.1}

\def\be{\begin{equation}}
\def\ee{\end{equation}}
\def\bi{\begin{itemize}}
\def\ei{\end{itemize}}
\def\ben{\begin{enumerate}}
\def\een{\end{enumerate}}

\newcommand{\bs}[1]{\boldsymbol{#1}}
\newcommand{\mb}[1]{\mathbf{#1}}

\begin{document}

\title{Phase-coherent mapping of gravitational-wave backgrounds using ground-based\\ 
laser interferometers}

\author{Joseph D.~Romano}
\affiliation{Department of Physics and Astronomy 
and Center for Gravitational-Wave Astronomy,
University of Texas at Brownsville, 
Brownsville, TX 78520, USA}

\author{Stephen R.~Taylor}
\affiliation{Jet Propulsion Laboratory, 
California Institute of Technology, 
4800 Oak Grove Drive, Pasadena, CA 91106, USA}

\author{Neil J.~Cornish}
\affiliation{Department of Physics, 
Montana State University, 
Bozeman, MT 59717, USA}

\author{Jonathan Gair}
\affiliation{Institute of Astronomy, 
University of Cambridge, 
Madingley Road, Cambridge, CB3 0HA, UK}

\author{Chiara M.~F.~Mingarelli}
\affiliation{TAPIR, California Institute of Technology, 
1200 E California Blvd., M/C 350-17, Pasadena, CA 91125, USA }
\affiliation{Max Planck Institute for Radio Astronomy, 
Auf dem H\"{u}gel 69, D-53121 Bonn, Germany }

\author{Rutger van~Haasteren}
\affiliation{Jet Propulsion Laboratory, 
California Institute of Technology, 
4800 Oak Grove Drive, Pasadena, CA 91106, USA }

\date{\today}

\begin{abstract}
We extend the formalisms developed in 
Gair~et~al.~\cite{Gair-et-al:2014} and 
Cornish and van~Haasteren~\cite{Cornish-vanHaasteren:2014}
to create maps of gravitational-wave backgrounds
using a network of ground-based laser interferometers.
We show that in contrast to pulsar timing arrays, 
which are insensitive to half of the gravitational-wave 
sky (the curl modes),
a network of ground-based interferometers is sensitive 
to both the gradient and curl components of the background.
The spatial separation of a network of interferometers,
or of a single interferometer at different times during 
its rotational and orbital motion around the Sun, allows 
for recovery of both components.
We derive expressions for the 
response functions of a laser interferometer in the small-antenna
limit, and use these expressions to calculate the overlap 
reduction function for a pair of interferometers.
We also construct maximum-likelihood estimates of the
$+$ and $\times$-polarization modes of the gravitational-wave
sky in terms of the response matrix for a network of 
ground-based interferometers, evaluated at discrete times 
during Earth's rotational and orbital motion around the Sun.
We demonstrate the feasibility of this approach for some 
simple simulated backgrounds (a single point source
and spatially-extended distributions having only grad
or curl components), calculating maximum-likelihood sky
maps and uncertainty maps based on the (pseudo)inverse 
of the response matrix.
The distinction between this approach and standard
methods for mapping gravitational-wave power is also
discussed.
\end{abstract}


\maketitle
\section{Introduction}
\label{s:intro}

Searches for anisotropic gravitational-wave backgrounds have 
typically been formulated in terms of the distribution of 
gravitational-wave power on the sky
(see, e.g.,~\cite{allen-ottewill, Cornish:2001hg, mitra-et-al, Thrane-et-al:2009, 
S5anisotropic:2011, Mingarelli:2013, TaylorGair:2013}).
The basic idea underlying these approaches 
is to use to cross-correlation measurements 
from two or more detectors to estimate the 
power in the gravitational-wave background as 
a function of sky position.
For a network of ground-based laser interferometers like 
LIGO~\cite{aLIGO:2015}, Virgo~\cite{aVirgo:2015}, 
etc., or space-based interferometers like LISA~\cite{Danzmann:1996},
eLISA~\cite{eLISA:2012, eLISA:2013} or 
BBO~\cite{Phinney-et-al:2004}, 
the motion of the detectors modulates the 
correlated gravitational-wave signal at harmonics of the 
Earth's daily rotational motion,
or the spacecrafts' yearly orbital motion around the Sun.
The time-varying signal carries information about the 
multipole moments characterizing the anisotropic 
distribution of power, which can be estimated using, e.g., 
maximum-likelihood methods~\cite{Thrane-et-al:2009}.

Recent papers by Gair~et~al.~\cite{Gair-et-al:2014} 
and Cornish and van~Haasteren~\cite{Cornish-vanHaasteren:2014}
describe an alternative approach for mapping the 
gravitational-wave sky, which can be used to recover 
both the amplitude and phase of 
the gravitational-wave signal at each sky position.
The analysis in \cite{Cornish-vanHaasteren:2014} is cast
in terms of the traditional plus and cross polarization 
components $\{h_+(f,\hat k)$, $h_\times(f,\hat k)\}$,
while in \cite{Gair-et-al:2014} 
the metric perturbations are decomposed in terms of 
spin-weighted or 
tensor (gradient and curl) spherical harmonics
$\{Y^G_{(lm)ab}(\hat k)$, $Y^C_{(lm)ab}(\hat k)\}$.
This latter decomposition is similar to that used to 
characterize the polarization of the cosmic microwave 
background~\cite{KKS:1997}, taking into account the symmetric, 
transverse-traceless nature of the metric 
perturbations $h_{ab}(f,\hat k)$.

Although the formalisms introduced in 
\cite{Gair-et-al:2014, Cornish-vanHaasteren:2014} 
are general, they were applied specifically to the case of 
gravitational-wave searches using pulsar timing arrays.
In contrast to the case of ground-based interferometers 
on a rotating, orbiting Earth or the orbiting LISA/eLISA 
spacecraft, a 
pulsar timing array operates effectively as a {\em static} 
galactic-scale gravitational-wave detector~\cite{foster-backer-1990}, 
with each
Earth-pulsar line-of-sight being the equivalent of a
one-way, one-arm interferometer with a common endpoint
at the solar system barycentre (SSB).
This is because the frequency range for pulsar timing 
measurements is such that the displacement of a detector
on Earth from the SSB is much smaller than the 
wavelength of the relevant gravitational waves, and hence 
the detector effectively resides at the SSB.
In addition, the times of arrival of the pulses from
a given set of pulsars at a radio telescope are 
typically converted to arrival times at the SSB, which
is a fixed origin.
A consequence of the static nature of pulsar-timing 
measurements is that
a pulsar timing array will never be able to observe
half of the gravitational-wave sky regardless of how 
many pulsars are included in the array~\cite{Gair-et-al:2014}.
In particular, the response of a pulsar timing array to 
curl modes is identically zero, as the gravitational 
contribution of such modes to the timing residual equals 
zero when integrated over the sky.

In this paper, we extend the formalisms developed in
\cite{Gair-et-al:2014, Cornish-vanHaasteren:2014} to the
case of ground-based interferometers like LIGO, Virgo, etc.
For simplicity we work in the {\em small-antenna} 
(or long-wavelength) limit, which is appropriate 
for such detectors.
We show that in contrast to pulsar timing arrays,
a network of ground-based interferometers is sensitive
to both the gradient and curl modes of the background.
The fact that the spatial separation of the detectors 
is the same order or greater than the radiation 
wavelength is sufficient to allow for recovery of both types of mode.
We demonstrate this by: 
(i) explicitly deriving analytic expressions for the 
gradient and curl response functions of a laser interferometer, 
and (ii) constructing maximum-likelihood estimates of the 
gravitational-wave sky for different types of simulated 
backgrounds.
The reconstruction of the sky maps is based on singular
value decomposition (SVD) of the whitened response 
matrix $\bar{\mb R}= \bar{\mb U}\bar{\bs\Sigma}\bar{\mb V}^\dag$, 
which maps the modes of 
the gravitational-wave background to the response of the 
individual interferometers, evaluated at discrete times 
during Earth's rotational and orbital motion around the sun.
The columns of $\bar{\mb U}$ and $\bar{\mb V}$ corresponding to the 
non-zero singular values of $\bar{\bs\Sigma}$ have the 
interpretation of {\em response range vectors} and 
{\em sky map basis vectors}, in terms of which the measured 
response and the maximum-likelihood sky map can be written 
as linear combinations~\cite{Cornish-vanHaasteren:2014}.
The recovered sky maps can be calculated in terms of 
either a pixel-based parametrisation, 
$\{h_+(f,\hat k_n), h_\times(f,\hat k_n)\}$, where $n$ labels the pixels on
the sky, or in terms of the gradient and curl spherical harmonic 
components $\{a^G_{(lm)}(f)$, $a^C_{(lm)}(f)\}$, where 
$(lm)$ labels the various multipole modes, up to some maximum
value $l_{\rm max}$.
 
The organization of the rest of the paper is as follows:
In Sec.~\ref{s:review}, we summarize key 
formulae related to the tensor spherical harmonic decomposition
approach described in \cite{Gair-et-al:2014}.
We derive, in Sec.~\ref{s:response}, analytic expressions 
for the gradient and curl response functions 
of a laser interferometer in the small-antenna limit.
(Details of the derivation are given in Appendix~\ref{s:derivation}.)
In Sec.~\ref{s:overlap}, we show that we can recover 
the standard overlap reduction functions using the analytic
expressions for the response functions derived in 
Sec.~\ref{s:response}, and 
in Sec.~\ref{s:rotation-orbit} we compare the effects
of Earth's rotational and orbital motion on the sky 
reconstruction.
We describe the map-making formalism in Sec.~\ref{s:maps} 
and demonstrate that a network of ground-based 
inteferometers can recover both the grad and curl 
components of a gravitational-wave background, by
constructing maximum-likelihood sky maps for some 
simple simulations.
We conclude in Sec.~\ref{s:discussion} with a brief summary
and discussion of the results, listing a few modifications
that might be needed in order to apply this formalism to 
real data.

\section{Brief review of tensor spherical harmonic decomposition}
\label{s:review}

For completeness, we summarize in this section 
some key formulae from the tensor spherical harmonic 
decomposition method described in \cite{Gair-et-al:2014}.
Interested readers are referred to that paper for 
more details.
\medskip
 
Any transverse-traceless tensor field on the sky 
can be decomposed as a superposition of gradients 
and curls of spherical harmonics:
\be
\begin{aligned}
Y^G_{(lm)ab}(\hat k) 
&= N_l 
\left[Y_{(lm);ab}(\hat k) - 
\frac{1}{2} g_{ab}  Y_{(lm);c}{}^{c}(\hat k) \right]\,,
\\
Y^C_{(lm)ab}(\hat k) 
&= \frac{N_l}{2} 
\left[Y_{(lm);ac}(\hat k)\epsilon^c{}_b 
+  Y_{(lm);bc}(\hat k) \epsilon^c{}_a \right]\,,
\end{aligned}
\ee
where semi-colon denotes covariant derivative, 
$g_{ab}$ is the metric tensor on the sphere, 
$\epsilon_{ab}$ is the Levi-Civita anti-symmetric tensor
\be
\epsilon_{ab} 
= \sqrt{g} \left( \begin{array}{cc}0&1\\-1&0\end{array}\right)\,,
\ee
and $N_l$ is a normalization constant
\be
N_l = \sqrt{\frac{2 (l-2)!}{(l+2)!}}\,.
\ee
In terms of the standard polarization tensors on 
the sky
\be
\begin{aligned}
e_{ab}^+(\hat k)&=
\hat\theta_a\hat\theta_b-\hat\phi_a\hat\phi_b\,,
\\
e_{ab}^\times(\hat k)&=
\hat\theta_a\hat\phi_b+\hat\phi_a\hat\theta_b \,,
\label{e:e+x}
\end{aligned}
\ee
we have
\begin{widetext}
\be
\begin{aligned}
Y^G_{(lm)ab}(\hat{k}) 
&= \frac{N_l}{2} \left[ W_{(lm)}(\hat{k}) e_{ab}^+(\hat k) 
+ X_{(lm)}(\hat{k}) e_{ab}^\times(\hat k) \right]\,, 
\\
Y^C_{(lm)ab}(\hat{k}) 
&= \frac{N_l}{2} \left[ W_{(lm)}(\hat{k})e_{ab}^\times(\hat k) 
- X_{(lm)}(\hat{k}) e_{ab}^+(\hat k) \right]\,,
\label{e:YGC}
\end{aligned}
\ee
where $W_{(lm)}(\hat k)$ and $X_{(lm)}(\hat k)$ can 
be written in terms of associated Legendre polynomials as
\begin{align}
W_{(lm)}(\hat{k}) 
&= +2 \sqrt{\frac{2l+1}{4\pi} \frac{(l-m)!}{(l+m)!}} 
\left\{-\left(\frac{l-m^2}{\sin^2\theta} 
+ \frac{1}{2} l(l-1) \right) P^m_l(\cos\theta) 
+ (l+m) \frac{\cos\theta}{\sin^2\theta} P^m_{l-1}(\cos\theta)
\right\} e^{im\phi}\,,
\label{e:Wlm}
\\
i X_{(lm)}(\hat{k}) 
&= -2 \sqrt{\frac{2l+1}{4\pi} \frac{(l-m)!}{(l+m)!}} 
\left\{\frac{m}{\sin^2\theta} 
\left[(l-1) \cos\theta P_l^m(\cos\theta) 
- (l+m) P^m_{l-1}(\cos\theta) \right]
\right\} e^{im\phi}\,,
\label{e:Xlm}
\end{align}
%
and are related to spin-2 spherical harmonics 
\cite{NewmanPenrose:1966, Goldberg:1967} through the equation
\begin{equation}
{}_{\pm2}Y_{(lm)}(\hat{k})
=\frac{N_l}{\sqrt{2}} \left[
W_{(lm)}(\hat{k})\pm i X_{(lm)}(\hat{k})\right]\,.
\label{e:Ypm2}
\end{equation}
In terms of the grad and curl spherical harmonics, a general 
gravitational-wave background can be decomposed as
\begin{equation}
h_{ab}(t,\vec x)
=
\int_{-\infty}^\infty {\rm d}f\>
\int_{S^2} {\rm d}^2 \Omega_{\hat{k}}\>
h_{ab}(f,\hat k)
e^{i 2\pi f(t-\hat k\cdot \vec x/c)}\,,
\label{e:hab(t,x)}
\end{equation}
where
%
\be
h_{ab}(f,\hat{k}) = \sum_{l=2}^{\infty} \sum_{m=-l}^{l} 
\left[ a^G_{(lm)}(f) Y^G_{(lm)ab}(\hat{k}) 
+ a^C_{(lm)}(f) Y^C_{(lm)ab}(\hat{k}) \right]\,.
\label{e:habGC}
\ee
\end{widetext}
The mode functions $a^G_{(lm)}(f)$, $a^C_{(lm)}(f)$ are
related to the more traditional $+$ and $\times$ 
polarisation components $h_+(f,\hat k)$, $h_\times(f,\hat k)$
defined by
\be
h_{ab}(f,\hat{k})\! = \!
h_+(f,\hat k)e^+_{ab}(\hat k)+
h_\times(f,\hat k)e^\times_{ab}(\hat k)\,,
\ee
via
\be
\begin{aligned}
h_+(f,\hat k)&=\!
\sum_{(lm)}\!
\frac{N_l}{2} \! \left[ a^G_{(lm)}(f) W_{(lm)}(\hat{k}) 
- a^C_{(lm)}(f) X_{(lm)} (\hat{k})\right],
\\
h_\times(f,\hat k)&=\!
\sum_{(lm)} \!
\frac{N_l}{2} \! \left[ a^G_{(lm)}(f) X_{(lm)}(\hat{k}) 
+ a^C_{(lm)}(f) W_{(lm)}(\hat{k}) \right],
\label{e:h+hx}
\end{aligned}
\ee
and
\be
\begin{aligned}
a^G_{(lm)}\!(f) \!= \! N_l  \!\! \int_{S^2} \!\! \! {\rm d}^2\Omega_{\hat k}\!
\left[ h_+\!(f,\hat k) W_{(lm)}^* \! (\hat k) 
\!+\! h_\times \! (f,\hat k) X_{(lm)}^*\! (\hat k)\right],
\\
a^C_{(lm)}\!(f)\! = \! N_l \!\! \int_{S^2}  \!\!\! {\rm d}^2\Omega_{\hat k}\!
\left[ h_\times \! (f,\hat k) W_{(lm)}^* \!(\hat k) 
\!-\! h_+ \! (f,\hat k) X_{(lm)}^*\! (\hat k)\right]\,,
\label{e:aGaC}
\end{aligned}
\ee
where we use the shorthand 
$\sum_{(lm)} \equiv \sum_{l=2}^\infty \sum_{m=-l}^l$.
Note that all the summations over $l$ start at $l=2$.

\section{Response function calculations}
\label{s:response}

The frequency-domain response of a laser interferometer 
to a gravitational-wave background is given by
\be
\tilde r(f) = \int_{S^2}{\rm d}^2\Omega_{\hat k}
\sum_{A}
R^A(f,\hat k)h_A(f,\hat k) \, ,
\ee
where $A=\{+,\times\}$ and
\be
R^A(f,\hat k) = 
\frac{1}{2}e^A_{ab}(\hat k)(u^au^b-v^av^b)
e^{-i2\pi f\hat k\cdot\vec x_0/c}\,.
\label{e:R+Rx}
\ee
Here $\hat u$, $\hat v$ are unit vectors along the arms
of the interferometer and $\vec x_0$ is the position vector
of the vertex of the interferometer at the time $t$ when
the measurement is made.
The above expression for $R^A(f,\hat k)$ is valid in the 
{\em small-antenna limit}, which is appropriate for 
ground-based interferometers like LIGO, Virgo, etc. (see
e.g., \cite{MingarelliSidery:2014} for this discussion in the
pulsar timing context).
The length of the interferometer arms do not enter the
expressions for the response functions in this limit.

Alternatively, if we expand the metric perturbations in 
terms of the gradient and curl modes 
$\{a^G_{(lm)}(f), a^C_{(lm)}(f)\}$, 
the response can be written as

\be
\tilde r(f) = \sum_{(lm)}
\sum_P R^P_{(lm)}(f)a^P_{(lm)}(f)\,,
\ee
\begin{widetext}
\noindent
where $P=\{G,C\}$ and
\be
\begin{aligned}
R^G_{(lm)}(f)
&=\frac{N_l}{2}
\int_{S^2}{\rm d}^2\Omega_{\hat k}\>
\left[R^+(f,\hat k)W_{(lm)}(\hat k) 
+ R^\times(f,\hat k)X_{(lm)}(\hat k)\right]\,,
\\
R^C_{(lm)}(f)
&=\frac{N_l}{2} 
\int_{S^2}{\rm d}^2\Omega_{\hat k}\>
\left[R^\times(f,\hat k)W_{(lm)}(\hat k) 
- R^+(f,\hat k)X_{(lm)}(\hat k)\right]\,.
\label{e:RGRC}
\end{aligned}
\ee
These integrals were evaluated in Appendix~D of \cite{Gair-et-al:2014} 
for the case of a reference frame which has $\vec x_0=\vec 0$:
\be
\begin{aligned}
R^G_{(lm)}(f)\big|_{\vec x_0=\vec 0}
&= \delta_{l2}\,\frac{4\pi}{5}\sqrt{\frac{1}{3}}
\left[Y_{2m}(\hat u) - Y_{2m}(\hat v)\right]\,,
\\
R^C_{(lm)}(f)\big|_{\vec x_0=\vec 0}
&= 0\,.
\label{e:RGRC_static}
\end{aligned}
\ee
Note that the curl response is identically zero in this frame, 
while the gradient response is independent of frequency and 
is non-zero only 
for the quadrupole components, $l=2$.
These results are qualitatively similar to those for a pulsar 
timing array, which also have zero curl response, and a 
frequency-independent grad response proportional to $Y_{lm}(\hat u)$, 
where $\hat u$ points in the direction to the pulsar.
In what follows, we will use the notation
\begin{equation}
F_{m}(\hat u,\hat v)\equiv 
\frac{4\pi}{5}\sqrt{\frac{1}{3}}
\left[Y_{2m}(\hat u) - Y_{2m}(\hat v)\right]
\end{equation}
to denote the particular combination of spherical harmonics
that appear in Eq.~(\ref{e:RGRC_static}).
Since only the quadrupole response is non-zero, the index
$m$ on $F_{m}$ is restricted to have values $m=0,\pm 1, \pm 2$.

For a single static interferometer, there is no loss in generality
in choosing a reference frame with the vertex of the interferometer 
located at the origin, with the response functions given as above.
But for a network of interferometers attached to a rotating and
orbiting Earth, such a coordinate choice is no longer possible.
If an interferometer is displaced from the origin by $\vec x_0$, 
one can show that
%
\be
\begin{aligned}
R^G_{(lm)}(f)
=&\sum_{m'=-2}^2 
\sum_{L=l-2}^{l+2}
\sum_{M=-L}^L 
F_{m'}(\hat u,\hat v)
4\pi (-i)^L j_L(\alpha)Y^*_{LM}(\hat x_0)
\\
&\times \frac{(-1)^{m'}}{2}
\sqrt{\frac{(2\cdot 2+1)(2l+1)(2L+1)}{4\pi}}
\left(\begin{array}{ccc}
2 & l & L\\
-m' & m & M 
\end{array}\right)
\left(\begin{array}{ccc}
2 & l & L\\
2 & -2 & 0 
\end{array}\right)
\left[(-1)^{l+L}+1\right]\,,
\\
R^C_{(lm)}(f)
=&\sum_{m'=-2}^2 
\sum_{L=l-2}^{l+2}
\sum_{M=-L}^L 
F_{m'}(\hat u,\hat v)
4\pi (-i)^L j_L(\alpha)Y^*_{LM}(\hat x_0)
\\
&\times \frac{(-1)^{m'}}{2i}
\sqrt{\frac{(2\cdot 2+1)(2l+1)(2L+1)}{4\pi}}
\left(\begin{array}{ccc}
2 & l & L\\
-m' & m & M 
\end{array}\right)
\left(\begin{array}{ccc}
2 & l & L\\
2 & -2 & 0 
\end{array}\right)
\left[(-1)^{l+L}-1\right]\,,
\label{e:RGRC_general}
\end{aligned}
\ee
\end{widetext}
where $\alpha \equiv 2\pi f|\vec x_0|/c$ 
and $j_L(\alpha)$ are spherical Bessel functions of order $L$.
These expressions are derived in Appendix~\ref{s:derivation}.
The two expressions in parentheses $(\ )$ for each response
function are Wigner 3-$j$ symbols
(see for example~\cite{Wigner:1959}, \cite{Messiah:1962}).
Note that, in general, the curl response is now non-zero, 
in contrast to the static single interferometer case.
In addition, both response functions depend on frequency 
via the quantity $\alpha$, which has the physical interpretation
of being $2\pi$ times the number of radiation wavelengths 
between the origin and the vertex of the interferometer.
Since the coordinate system for the response functions in
(\ref{e:RGRC_static}) 
was not chosen in any particular 
orientation relative to the unit vectors $\hat u$ and $\hat v$,
Eqs.~(\ref{e:RGRC_general}) are valid 
in an arbitrary translated and rotated coordinate 
system, provided we use the angles for $\hat u$, $\hat v$,
and $\hat x_0$ as calculated in the rotated frame.

\section{Recovery of standard overlap reduction functions}
\label{s:overlap}

Given the above expressions for $R^P_{(lm)}(f)$ one can calculate
the overlap reduction function $\Gamma_{12}(f)$ for a pair of
interferometers for gravitational-wave backgrounds having different 
statistical properties.
For example, for a statistically isotropic background with 
$C^{GG}_l = C^{CC}_l\equiv C_l$ and $C^{GC}_l=0=C^{CG}_l$,
it was shown in \cite{Gair-et-al:2014} that 
\be
\Gamma_{12}(f) = \sum_{l=2}^\infty C_l \Gamma_{12,l}(f)\,,
\ee
where
\be
\Gamma_{12,l}(f) = 
\sum_{m=-l}^l  R^P_{1(lm)}(f) R^{P*}_{2(lm)}(f)\,.
\ee
This is most-easily evaluated
in a reference frame in which the vertex of one interferometer
is located at the origin, and the vertex of the second 
interferometer is located along the $z$-axis of this frame.
This so-called {\em computational frame} is related to the 
cosmic reference frame located at the solar system barycentre 
via a translation by the position vector $\vec x_1$ and a 
rotation ${\cal R}\equiv {\cal R}_y(\theta_0){\cal R}_z(\phi_0)$
such that $\Delta \vec x\equiv\vec x_2-\vec x_1$ is directed along 
the $z$-axis. 
(Here $(\theta_0,\phi_0)$ are the polar and azimuthal 
angles of $\Delta \vec x$ with respect to the cosmic frame.)
In the computational frame
\be
R^P_{1(lm)}(f) = \delta^{PG}\delta_{l2}\,
F_{m}({\cal R}\hat u_1,{\cal R}\hat v_1)\,,
\ee
where the arguments of $F_{m}$ are the polar and azimuthal angles
of $\hat u_1$ and $\hat v_1$ with respect to 
the computational frame. 
This form for $R^P_{1(lm)}(f)$ implies
\be
\Gamma_{12,l}(f) = 
\delta_{l2}\sum_{m=-2}^2  
F_{m}({\cal R}\hat u_1,{\cal R}\hat v_1)R^{G*}_{2(2m)}(f)\,,
\ee
which in turn implies
\be
\begin{aligned}
\Gamma_{12}(f) 
&= C_2\,\Gamma_{12,2}(f)
\\
&= C_2\sum_{m=-2}^2 
F_{m}({\cal R}\hat u_1,{\cal R}\hat v_1)R^{G*}_{2(2m)}(f)\,.
\end{aligned}
\ee
Thus, the only non-zero contribution to the overlap reduction
function comes from the quadrupole gradient terms.
Using Eq.~(\ref{e:RGRC_general}), it follows that 
\begin{widetext}
\begin{align}
R^G_{2(2m)}(f) 
=F_{m}({\cal R}\hat u_2,{\cal R}\hat v_2)
\bigg[j_0(\alpha) 
&+(-1)^{m+1}(2\cdot 2+1)^2
\left(\begin{array}{ccc}
2 & 2 & 2\\
-m & m & 0
\end{array}\right)
\left(\begin{array}{ccc}
2 & 2 & 2\\
2 & -2 & 0
\end{array}\right)
j_2(\alpha)
\nonumber\\
&+(-1)^m(2\cdot 2+1)(2\cdot 4 +1)
\left(\begin{array}{ccc}
2 & 2 & 4\\
-m & m & 0
\end{array}\right)
\left(\begin{array}{ccc}
2 & 2 & 4\\
2 & -2 & 0
\end{array}\right)
j_4(\alpha)\bigg]\,,
\end{align}
where $\alpha \equiv 2\pi f|\Delta\vec x|/c$.
Thus,
\begin{align}
\Gamma_{12}(f) 
= C_2\left[A j_0(\alpha) + B j_2(\alpha) + C j_4(\alpha)\right]\,,
\label{e:Gamma12}
\end{align}
where
\be
\begin{aligned}
A&= \sum_{m=-2}^2
F_{m}({\cal R}\hat u_1,{\cal R}\hat v_1)
F^*_m({\cal R}\hat u_2,{\cal R}\hat v_2)\,,
\\
B&= \sum_{m=-2}^2
F_{m}({\cal R}\hat u_1,{\cal R}\hat v_1)
F^*_m({\cal R}\hat u_2,{\cal R}\hat v_2)
(-1)^{m+1}
(2\cdot 2+1)^2
\left(\begin{array}{ccc}
2 & 2 & 2\\
-m & m & 0
\end{array}\right)
\left(\begin{array}{ccc}
2 & 2 & 2\\
2 & -2 & 0
\end{array}\right)\,, 
\\
&=\frac{5}{7} \sum_{m=-2}^2F_{m}({\cal R}\hat u_1,{\cal R}\hat v_1)
F^*_m({\cal R}\hat u_2,{\cal R}\hat v_2)
(-1)^{2m+1}(m^2-2) \,,
\\
C&= \sum_{m=-2}^2
F_{m}({\cal R}\hat u_1,{\cal R}\hat v_1)
F^*_m({\cal R}\hat u_2,{\cal R}\hat v_2)
(-1)^m(2\cdot 2+1)(2\cdot 4 +1)
\left(\begin{array}{ccc}
2 & 2 & 4\\
-m & m & 0
\end{array}\right)
\left(\begin{array}{ccc}
2 & 2 & 4\\
2 & -2 & 0
\end{array}\right)\,, 
\\
&=\frac{12}{7} \sum_{m=-2}^2
F_{m}({\cal R}\hat u_1,{\cal R}\hat v_1)
F^*_{m}({\cal R}\hat u_2,{\cal R}\hat v_2)
\frac{(-1)^m}{(2-m)!(2+m)!}\,.
\end{aligned}
\ee
\end{widetext}
\noindent
Here we have used the definition of the Wigner 3-$j$ sycalol to simplify
the expressions for $B$ and $C$.

\begin{figure}
\begin{center}
\includegraphics[angle=0, width=.5\textwidth]{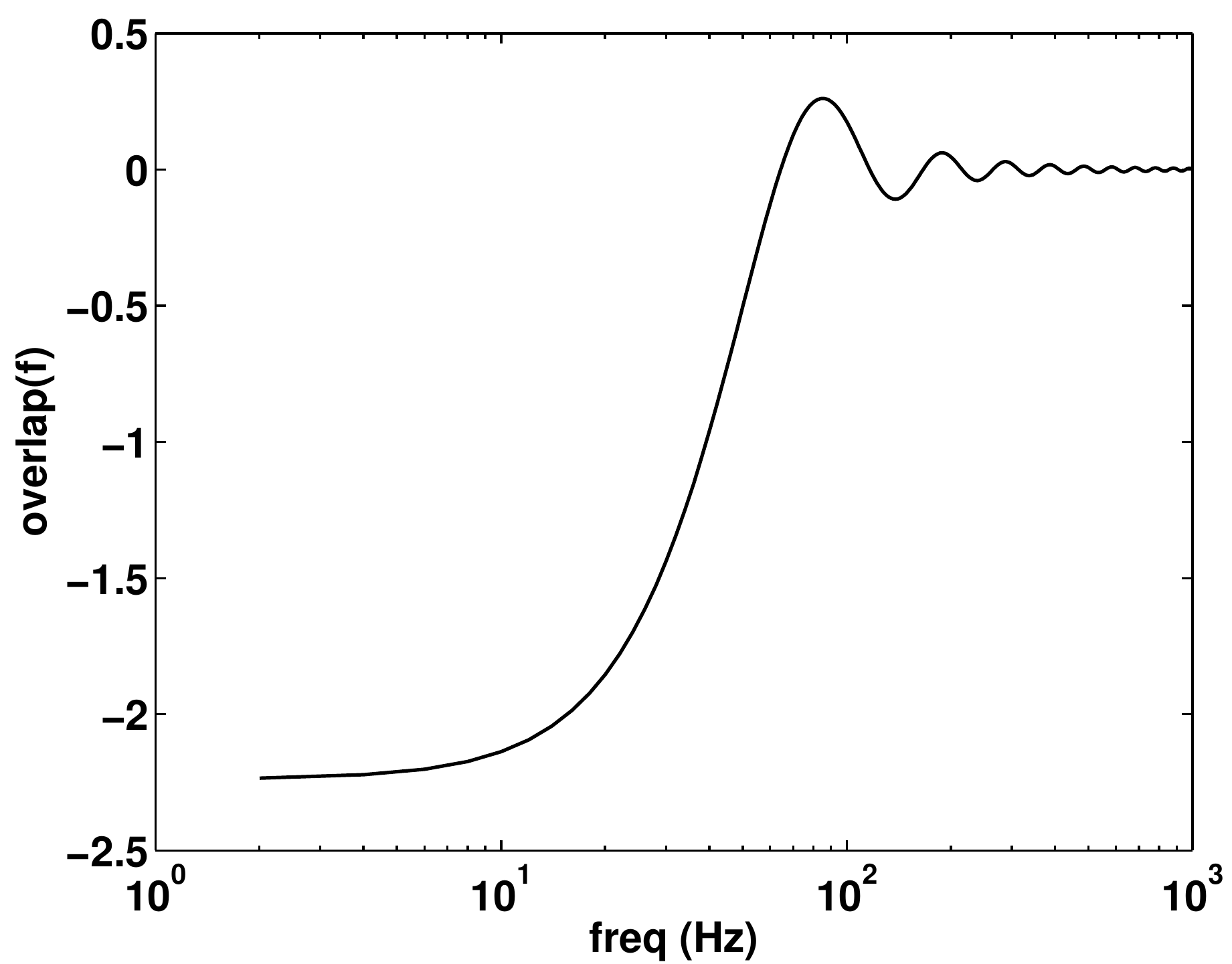}
\caption
{Overlap reduction function $\Gamma_{12}(f)$ for 
an unpolarized statistically isotropic background
for the LIGO Hanford-LIGO Livingston detector pair,
plotted on a logarithmic frequency axis.}
\label{f:overlapH1L1}
\end{center}
\end{figure}

For an unpolarised, isotropic and uncorrelated background,
the above expression for $\Gamma_{12}(f)$ in terms of 
spherical Bessel functions is similar in form to 
expressions for the overlap reduction function in 
Refs.~\cite{Flanagan:1993} and \cite{Allen:1999}.
The difference is that in those papers the expansion is
in terms of $j_0(\alpha)$, $j_1(\alpha)/\alpha$, and
$j_2(\alpha)/\alpha^2$, while the above expansion is in
terms of $j_0(\alpha)$, $j_2(\alpha)$, and $j_4(\alpha)$.
These two expansions can be related using the recurrence 
relation
\be
j_{l+1}(\alpha) = \frac{2l+1}{\alpha}\,j_l(\alpha) - j_{l-1}(\alpha)\,,
\ee
for which
\begin{multline}
\Gamma_{12}(f) =
C_2\bigg[
(A-B+C)\,j_0(\alpha) 
\\
+ (3B-10C)\,\frac{j_1(\alpha)}{\alpha}
+ 35 C\,\frac{j_2(\alpha)}{\alpha^2}
\bigg]\,.
\end{multline}
A plot of the corresponding overlap reduction function
for an unpolarised statistically isotropic background is
shown in Fig.~\ref{f:overlapH1L1} for the LIGO Hanford-LIGO 
Livingston detector pair.

\section{Rotational and orbital motion}
\label{s:rotation-orbit}

As mentioned in Sec.~\ref{s:intro},
previous analyses for anisotropic stochastic backgrounds using 
ground-based interferometers have been formulated in terms of the 
distribution of gravitational-wave power on the 
sky~\cite{allen-ottewill, Cornish:2001hg, mitra-et-al, Thrane-et-al:2009, 
S5anisotropic:2011, Mingarelli:2013, TaylorGair:2013}.
In addition, these approaches typically use cross-correlation 
measurements between pairs of detectors as the input data.
As such, these analyses are {\em insensitive} to the phase 
of the gravitational-wave background at different 
spatial locations.
Recall that the position-dependent phase information
appears explicitly in the 
Fourier components of the metric perturbations, 
$h_{ab}(f,\hat k)e^{-i2\pi f \hat k\cdot\vec x/c}$
(cf.~Eq.~(\ref{e:hab(t,x)})).
It also appears in the response functions 
$R^A(f,\hat k)$ and $R^P_{(lm)}(f)$,
cf.~Eqs.~(\ref{e:R+Rx}) and (\ref{e:RGRC_general}),
where $\vec x_0$ is the location of the 
detector at the time $t$ at which the measurement is made.
For cross-correlation measurements between a pair of 
detectors, the correlated response depends on the location
of the detectors $\vec x_1$ and $\vec x_2$ only via their difference
$\Delta\vec x\equiv \vec x_2-\vec x_1$, which is independent 
of choice of origin.
Thus, the correlated response repeats after one sidereal day of
observation.
This means that the orbital motion of the Earth is 
effectively irrelevant for such analyses; 
only the rotational motion counts.
In fact, one can {\em fold} several days of observed data to a 
single sidereal day, and perform the analysis on the 
folded data~\cite{ain-et-al:2015}.
This has obvious benefits in regards to the reduction of data 
volume and the computational cost of the analysis.

In contrast, the goal of our analysis is to recover both the 
amplitude {\em and} phase of the gravitational-wave background 
at each point on the sky, based on a likelihood function that is 
not tied to cross-correlated data.
As we shall see below, for such an analysis, the spatial locations
of the detectors are as important as their relative orientations.
Since our detectors are ground-based interferometers that 
orbit the Sun with the Earth, it is natural to reference our 
response functions and reconstructed sky maps back to the SSB.
As such, we will take the origin of coordinates for our 
calculations to be the SSB.
The detector locations are thus referenced from there.

Due to the Earth's rotational and orbital motion around the Sun, 
a single interferometer actually defines a set of {\em virtual} 
interferometers located along a quasi-circular ring $1~{\rm AU}$
from the SSB.
The Doppler shift in the observed frequency due to Earth's
velocity with respect to the SSB is not important for searches 
for broad-band gravitational-wave backgrounds, since
$v/c\sim 10^{-4}$ introduces frequency shifts of at most 
$\delta f\sim {\rm few}\times 10^{-1}~{\rm Hz}$ in the 
frequency band relevant for ground-based interferometers.
Thus, the rotational and orbital motion of the Earth 
synthesizes a set of {\em static} virtual interferometers, 
each observing the gravitational-wave background from a 
different spatial location and with a different orientation.

To compare the effects of rotational and orbital motion, 
we calculate the time-scale over which measurements made 
by different virtual interferometers are correlated.
The relevant quantity is 
$\alpha = 2\pi f |\Delta\vec x|/c$,
which appears in expressions for 
the overlap reduction function $\Gamma_{12}(f)$,
e.g.~(\ref{e:Gamma12}). 
But here $\Delta \vec x\equiv \vec x_0(t_2)-\vec x_0(t_1)$
is the spatial separation between the vertices of 
two virtual interferometers, defined by 
the vertex of a single (real) interferometer at 
times $t_1$ and $t_2$; and
$f$ is the frequency of a gravitational wave.
The above relation can be turned into a correlation 
time-scale by writing $|\Delta\vec x|= v\Delta t$, 
where $v$ is the average speed of the interferometer 
(due to Earth's rotational or orbital motion) 
over the time interval $\Delta t\equiv t_2-t_1$,
and then finding that value 
of $\Delta t$ for which $\alpha=\pi$:
\be
t_{\rm corr} = \frac{c}{2vf}\,.
\label{e:tcorr}
\ee
This corresponds to a spatial separation 
$|\Delta\vec x|=\lambda/2$, where
$\lambda=c/f$ is the wavelength of the gravitational wave.
For durations $\Delta t\lesssim t_{\rm corr}$, measurements taken 
by the two virtual interferometers are correlated;
for durations $\Delta t\gtrsim t_{\rm corr}$, the 
measurements are uncorrelated
with one another. This is justified by noting 
that two detectors will (on-average) be driven in 
coincidence by a 
gravitational wave propagating along their
separation vector whenever its wavelength is
more than twice the separation between the detectors.
This argument~\cite{Allen:1999} provides a rough lower bound 
on the decorrelation timescale of the detectors,
which will actually be slightly 
larger since we must average over all propagation directions of 
the gravitational waves when
considering a stochastic background.

\begin{figure*}[hbtp!]
\begin{center}
\hspace{0.04\textwidth}
\includegraphics[angle=0, width=.45\textwidth]{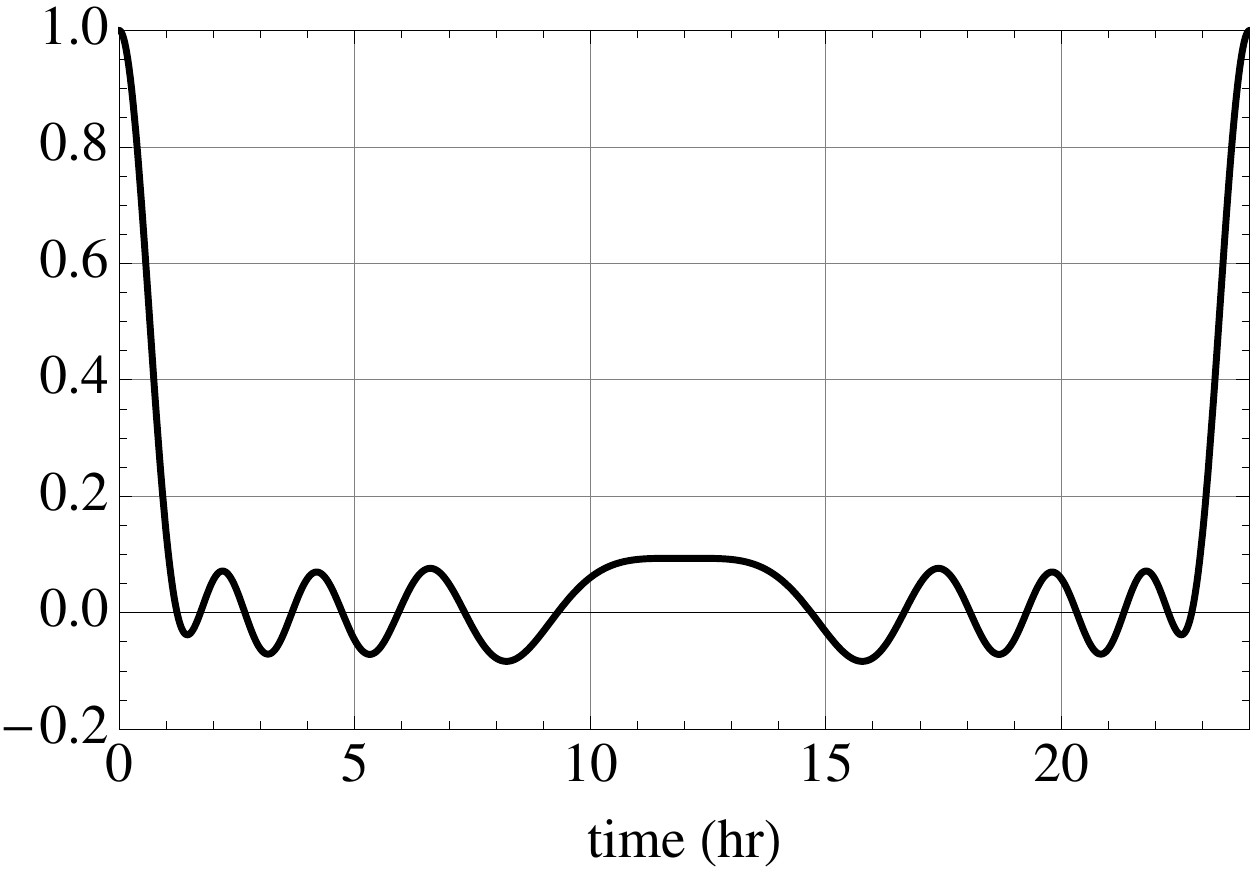}
\hspace{0.04\textwidth}
\includegraphics[angle=0, width=.45\textwidth]{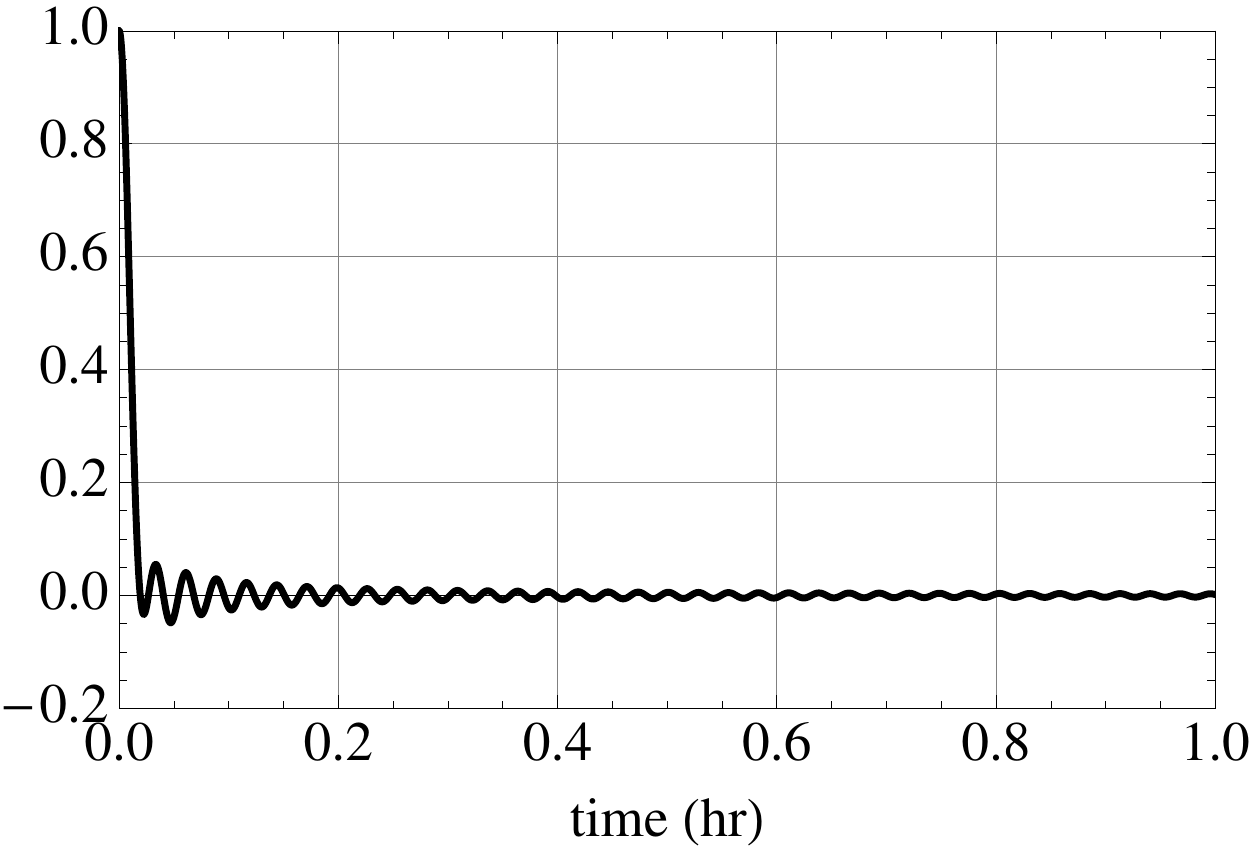}
\caption
{Overlap reduction function at $f=100~{\rm Hz}$ for two
virtual interferometers as a function of time. 
The left-hand plot is for a set of virtual interferometers 
located on Earth's equator, associated with Earth's
daily rotational motion.
The virtual interferometers have one arm pointing North
and the other pointing East.
There is no orbital motion for this case, as the 
center of the Earth is fixed at the SSB.
The right-hand plot is for a set of virtual interferometers
at $1~{\rm AU}$ from the SSB, associated with
Earth's yearly orbital motion.
There is no rotational motion for this case, as the 
interferometers are located at the center of the 
Earth in its orbit around the Sun, with the orientation 
of the interferometer arms unchanged by the orbital motion.}
\label{f:overlap_vs_time}
\end{center}
\end{figure*}

For a gravitational wave with frequency $f=100~{\rm Hz}$ 
($\lambda=3\times 10^6~{\rm m}$)
and $v=2\pi R_E/(1~{\rm day})\approx 500~{\rm m/s}$,
which is relevant for Earth's daily rotational motion, 
we find
\be
t_{\rm corr} \approx 3000~{\rm s}
\quad({\rm rotational\ motion})\,.
\ee
For $v=2\pi R_{ES}/(1~{\rm yr})\approx 3\times 10^4~{\rm m/s}$,
which is relevant for Earth's yearly orbital motion, 
we find
\be
t_{\rm corr} \approx 50~{\rm s}
\quad({\rm orbital\ motion})\,.
\ee
Figure~\ref{f:overlap_vs_time} is the overlap reduction 
function $\Gamma_{12}(f)$ for an unpolarized isotropic background,
evaluated at $f=100~{\rm Hz}$,
for two virtual interferometers as a function of time. The left panel
is for a set of virtual interferometers synthesized by the daily
rotation of a detector positioned at the Earth's equator, with no
orbital motion. One can see that the detector
decorrelates on a timescale of $\sim 1$ hour, and recorrelates after
$24$ hours whenever it returns to its starting position. If we switch
off daily rotation and synthesize a set of virtual interferometers
from the orbital motion of the Earth around the Sun, then we get the
overlap reduction function in the right panel. Since the orbital
velocity of the Earth around the Sun is much larger than the velocity
of a detector on the surface of the Earth, the virtual interferometers
build up a larger separation baseline on a shorter timescale. Hence
the overlap reduction
function goes to zero much more rapidly in this case and will only
recorrelate after $1$ year. 

\begin{figure*}[hbtp!]
\begin{center}
\includegraphics[angle=0, width=.48\textwidth]{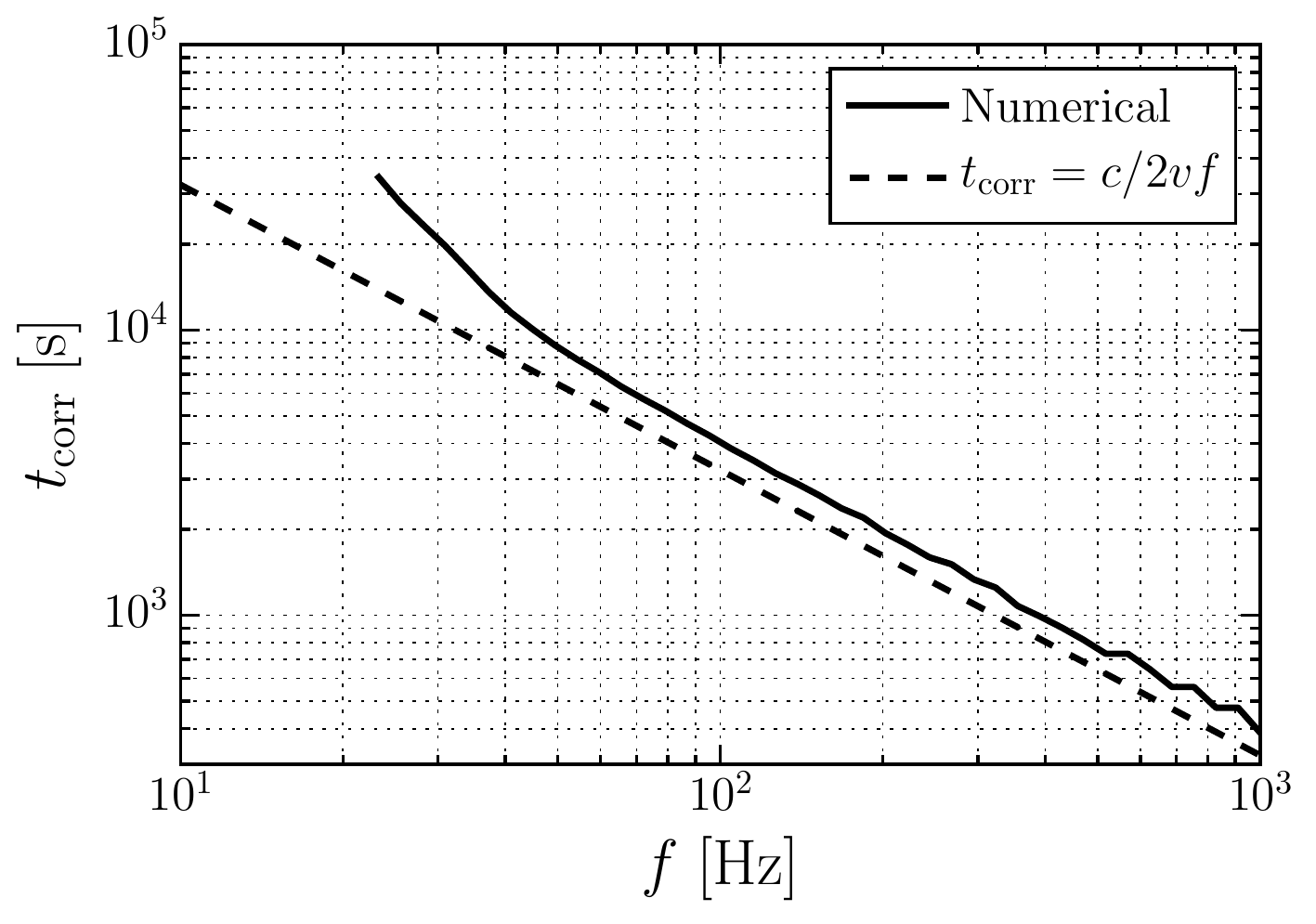}
\includegraphics[angle=0, width=.48\textwidth]{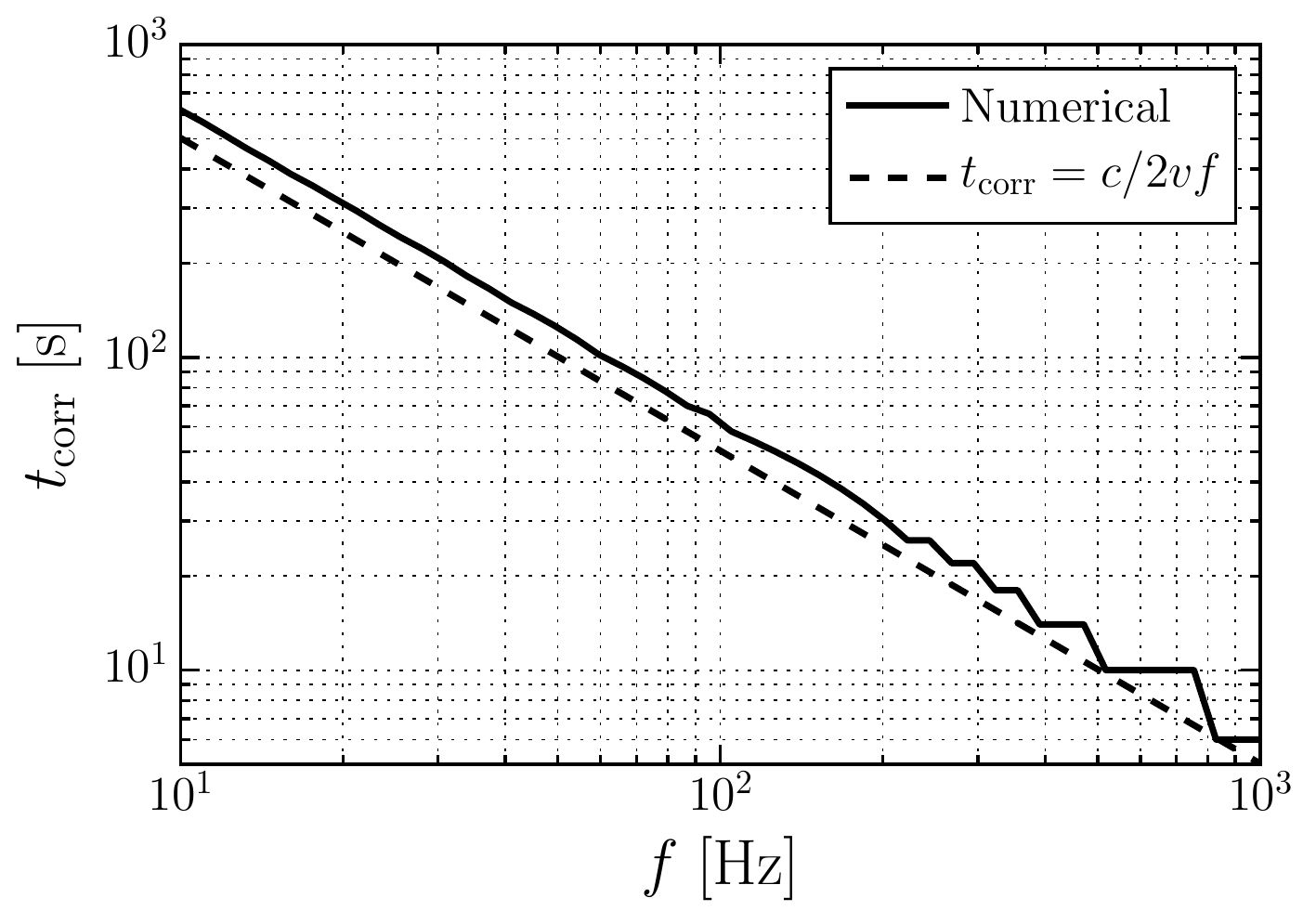}
\caption{Analysis of Fig.\ \ref{f:overlap_vs_time} repeated for
  $10\leq f \leq1000$ Hz. The plots show the first
  time the overlap reduction function between virtual interferometers
  goes to zero for daily rotation ({\it left panel}) and orbital motion
  ({\it right panel}). The solid line in the left panel does not
  extend fully to $10$ Hz since the overlap reduction function does
  not go to zero at low frequencies. A rough indication of when the overlap
  reduction function should go to zero is given by considering that a
  pair of detectors should be driven in coincidence by a passing 
  gravitational wave
  when the wavelength is more than twice the separation
  between the detectors. This defines a lower bound on the
  decorrelation timescale of the virtual interferometers, shown
  by a dashed line.}
\label{f:decorr-times}
\end{center}
\end{figure*}

We investigate this decorrelation
timescale by numerically computing the overlap reduction functions for daily
rotation and orbital motion at a variety of gravitational-wave frequencies. 
The times at which the detectors first decorrelate are shown in
Fig.~\ref{f:decorr-times}, with the lower bounds given by Eq.~(\ref{e:tcorr}). The
decorrelation timescale does indeed obey a simple $1/f$ scaling, and
at $100$ Hz the timescale for daily rotation and orbital motion are
actually $\sim 67~{\rm min}\ (=4020~{\rm s})$ and $60~{\rm s}$, respectively.
Therefore orbital motion of the Earth around the Sun
will rapidly synthesize a large network of independent virtual
interferometers from the motion of a single detector, with a resolving power 
that increases on a relatively short timescale.

\section{Map making}
\label{s:maps}

In this section, we extend the map-making formalism of
\cite{Gair-et-al:2014,Cornish-vanHaasteren:2014}
to data taken by 
a network of ground-based interferometers.
The key observation is that the time-dependent 
ground-based interferometer analysis can be mapped
to a static PTA-like analysis, for a set of
static virtual interferometers in a quasi-circular
ring 1~AU from the SSB.
Unlike the static PTA 
analysis~\cite{Gair-et-al:2014,Cornish-vanHaasteren:2014},
the virtual interferometers are not all centered 
at the same location, but see the sky from different 
locations due to the Earth's rotational and orbital motion. 
This allows for recovery of both the grad and curl 
components of the background, as discussed in Sec.~\ref{s:response} 
in terms of the response functions.
We shall demonstrate this explicitly via 
maximum-likelihood recovered sky maps in 
Sec.~\ref{s:simulations} below.

\subsection{Response vector}
\label{s:response-vector}

As described in Sec.~\ref{s:response}, the 
Fourier-domain response of detector $I$ to 
a gravitational-wave background 
is
\be
\begin{aligned}
\tilde r_I(f)
&=\int_{S^2} {\rm d}^2\Omega_{\hat k}\>\sum_A
R_I^A(f,\hat k)h_A(f,\hat k)
\quad{\rm or}
\\
\tilde r_I(f)
&=\sum_{(lm)}\sum_P
R^P_{I(lm)}(f)
a^P_{(lm)}(f)\,,
\end{aligned}
\ee
where the response functions 
$R^A_I(f,\hat k)$, $A=\{+,\times\}$ and 
$R^P_{I(lm)}(f)$, $P=\{G,C\}$ are given by
Eqs.~(\ref{e:R+Rx}) and (\ref{e:RGRC_general}).
We write this response abstractly as
\be
{\mb r} = {\mb R} {\mb h}\,,
\ee
where ${\mb h}$ denotes the components of the
gravitational-wave background in either the pixel
or spherical harmonic basis, and ${\mb R}$ denotes
the corresponding response function in that basis.
The response function ${\mb R}$ acts on ${\mb h}$ 
via a sum over polarizations
and an integration over the sky, or a sum over
polarizations and a sum over spherical harmonic 
components.

When performing the data analysis to produce 
maps of the gravitational-wave sky, we need to 
discretize both the map (in terms of pixels
or spherical harmonic components) and the observed 
data.
This leads to a time-frequency decomposition where
the data are broken up into segments of duration $\tau$, 
which should be short compared to the 
timescale over which the orientation of the detectors 
changes appreciably. Since the peak sensitivity of the advanced
ground-based interferometers is $\sim\!100$ Hz, we take the minimum
segment duration to be the time required for a detector to
decorrelate from itself under orbital motion, thus synthesizing an
independent virtual detector (e.g., $\tau \approx 60~{\rm s}$). The
longest segment duration is the time beyond which the Earth's rotation
will have appreciably changed the antenna response pattern orientation
(e.g., $\tau \approx 2048~{\rm s}$).
Each segment of data is then discrete Fourier 
transformed, yielding a finite number of components
for the vectors ${\mb r}$ and ${\mb h}$.
In the following, we will denote the discrete (positive)
frequencies by $f_j$, where $j=1,2,\cdots,N_f$; 
the sky pixels by $\hat k_n$, where $n=1,2,\cdots,N_{\rm pix}$;
and the spherical harmonic components of the sky by $(lm)$, 
where $l=0,1,\cdots,l_{\rm max}$, and $-l\leq m\leq l$.
The detectors 
(interferometers) are labelled
by the index $I=1,2,\cdots, N_d$, and the time
segments recorded by detector $I$ as $t_{Ii}$, 
where $i=1,2,\cdots, N_{I}$.
Combining the responses from all detectors, we have
\be
\begin{aligned}
{\mb r}\equiv 
&\{\tilde r_{Ii}(f_j)\}\,,
\\
{\mb h}\equiv
&\{h_+(f_j,\hat k_n),h_\times(f_j,\hat k_n)\}
{\rm\ or\ }
\{a^G_{(lm)}(f_j),a^C_{(lm)}(f_j)\}\,,
\\
{\mb R}\equiv
&\{R_{Ii}^+(f_j,\hat k_n),R_{Ii}^\times(f_j,\hat k_n)\}
\\
&\hspace{.75in}
{\rm\ or\ }\
\{R_{Ii(lm)}^G(f_j),R_{Ii(lm)}^C(f_j)\}\,.
\end{aligned}
\ee
Written this way, the time-dependent ground-based 
interferometer analysis is mapped to a static 
PTA-like analysis, for a set of virtual interferometers 
synthesized by the Earth's rotational and orbital motion 
around the Sun.
The response vector ${\mb  r}$ has 
$N \equiv N_f\sum_I N_{I}$ complex components.
In contrast to a cross-correlation analysis, we do not require 
that all the detectors have data for the same time periods.
When the number of time segments $N_t$ is the same for all 
$N_d$ detectors, then $N = N_t N_d N_f$.
Similarly, ${\mb h}$ is a complex-valued vector of dimension
$M\equiv 2N_{\rm pix} N_f$ (pixel basis) or $M\equiv 2N_m N_f$ (spherical harmonic basis),
where $N_m = (l_{\rm max}+1)^2 - 4$ is the number of spherical
harmonic $(lm)$ modes out to $l_{\rm max}$. 
(The $-4$ in the last expression is because summations over 
$l$ start at $l=2$).
The response function  ${\mb R}$ is thus represented by 
a complex-valued matrix of dimension $N\times M$.
Since the frequency components of ${\mb r}$ and ${\mb h}$ 
are identical, the frequency transformation part of 
${\mb R}$ is simply the identity matrix $\mathds{1}_{N_f\times N_f}$.

To simplify the discussion for the remainder of this 
section, we will work in the pixel basis.
The subsequent calculations are formally identical in both bases, 
and the resulting sky maps are effectively the  same, 
provided $l_{\rm max}$ is chosen so the the total 
number of modes $N_m$ in the spherical
harmonic basis is of the same order as the number of 
pixels $N_{\rm pix}$.

\subsection{Maximum-likelihood estimation}
\label{s:ML-estimation}

Using the above notation, the measured data can be represented 
by an $N$-dimensional complex vector ${\mb d}\equiv\{\tilde d_{Ii}(f_j)\}$,
with contributions, in general, from both the 
gravitational-wave signal ${\mb r}$ and detector noise ${\mb n}$:
\be
{\mb d} 
= {\mb r}+{\mb n}
= {\mb R}{\mb h} + {\mb n}\,.
\label{e:data}
\ee
If we assume that the noise is Gaussian-stationary, 
then we can represent it by an $N\times N$ (Hermitian, positive definite) 
covariance matrix ${\mb C}$, whose components 
are given by 
\be
C_{Iij,I'i'j'}\equiv
\langle \tilde n_{Ii}(f_j)\tilde n^*_{I'i'}(f_{j'})\rangle 
= \frac{1}{2\delta f}\delta_{ii'}\delta_{jj'}C_{II'}(f_j)\,,
\ee
where $\delta f$ is the frequency resolution.
(Past analyses for stochastic backgrounds using 
ground-based interferometers have typically used
$\delta f=0.25~{\rm Hz}$, which is much greater than the
$1/\tau\sim 0.001~{\rm Hz}$ frequency resolution 
associated with the duration of the short-term Fourier 
transform, see e.g.,~\cite{Thrane-et-al:2009, S1Isotropic}.
This amounts to working with a {\em coarse-grained} 
frequency series, obtained by averaging over neighboring 
frequency bins.)
If we further assume that the noise is uncorrelated between 
different detectors, then
\be
C_{II'}(f) = \delta_{II'}S_I(f)\,,
\ee
where $S_{I}(f)$ is the (one-sided) power spectral 
density of the noise in detector $I$.
In terms of these quantities, the likelihood function for 
the data is
\be
p({\mb d}|{\mb C},{\mb h}) 
\propto
\exp\left[-
({\mb d}-{\mb R}{\mb h})^\dagger
{\mb C}^{-1}
({\mb d}-{\mb R}{\mb h})\right]\,,
\label{e:likelihood}
\ee
which is a multivariate Gaussian distribution for 
the noise.
Note that there is no factor of $1/2$ in the exponential as 
the matrix sum is over only positive-frequency components.
Given the likelihood function, we can now use either
Bayesian inference or frequentist maximum-likehood statistics
to estimate the model parameters.
The latter is relatively simple to do if we fix the noise, 
since the signal parameters enter linearly in the 
likelihood in Eq. (\ref{e:likelihood}).

Maximizing the likelihood with respect to ${\mb h}$ leads to
\be
{\mb h}_{\rm ML}
=({\mb R}^\dagger{\mb C}^{-1}{\mb R})^{-1}
{\mb R}^\dagger{\mb C}^{-1}{\mb d}
\label{e:hML}
\ee
for the recovered map.
This is only a formal expression, however, 
since the {\em Fisher matrix}, 
\be
{\mb F}\equiv 
{\mb R}^\dagger{\mb C}^{-1}{\mb R}\,,
\label{e:Fisher}
\ee
is not invertible in general, since ${\mb R}$ may have
not have full column rank. This occurs if the 
number of data points $N$ is less than the number of
modes $M$ that we are trying to recover, or if
the response matrix has null (or nearly null) 
directions---i.e., 
particular gravitational-wave skies have ${\mb h}_{\rm null}$
to which the network of detectors is effectively blind. This is discussed further 
in Sec. \ref{s:basis_vectors}.
Thus, calculating ${\mb h}_{\rm ML}$ will, in general, 
require some form of regularization~\cite{Press92a}.

To do the regularization, it is simplest to work with
the whitened data $\bar{\mb d}\equiv {\mb L}^\dagger{\mb d}$ 
and whitened response matrix
$\bar{\mb R}\equiv {\mb L}^\dagger{\mb R}$,
where ${\mb L}$ is a lower triangular matrix defined by the
Cholesky decomposition of the inverse covariance matrix,
${\mb C}^{-1} = {\mb L}{\mb L}^\dagger$.
(An alternative approach, based on the unwhitened 
response matrix ${\mb R}$, is described in App.~\ref{s:whitening}.)
In terms of the whitened quantities, we have 
${\mb F}=\bar{\mb R}^\dagger \bar{\mb R}$ and 
\be
{\mb h}_{\rm ML}
=(\bar{\mb R}^\dagger\bar{\mb R})^{-1}
\bar{\mb R}^\dagger\bar{\mb d}
\equiv \bar{\mb R}^+\bar{\mb d}\,,
\label{e:hML_Rbar}
\ee
where
\be
\bar{\mb R}^+
\equiv
(\bar{\mb R}^\dagger\bar{\mb R})^{-1}\bar{\mb R}^\dagger
\ee
is the so-called {\em pseudo-inverse} of $\bar{\mb R}$.
As before, this is just a formal expression as the $M\times M$ matrix
$\bar{\mb R}^\dagger\bar{\mb R}$ is not invertible in general.
However, it is {\em always} possible to define the 
pseudo-inverse $\bar{\mb R}^+$ in terms of the 
singular value decomposition (SVD) of $\bar{\mb R}$:
\be
\bar{\mb R} = \bar{\mb U}\bar{\bs\Sigma}\bar{\mb V}^\dagger\,,
\label{e:svd(Rbar)}
\ee
where $\bar{\mb U}$ and $\bar{\mb V}$ are $N\times N$ and 
$M\times M$ unitary 
matrices, and $\bar{\bs\Sigma}$ is an $N\times M$ 
rectangular matrix with (real, non-negative) singular 
values $\bar\sigma_k$ along the diagonal, and zeros 
everywhere else.
(Without loss of generality, we can assume that
the singular values are arranged from largest to 
smallest along the diagonal.)
Then
\be
\bar{\mb R}^+ = \bar{\mb V}\bar{\bs\Sigma}^+\bar{\mb U}^\dagger\,,
\label{e:Rbar+}
\ee
where $\bar{\bs\Sigma}^+$ is defined by taking the reciprocal 
of each non-zero singular value of $\bar{\bs\Sigma}$, leaving 
the zeros in place, and then transposing the matrix.
In terms of the SVD of $\bar{\mb R}$,
the maximum-likelihood estimator can be written as
\be
{\mb h}_{\rm ML} = 
\bar{\mb R}^+\bar{\mb d}
=\bar{\mb V}\bar{\bs\Sigma}^+\bar{\mb U}^\dagger
\bar{\mb d}\,.
\label{e:hML_svd}
\ee
The expected value and variance of ${\mb h}_{\rm ML}$ 
are given by
\be
\begin{aligned}
\langle{\mb h}_{\rm ML}\rangle 
&=\bar{\mb R}^+\bar{\mb R}{\mb h}\,,
\\
var({\mb h}_{\rm ML})
&\equiv
\langle{\mb h}_{\rm ML}{\mb h}_{\rm ML}^\dagger\rangle
-\langle{\mb h}_{\rm ML}\rangle
\langle{\mb h}_{\rm ML}^\dagger\rangle
= \bar{\mb R}^+(\bar{\mb R}^+)^\dagger\,,
\label{e:var2}
\end{aligned}
\ee
where the expression for the variance assumes 
that the gravitational-wave signal is {\em weak}
compared to the detector noise.
Although not explicit in the last expression, 
the variance of ${\mb h}_{\rm ML}$
{\em does} depend on the noise ${\mb C}$, since
$\bar{\mb R}={\mb L}^\dagger {\mb R}$ and
${\mb C}^{-1}={\mb L}{\mb L}^\dagger$.

If the non-zero singular values of $\bar{\bs\Sigma}$ vary over 
several orders of magnitude, it is usually necessary to set 
to zero (by hand) all non-zero singular values less than or 
equal to some minimum value $\bar\sigma_{\rm min}$
(e.g., $10^{-5}$ times the largest non-zero singular value).
This reduces the noise in the maximum-likelihood 
reconstruction, which 
is dominated by those modes that we are least sensitive to.
So in what follows, when we speak of non-zero singular values of 
$\bar{\bs\Sigma}$, we will actually mean the singular values 
$\bar\sigma_k$ satisfying $\bar\sigma_k \ge \bar\sigma_{\rm min}$.

\subsection{Sky map basis vectors}
\label{s:basis_vectors}

Expression (\ref{e:hML_svd}) for the maximum-likelihood
estimate has several nice 
geometrical properties~\cite{Cornish-vanHaasteren:2014}.
In particular, the columns of $\bar{\mb U}$ and $\bar{\mb V}$ 
corresponding to the non-zero singular values of $\bar{\bs\Sigma}$ 
have the interpretation of {\em response range vectors} 
and {\em sky map basis vectors}
respectively, in terms of which 
the measured response $\bar{\mb R}{\mb h}$
and the maximum-likelihood estimate
${\mb h}_{\rm ML}$ can be written as linear combinations.
To see this, let $\bar{\mb u}_{(k)}$ and $\bar{\mb v}_{(k)}$ 
denote the $k$th columns of $\bar{\mb U}$ and $\bar{\mb V}$, and 
let $r\le {\rm min}(N,M)$ be the number of non-zero singular 
values of $\bar{\bs\Sigma}$.
Then it follows from Eqs.~(\ref{e:svd(Rbar)}) 
and (\ref{e:hML_svd}) that
\be
\begin{aligned}
&\bar{\mb R}{\mb h} = \sum_{k=1}^{r} 
\bar\sigma_k (\bar{\mb v}_{(k)}\cdot{\mb h})\bar{\mb u}_{(k)}\,,
\\
&{\mb h}_{\rm ML} = \sum_{k=1}^r 
\bar\sigma_k^{-1} (\bar{\mb u}_{(k)}\cdot \bar{\mb d}) 
\bar{\mb v}_{(k)}\,,
\end{aligned}
\ee
where dot product of two vectors ${\mb a}$ 
and ${\mb b}$ is defined by 
${\mb a}\cdot {\mb b}\equiv{\mb a}^\dagger{\mb b}$.
If we expand $\bar{\mb d}=\bar{\mb R}{\mb h}+\bar{\mb n}$, 
then we can also write 
\be
{\mb h}_{\rm ML}
=\sum_{k=1}^r 
(\bar{\mb v}_{(k)}\cdot {\mb h}) \bar{\mb v}_{(k)}
+\bar{\mb R}^+\bar{\mb n}\,.
\label{e:hML_projection}
\ee
Note that this last expression for ${\mb h}_{\rm ML}$
involves the projections of ${\mb h}$ onto 
$\bar{\mb v}_{(k)}$ for only the non-zero singular 
values of $\bar{\bs\Sigma}$.

It is important to discuss in some detail those cases 
where there are fewer data points than modes we are trying 
to recover (i.e., $N<M$), or if there are certain modes of
the gravitational-wave background that our response matrix
is insensitive to.
For either of these two cases, the system of equations
$\bar{\mb d}=\bar{\mb R}{\mb h}$ 
is {\em under-determined}, 
which implies that there exist {\em multiple} solutions 
for the recovered gravitational-wave background:
\be
{\mb h}
=\bar{\mb R}^+\bar{\mb d}
+(\mathds{1}_{M\times M}-
\bar{\mb R}^+\bar{\mb R}){\mb h}_{\rm arb}\,,
\ee
where ${\mb h}_{\rm arb}$ represents an 
{\em arbitrary} gravitational-wave background.
The particular solution that we have chosen for 
${\mb h}_{\rm ML}$ (given by Eq.~(\ref{e:hML_svd})
or (\ref{e:hML_projection})) ignores the
term proportional to ${\mb h}_{\rm arb}$,
setting to zero those modes that we are insensitive to.
Our solution also sets to zero the variance of 
these modes, as can be 
seen from the expression for $var({\mb h}_{\rm ML})$ 
given in Eq.~(\ref{e:var2}):
\be
var({\mb h}_{\rm ML}) 
= \bar{\mb R}^+(\bar{\mb R}^+)^\dagger
= \bar{\mb V}\bar{\bs\Sigma}^+(\bar{\bs\Sigma}^+)^\dagger \bar{\mb V}^\dagger\,,
\ee
which can be diagonalized by a similarity transformation 
involving $\bar{\mb V}^\dagger$.
This yields 
$\bar{\bs\Sigma}^+(\bar{\bs\Sigma}^+)^\dagger$, which
has $M-r$ zeros along its diagonal.

In a Bayesian formulation of the problem, things will be
different, however, as one must also specify {\em prior} 
probability distributions for the signal parameters, in 
addition to the likelihood function (\ref{e:likelihood}).
For a signal parameter (or a combination of signal 
parameters) corresponding to a mode of the background that 
the detectors are insensitive to, the marginalized posterior 
will simply recover the prior distribution on this parameter 
(or combination of parameters), since the data are 
completely uniformative about this mode.
This is more in line with what we would expect for a mode 
that is unconstrained by the data, but such an analysis 
requires the specification of prior probability distributions 
which frequentist estimators, like ${\mb h}_{\rm ML}$, do not provide.
We therefore choose to construct our maximum-likelihood estimator
such that it sets the modes that we are insensitive to equal 
to zero, and acknowledge the fact that we cannot say anything 
about them with our experiment.

\subsection{Sky maps, uncertainty maps, and SNRs}
\label{s:skymaps}

To construct sky maps from the maximum-likelihood estimator
${\mb h}_{\rm ML}$, we need to either restrict attention to a
particular discrete frequency $f_j$ or perform an
average over the different frequency components.
In either case, the dimensionality of ${\mb h}_{\rm ML}$ 
reduces to $2N_{\rm pix}$, 
corresponding to the $+$ and $\times$ 
components of the estimated gravitational-wave background at 
each pixel on the sky.
Uncertainty maps for these estimates are given by the square-root 
of the diagonal elements of the variance estimate given in 
(\ref{e:var2}),
\be
{\bs\sigma}_{\rm ML} =\sqrt{{\rm diag}\left[var({\mb h}_{\rm ML})\right]}\,.
\label{e:sigma_map}
\ee
Similarly, we can construct signal-to-noise ratio (SNR) 
maps by simply dividing the 
estimates of $h_+$ and $h_\times$ at each pixel on the sky by
the corresponding values of ${\bs\sigma}_{\rm ML}$.
Examples of such maps are given in Sec.~\ref{s:simulations}.


\begin{table*}[htpb!]
\begin{tabular}{l c c c c}
\hline
Interferometer & Longitude & Latitude & Orientation
& Spectral density \\
\hline
LIGO, Hanford & $119^\circ$ $24'$ $27.6''$ W & $46^\circ$ $27'$ $18.5''$ N & $279.0^\circ$ & $1.591\times 10^{-47}~{\rm Hz}^{-1}$\\
LIGO, Livingston & $90^\circ$ $46'$ $27.3''$ W & $30^\circ$ $33'$ $46.4''$ N & $208.0^\circ$ & $1.591\times 10^{-47}~{\rm Hz}^{-1}$\\
Virgo, Italy & $10^\circ$ $30'$ $16''$ E & $43^\circ$ $37'$ $53''$ N & $333.5^\circ$ & $2.063\times 10^{-47}~{\rm Hz}^{-1}$\\
KAGRA, Japan & $137^\circ$ $10'$ $48''$ E & $36^\circ$ $15'$ $00''$ N & $20.0^\circ$ & $9.320\times 10^{-48}~{\rm Hz}^{-1}$\\
INDIGO, India & $74^\circ$ $02'$ $59''$ E & $19^\circ$ $05'$ $47$ N & $270.0^\circ$ & $1.591\times 10^{-47}~{\rm Hz}^{-1}$\\
AIGO, Australia & $115^\circ$ $42'$ $51''$ E & $31^\circ$ $21'$ $29''$ S & $45.0^\circ$ & $1.591\times 10^{-47}~{\rm Hz}^{-1}$\\
\hline
\end{tabular}
\caption{Geographic information for ground-based 
interferometers used in our simulations, adapted from \cite{schutz:2011}.
Orientation is the angle that the bisector of the two interferometer 
arms makes with geographic North (positive for directions pointing East of North).
All interferometers are assumed to have $90^\circ$ opening angle
between the two arms.
Spectral density is the value of the one-sided noise power spectrum 
$S_n(f)$ for the corresponding interferometer, evaluated at $f=100~{\rm Hz}$.
The values of $S_n(f)$ for LIGO, Virgo, and KAGRA are taken from design sensitivity
documents and publicly accessible data \citep{aligo-noise-curve,advirgo-noise-curve,kagra-noise-curve}; 
the values for INDIGO and AIGO are taken to be the same as those for 
the LIGO interferometers, as they are in the initial planning stages.}
\label{t:ifos}
\end{table*}

\subsection{Simulations}
\label{s:simulations}

We now illustrate the mapping procedure described above by
constructing maximum-likelihood estimates of the real 
and imaginary parts of $h_+(f,\hat k)$ and $h_\times(f,\hat k)$
for three different simulated gravitational-wave 
backgrounds: a point source and a spatially-extended 
backgrounds having only gradient or curl modes.
For simplicity, we consider only a single frequency
component $f=100~{\rm Hz}$, and we pixelize the sky using
a HEALPix~\cite{HEALPix} grid containing $N_{\rm pix}=768$ pixels.
(The sky map vectors ${\mb h}_{\rm ML}$ and ${\mb h}_{\rm inj}$
thus have dimension $M=2N_{\rm pix}=1536$.)
We will work primarily with a network of $N_d=6$ detectors,
comprising both the existing and planned large-scale,
ground-based 
laser interferometers LIGO-Hanford, LIGO-Livingston, 
Virgo, KAGRA, INDIGO, and AIGO.
(Relevant information for each interferometer is given in
Table~\ref{t:ifos}, which is adapted from \cite{schutz:2011}.)
For comparison, we will also consider a reduced network
having just $N_d=3$~detectors 
(LIGO-Hanford, LIGO-Livingston, and Virgo), 
which is more realistic for the near future.
The measured data will be given in the 
frequency domain, corresponding to short-term Fourier
transforms of time segments of duration 
$\tau=(1~{\rm sidereal\ day})/60\approx 1436~{\rm s}$.
The simulations for the 6-detector network will have 
a total of $N_t=400$ samples for each interferometer, 
corresponding to 6.67~days of simulated data.
The simulations for the 3-detector network will have
either $N_t=400$ or $N_t=800$ samples for each interferometer,
corresponding to either 6.67~days or 13.33~days of 
simulated data.
The data and response vectors ${\mb d}$ and ${\mb r}$ 
will thus have dimensions $N=N_dN_t=2400$ 
for the 6-detector network, and $N=1200$ or 2400 
for the two 3-detector networks.

The detector noise will be described by an $N\times N$ 
block-diagonal covariance matrix, whose $N_d$ 
blocks (corresponding to the $N_d$ detectors in 
the network) are each proportional to the unit matrix 
$\mathds{1}_{N_t\times N_t}$.
The proportionality constants are the values of the 
one-sided power spectral densities $S_I(f)$,
$I=1,2,\cdots, N_d$ evaluated at $f=100~{\rm Hz}$
(see the last column of Table~\ref{t:ifos}) divided
by $4\delta f$, where $\delta f$ is the size of the
frequency bins.
The factor of 4 is due to the use of one-sided power
spectral densities (one factor of 2) and the summation
over only positive-frequency bins (the other factor of 2).
For our simulations, we take $\delta f=0.25~{\rm Hz}$,
as is common for stochastic background searches using
ground-based interferometers~\cite{S1Isotropic}.
The real and imaginary parts of the noise vector 
${\mb n}$ are generated by randomly drawing independent 
samples from a multivariate Gaussian distribution defined
by this block-diagonal matrix.

\begin{figure*}[hbtp!]
\begin{center}
\includegraphics[trim=3cm 4cm 3cm 2.5cm, clip=true, angle=0, width=0.24\textwidth]{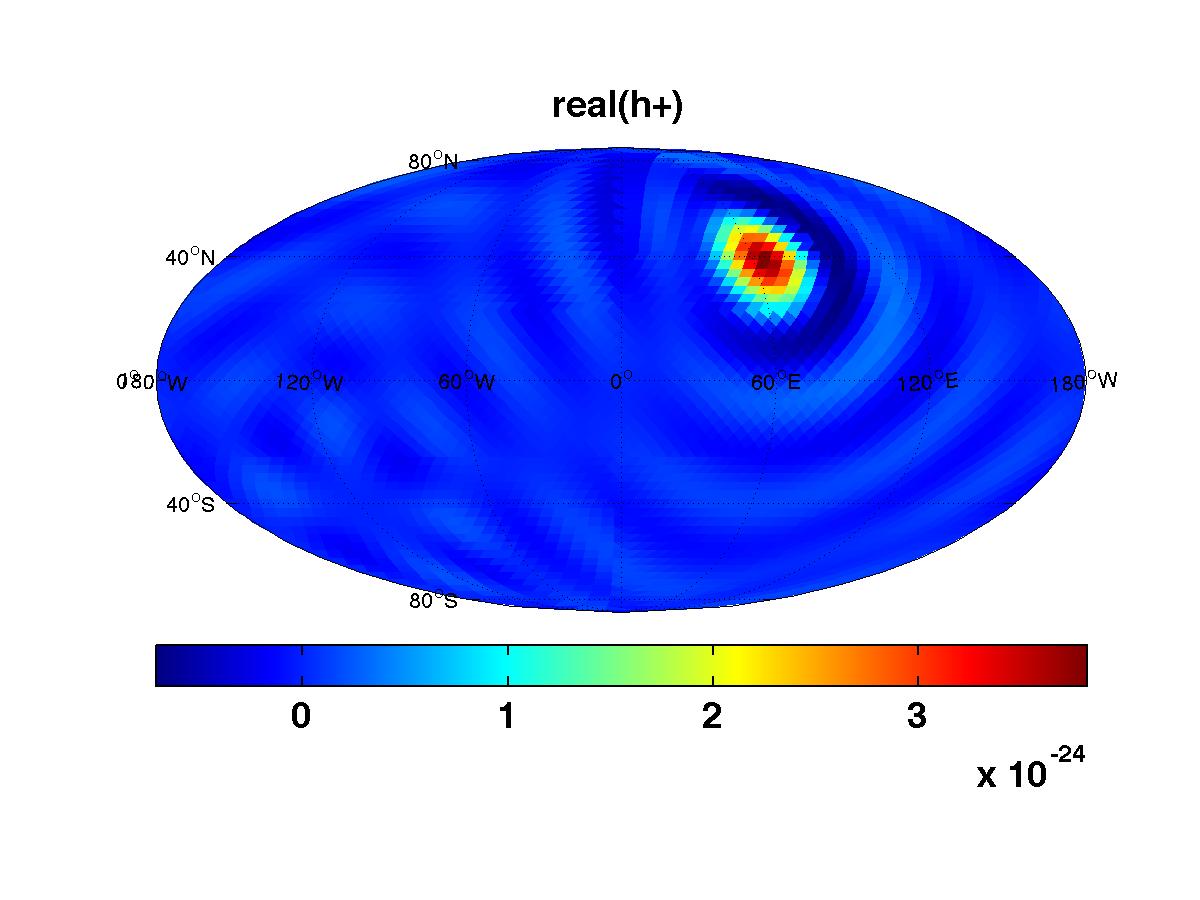}
\includegraphics[trim=3cm 4cm 3cm 2.5cm, clip=true, angle=0, width=0.24\textwidth]{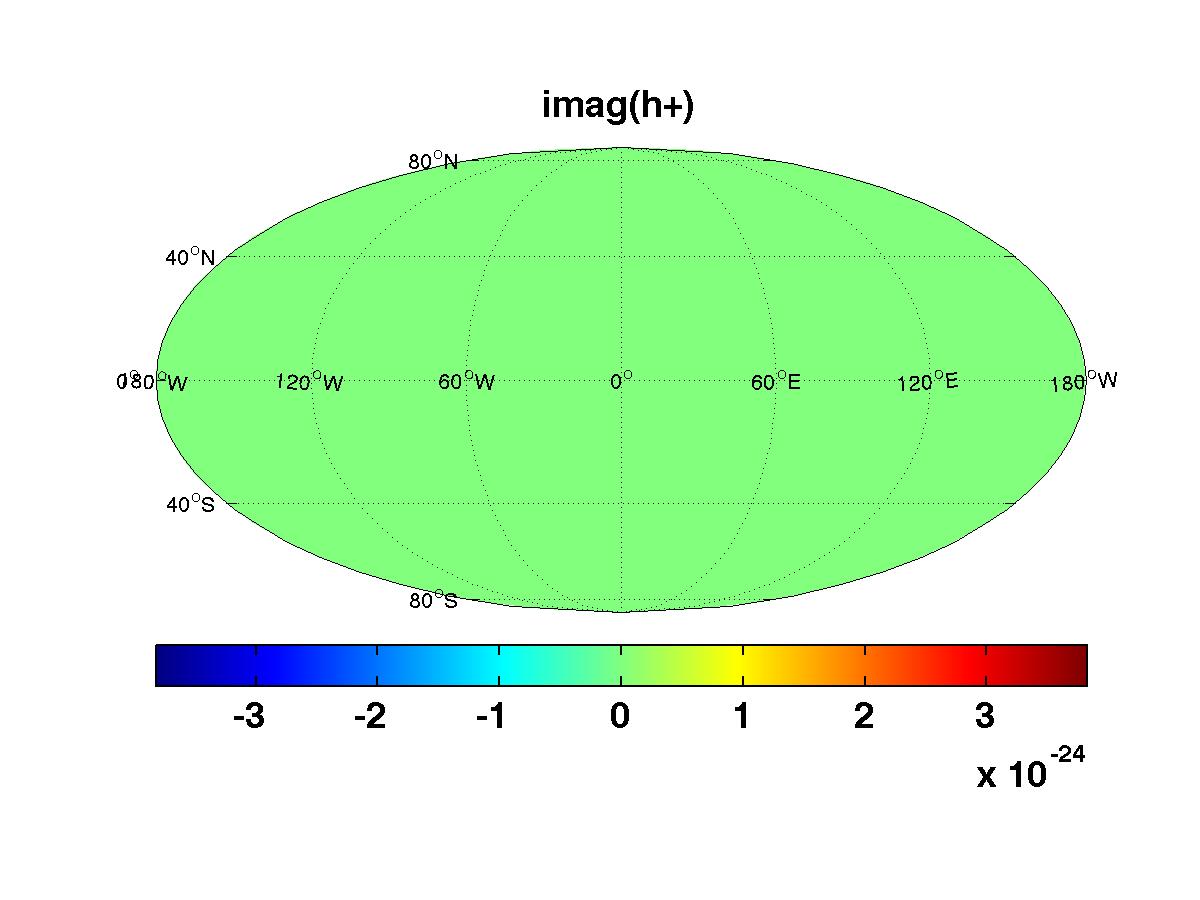}
\includegraphics[trim=3cm 4cm 3cm 2.5cm, clip=true, angle=0, width=0.24\textwidth]{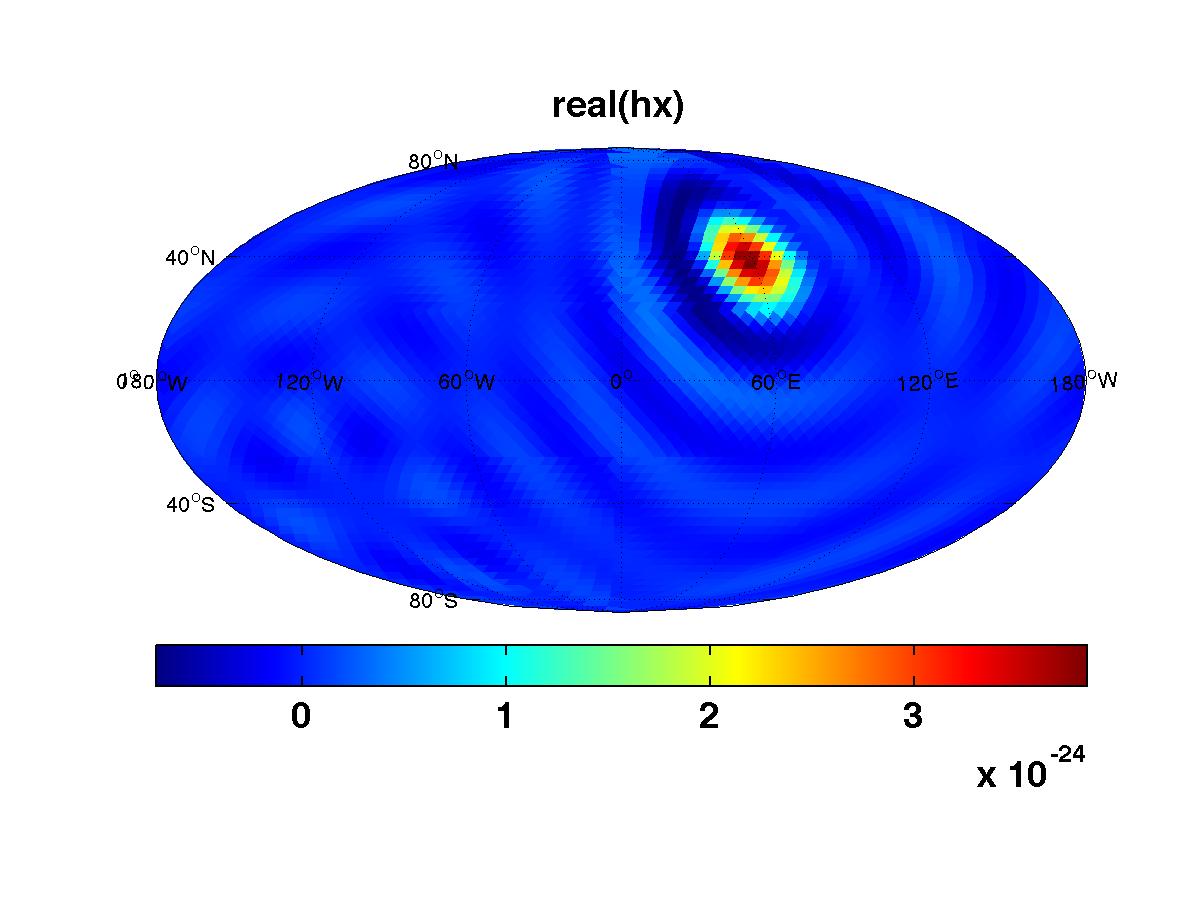}
\includegraphics[trim=3cm 4cm 3cm 2.5cm, clip=true, angle=0, width=0.24\textwidth]{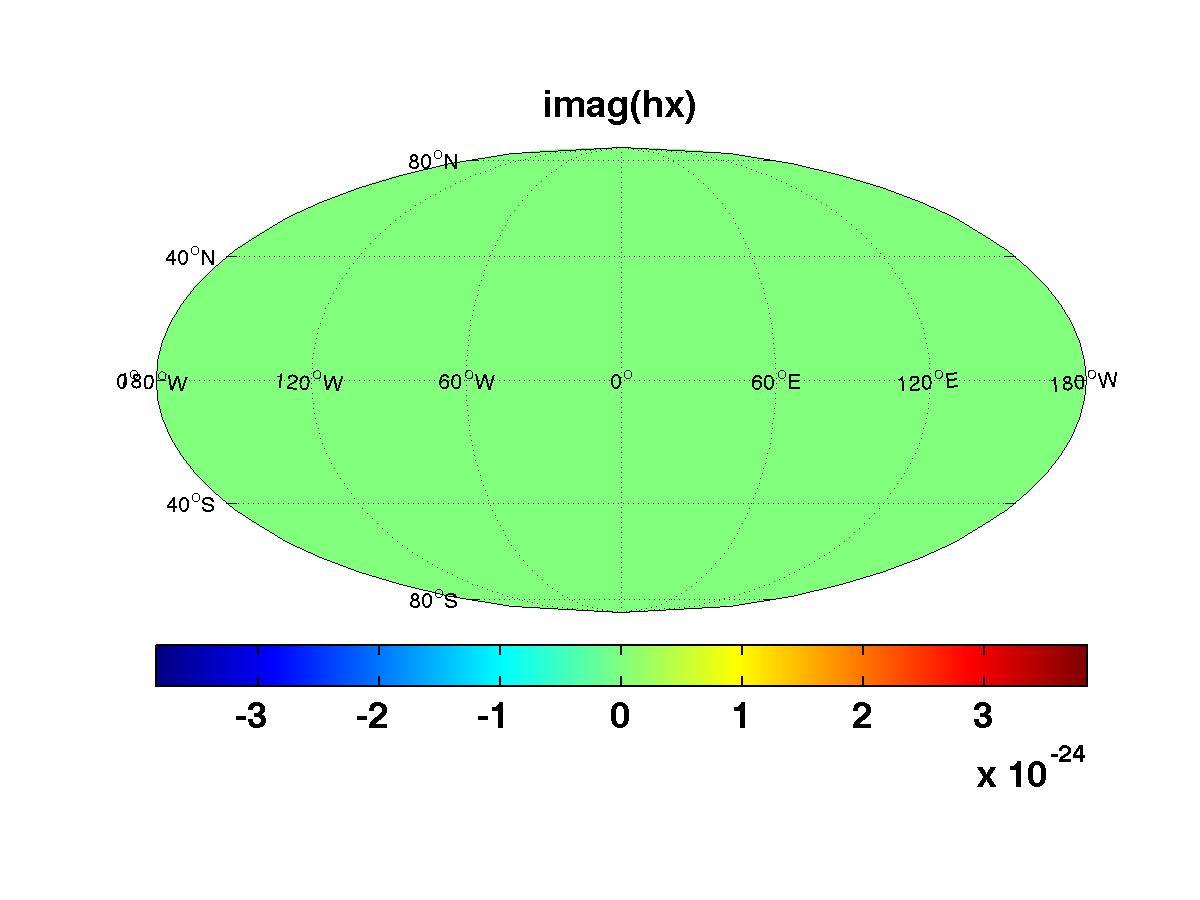}
\includegraphics[trim=3cm 4cm 3cm 2.5cm, clip=true, angle=0, width=0.24\textwidth]{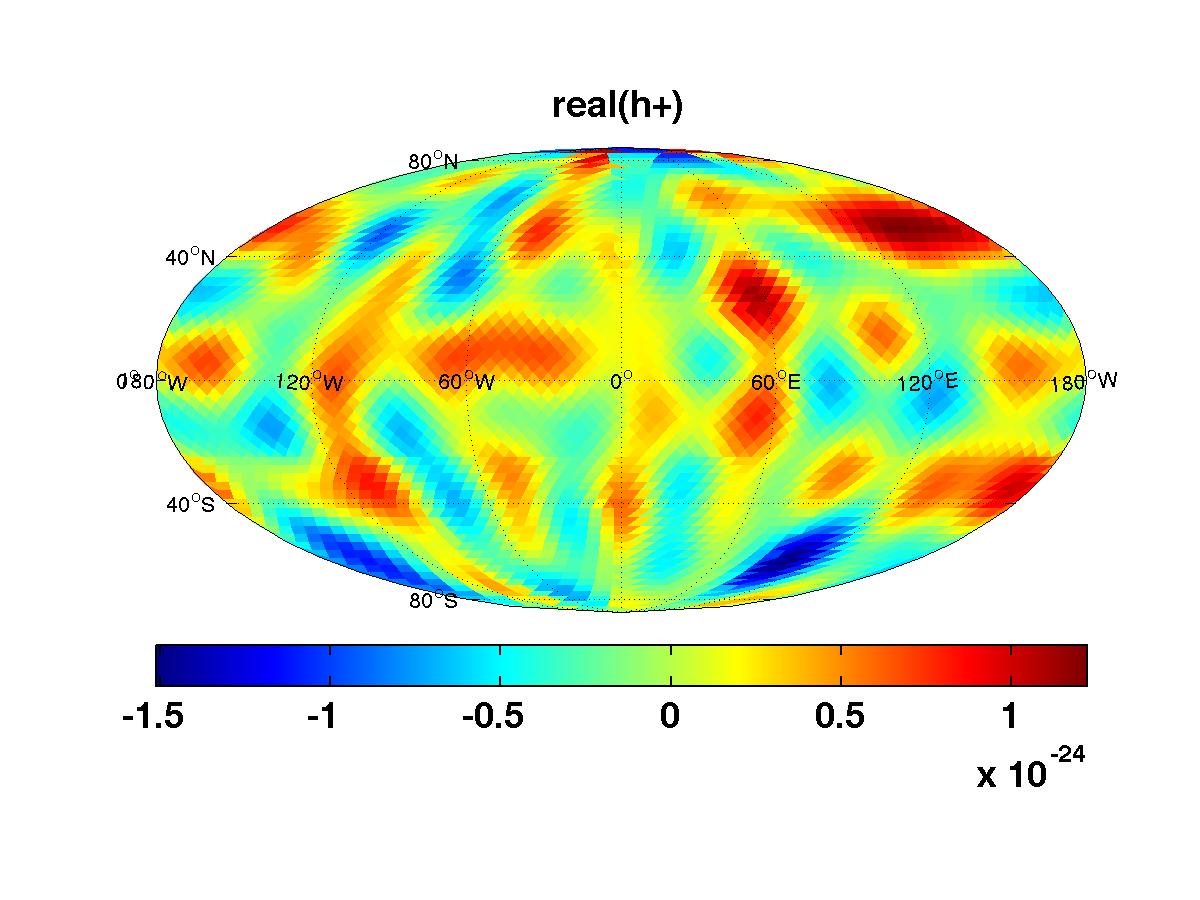}
\includegraphics[trim=3cm 4cm 3cm 2.5cm, clip=true, angle=0, width=0.24\textwidth]{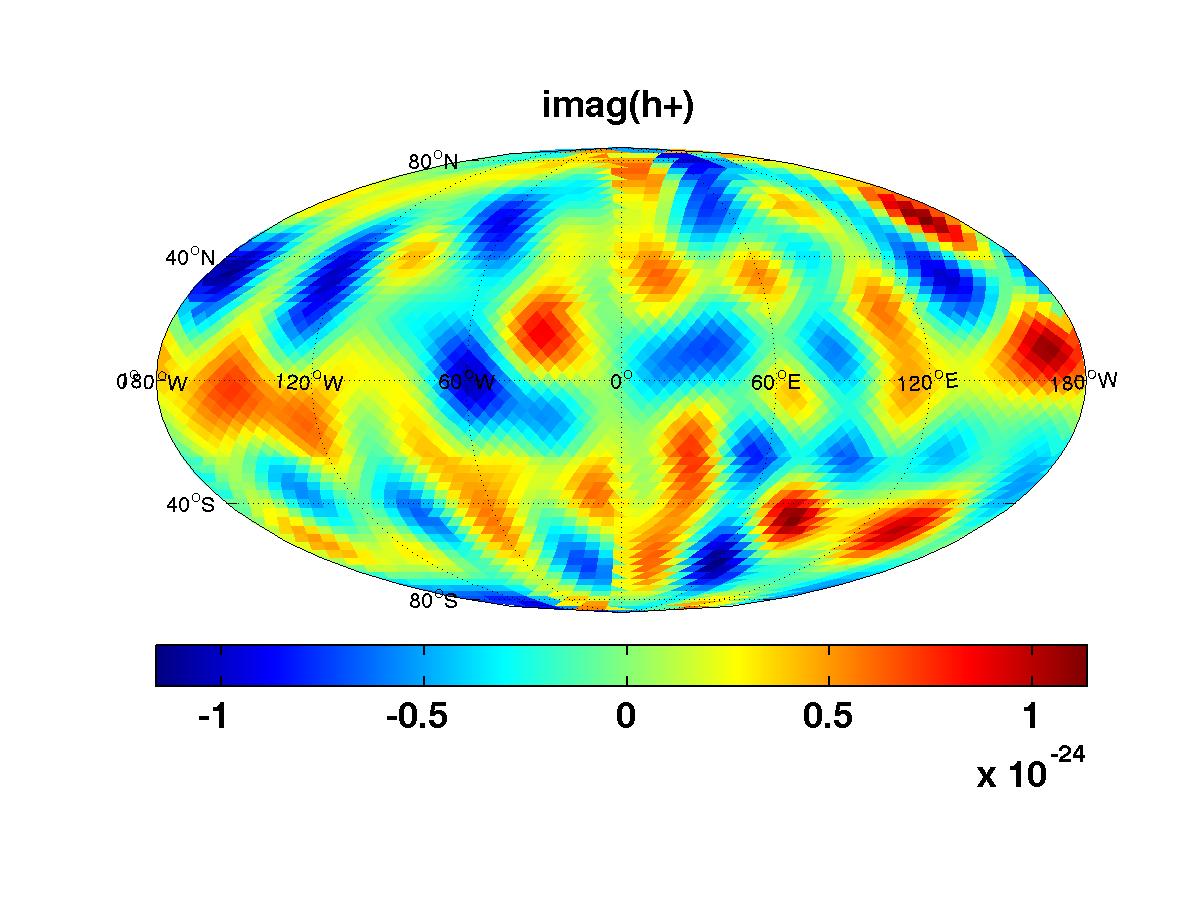}
\includegraphics[trim=3cm 4cm 3cm 2.5cm, clip=true, angle=0, width=0.24\textwidth]{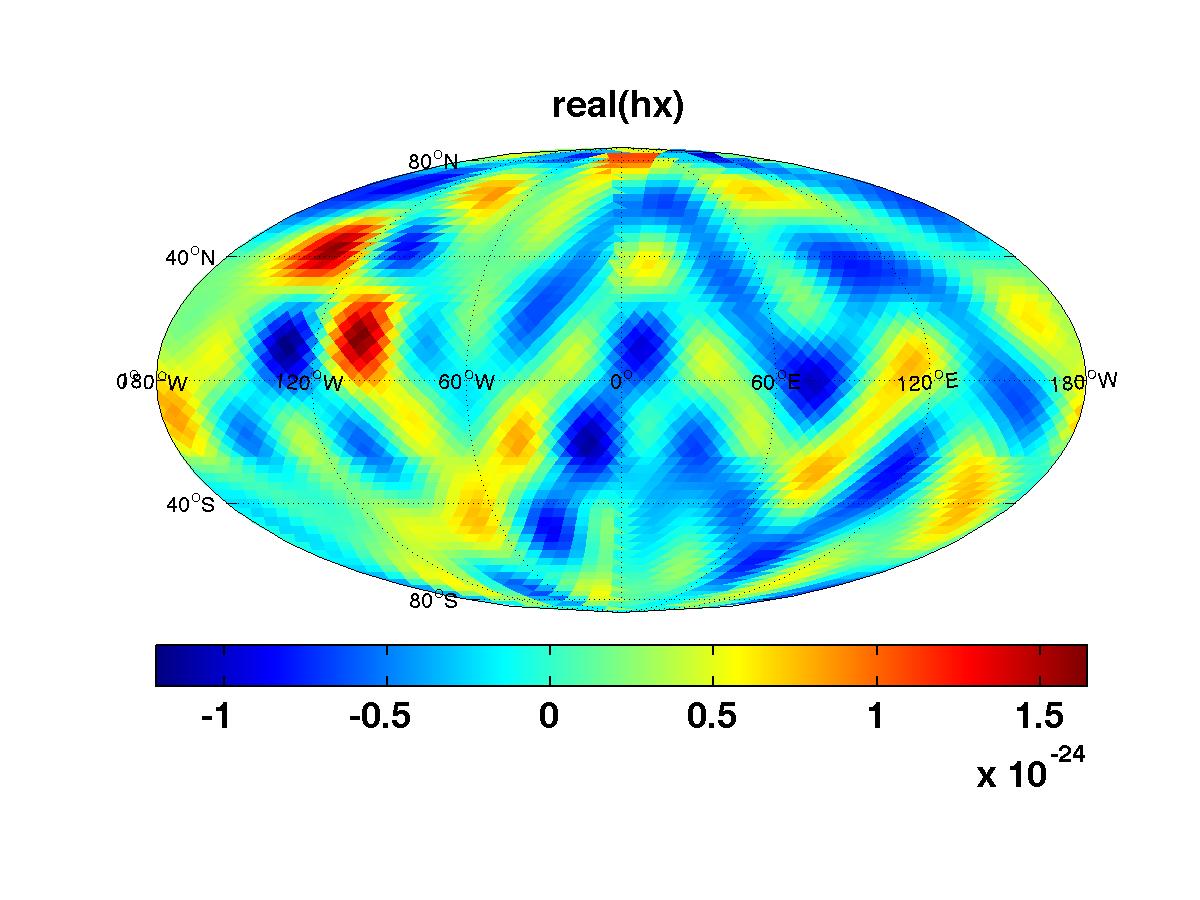}
\includegraphics[trim=3cm 4cm 3cm 2.5cm, clip=true, angle=0, width=0.24\textwidth]{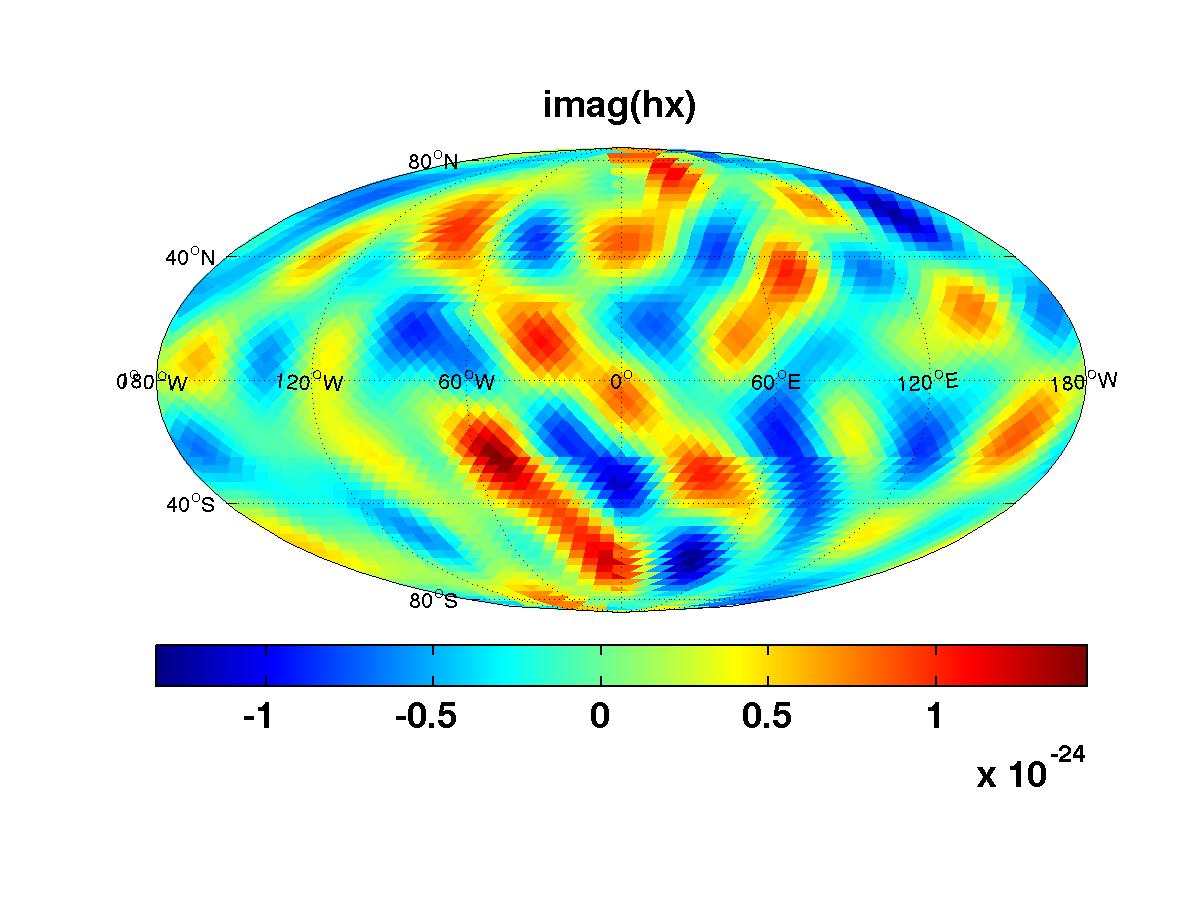}
\includegraphics[trim=3cm 4cm 3cm 2.5cm, clip=true, angle=0, width=0.24\textwidth]{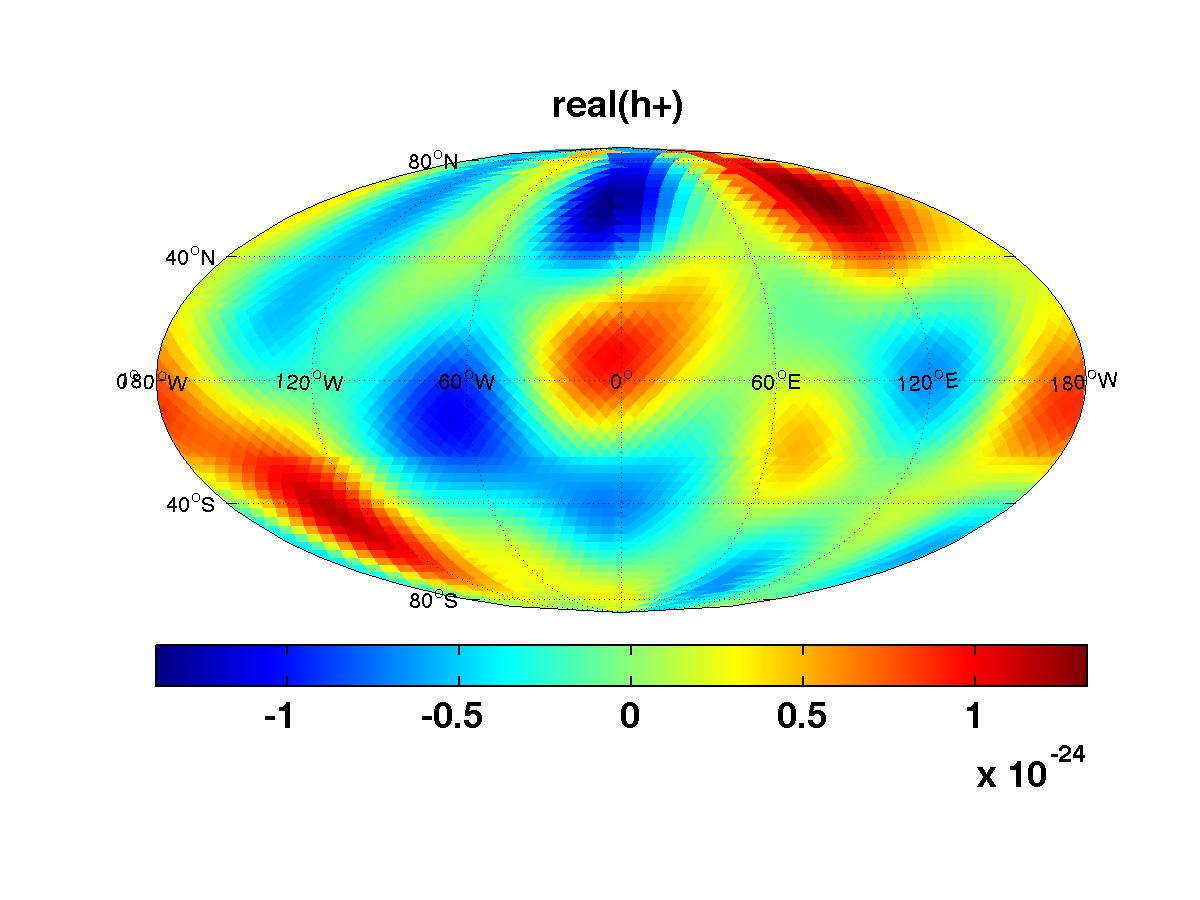}
\includegraphics[trim=3cm 4cm 3cm 2.5cm, clip=true, angle=0, width=0.24\textwidth]{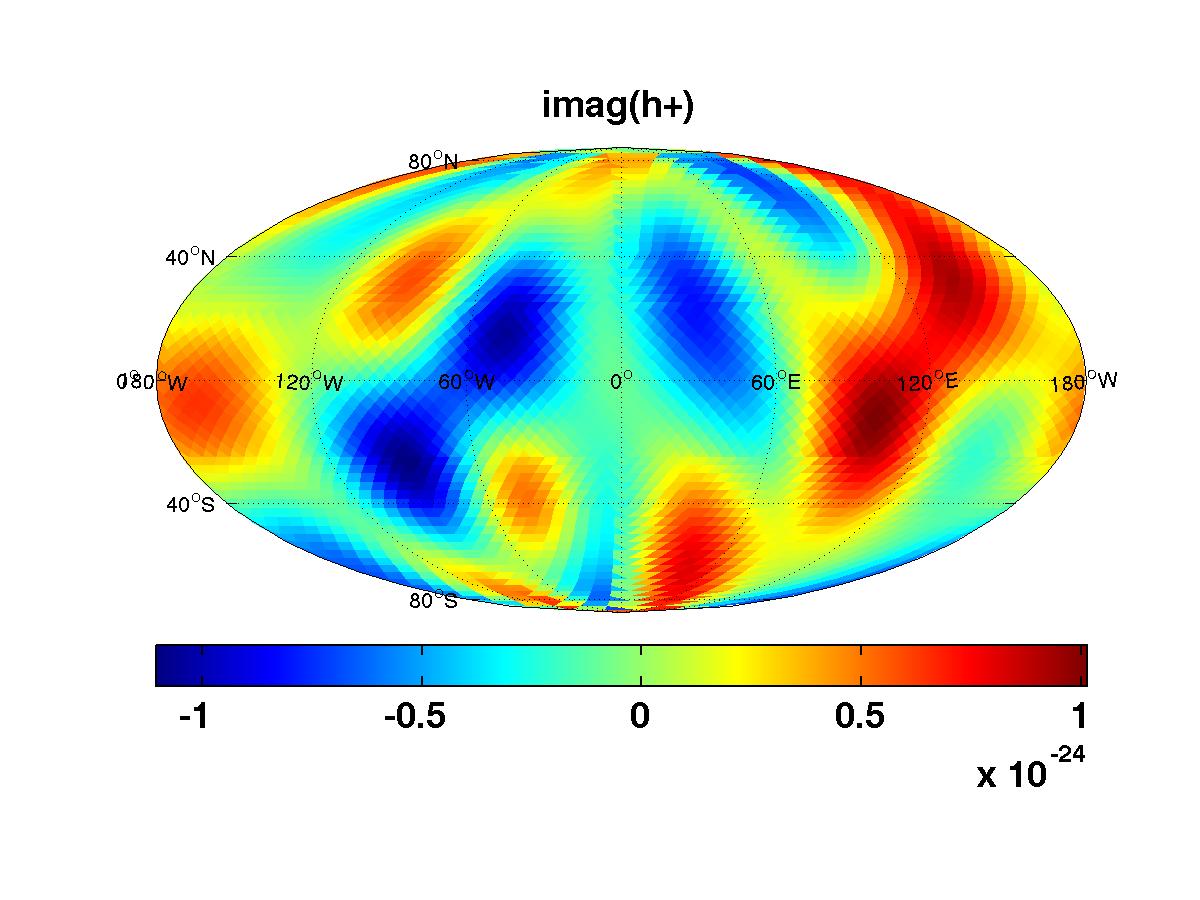}
\includegraphics[trim=3cm 4cm 3cm 2.5cm, clip=true, angle=0, width=0.24\textwidth]{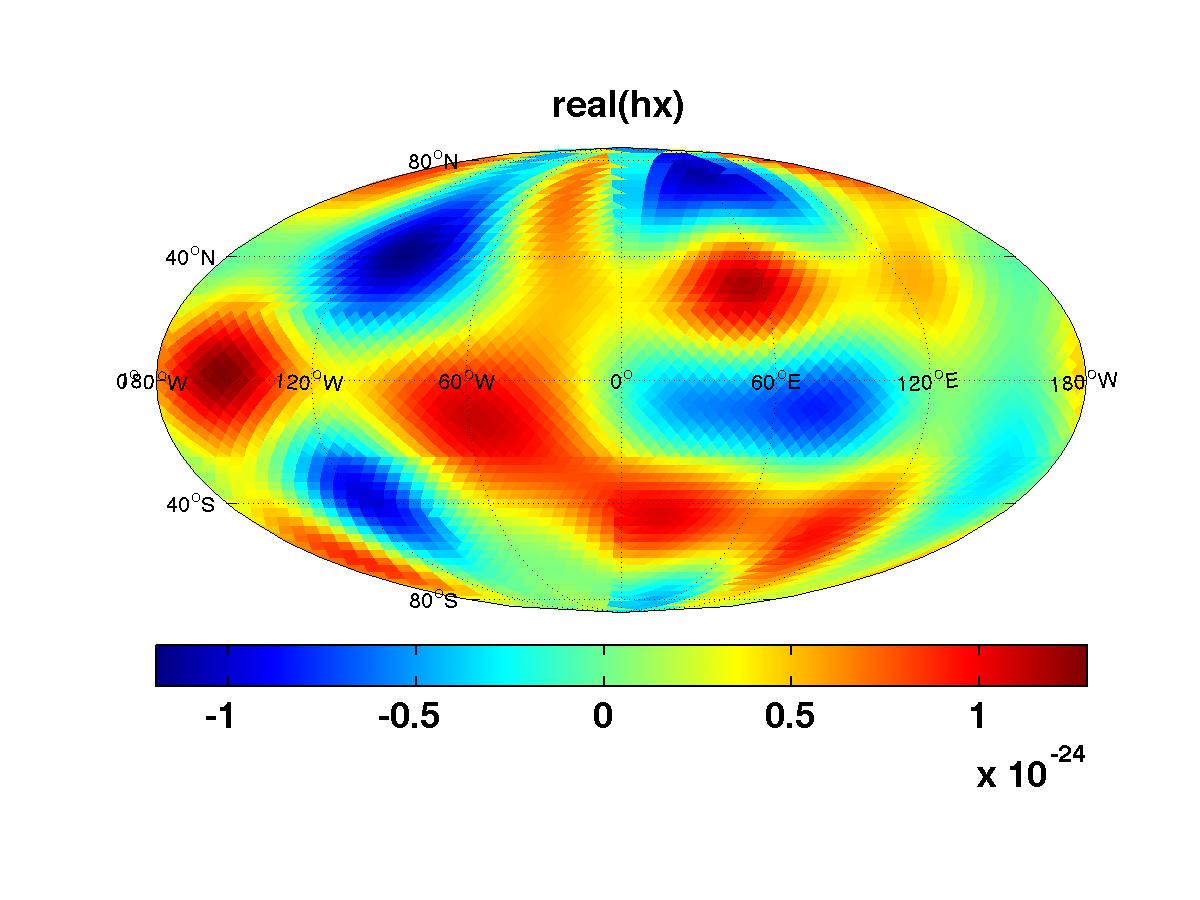}
\includegraphics[trim=3cm 4cm 3cm 2.5cm, clip=true, angle=0, width=0.24\textwidth]{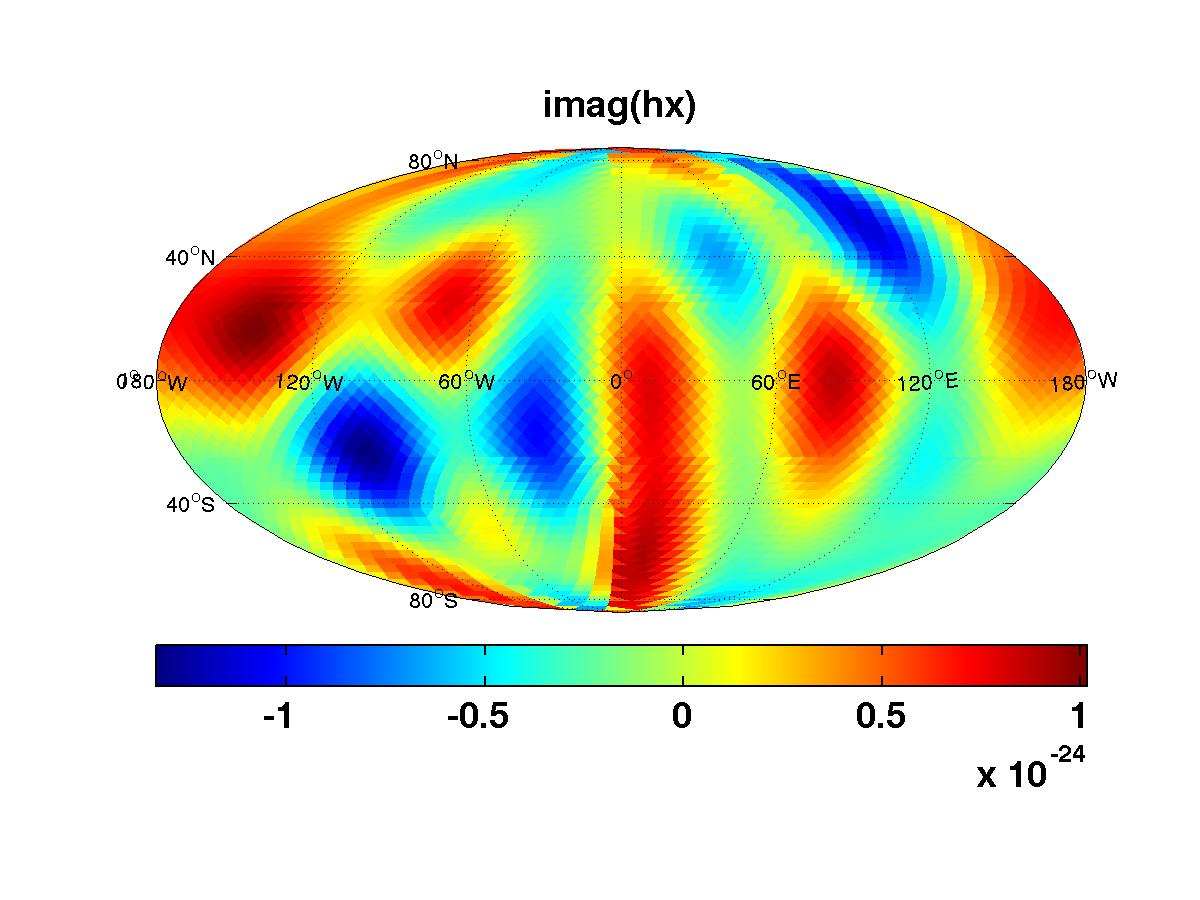}
\caption{Simulated maps (at a single fixed frequency) for three different 
anisotropic gravitational-wave backgrounds:
(i) point source located at $40^\circ$ latitude, $60^\circ$ longitude having
$l_{\rm max}=10$ (first row);
(ii) grad-only statistically-isotropic background with $C_l={\rm const}$ for 
$2\le l\le 10$ (second row);
(iii) curl-only statistically-isotropic background with $C_l={\rm const}$ for 
$2\le l\le 5$.
The four columns correspond to the real and imaginary parts of $h_+$ and $h_\times$.}
\label{f:simulated_maps}
\end{center}
\end{figure*}

The simulated gravitational-wave backgrounds
will consist of a point source and two 
spatially-extended distributions.
The point source is not an ideal point source, but more
of a Gaussian `blob', 
since we generate it out to only $l_{\rm max}=10$ 
(see the first row of maps in Fig.~\ref{f:simulated_maps}).
Nonetheless, it serves its purpose as being a 
simple yet extreme example of an anisotropic 
background for both $h_+$ and $h_\times$.
(We also considered single-pixel point sources and 
found similar results, but the corresponding sky
maps are not very clear.)
The two spatially-extended backgrounds are a 
grad-only statistically isotropic background 
with equal contributions for 
multipoles $2\le l\le 10$, and a curl-only 
statistically isotropic background with equal
contributions for multipoles $2\le l \le 5$ (see the second and third rows of Fig.~\ref{f:simulated_maps}).
These last two backgrounds were also considered 
in \cite{Gair-et-al:2014} in the context of pulsar 
timing arrays.
It is interesting to compare the recovered sky maps 
for the ground-based and pulsar
timing analyses, especially for the curl-only
background, which cannot be recovered using 
timing residual data from a pulsar timing array.
\begin{figure}[hbtp!]
\begin{center}
\includegraphics[angle=0, width=.49\textwidth]{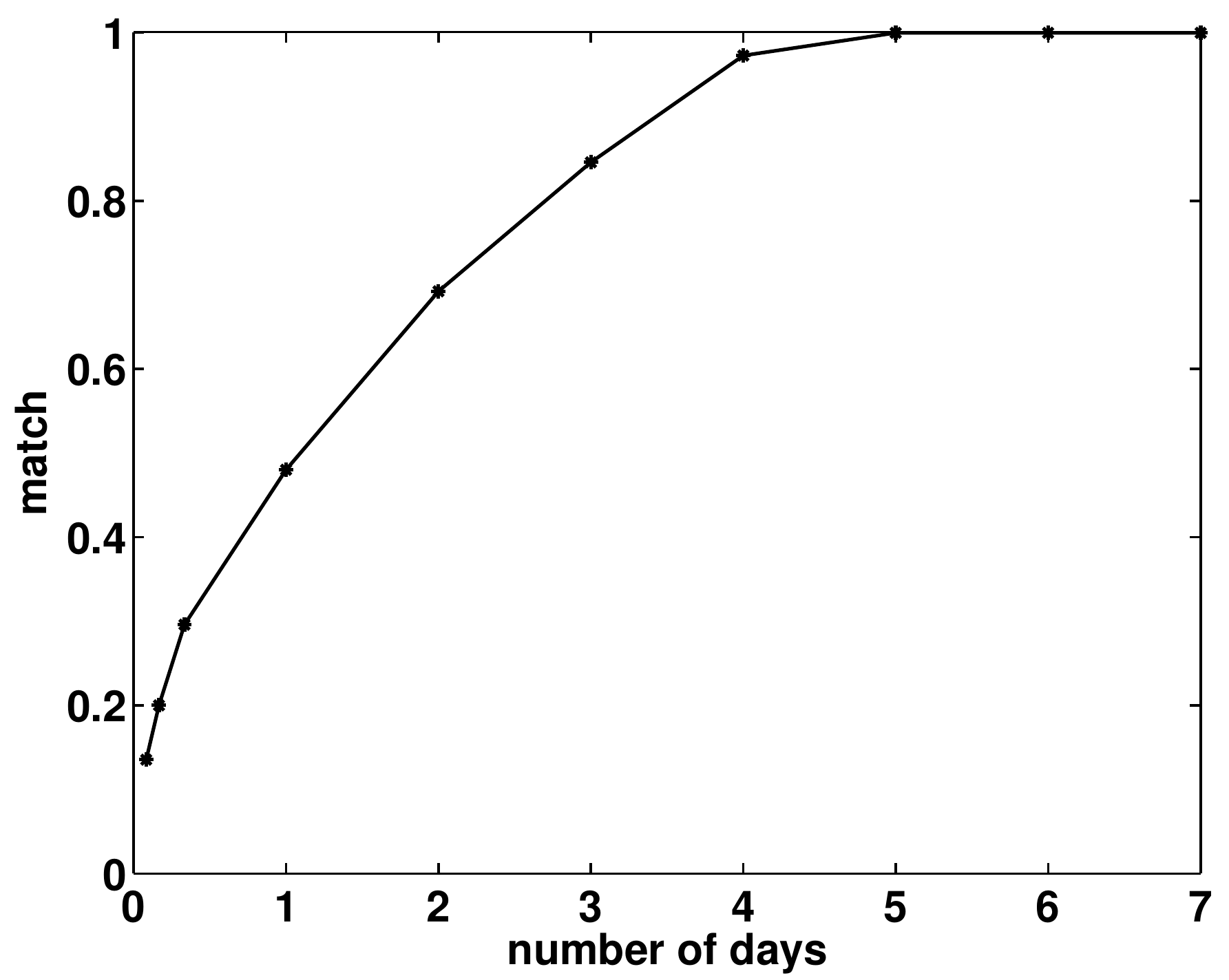}
\caption
{Match as a function of number of days of observation 
for the 6-detector network and a noiseless point-source injection.
A match value equal to 1 corresponds to perfect recovery.}
\label{f:match_vs_time}
\end{center}
\end{figure}
The amplitudes of the injected gravitational-wave backgrounds 
were chosen to give reasonable recoveries after just a few 
days of simulated data.
For the values of the noise spectral densities $S_I(f)$ given 
in Table~\ref{t:ifos}, we found that an amplitude 
$A=4\times 10^{-25}$ was sufficient for the three different
backgrounds.
(If we used a smaller value of $A$, we would have had to 
integrate for a longer period of time.)
The faithfulness of the recovery is measured by calculating
the {\em match} between the injected and 
maximum-likelihood-recovered sky maps,
\be
\mu
\equiv 
\frac{\frac{1}{2}({\mb h}_{\rm inj}\cdot{\mb h}_{\rm ML} 
+ {\mb h}_{\rm ML}\cdot{\mb h}_{\rm inj})}
{\sqrt{{\mb h}_{\rm inj}\cdot{\mb h}_{\rm inj}}
\sqrt{{\mb h}_{\rm ML}\cdot{\mb h}_{\rm ML}}}\,.
\ee
This is just the coherence between the two maps.
We will also construct uncertainty maps and SNR maps
to evaluate how well we can recover the injections.

Figure~\ref{f:match_vs_time} is a plot of the 
match as a function of the number of days 
of observation for the 6-detector network and 
a {\em noiseless} point-source injection.
The match increases as the total observation time 
increases as expected.
Note we get perfect match 
after 5 days of observation.
This follows from the fact that the total number of 
data points taken by the 6-detector network 
over 5 days is given by 
$N=5~{\rm days}\times 60~{\rm samples}/{\rm day}\times 6=1800$,
which is greater than the number of modes 
$M=2N_{\rm pix}=1536$ we are trying to recover.
Thus, in the absence of noise we have (more than) 
enough information to completely recover the 
injected background after 5 days of observation.
We would have complete recovery for the two other
simulated backgrounds as well.

\begin{figure*}[hbtp!]
\begin{center}
\includegraphics[trim=3cm 4cm 3cm 2.5cm, clip=true, angle=0, width=0.24\textwidth]{point_realhp}
\includegraphics[trim=3cm 4cm 3cm 2.5cm, clip=true, angle=0, width=0.24\textwidth]{point_imaghp}
\includegraphics[trim=3cm 4cm 3cm 2.5cm, clip=true, angle=0, width=0.24\textwidth]{point_realhc}
\includegraphics[trim=3cm 4cm 3cm 2.5cm, clip=true, angle=0, width=0.24\textwidth]{point_imaghc}
\includegraphics[trim=3cm 4cm 3cm 2.5cm, clip=true, angle=0, width=0.24\textwidth]{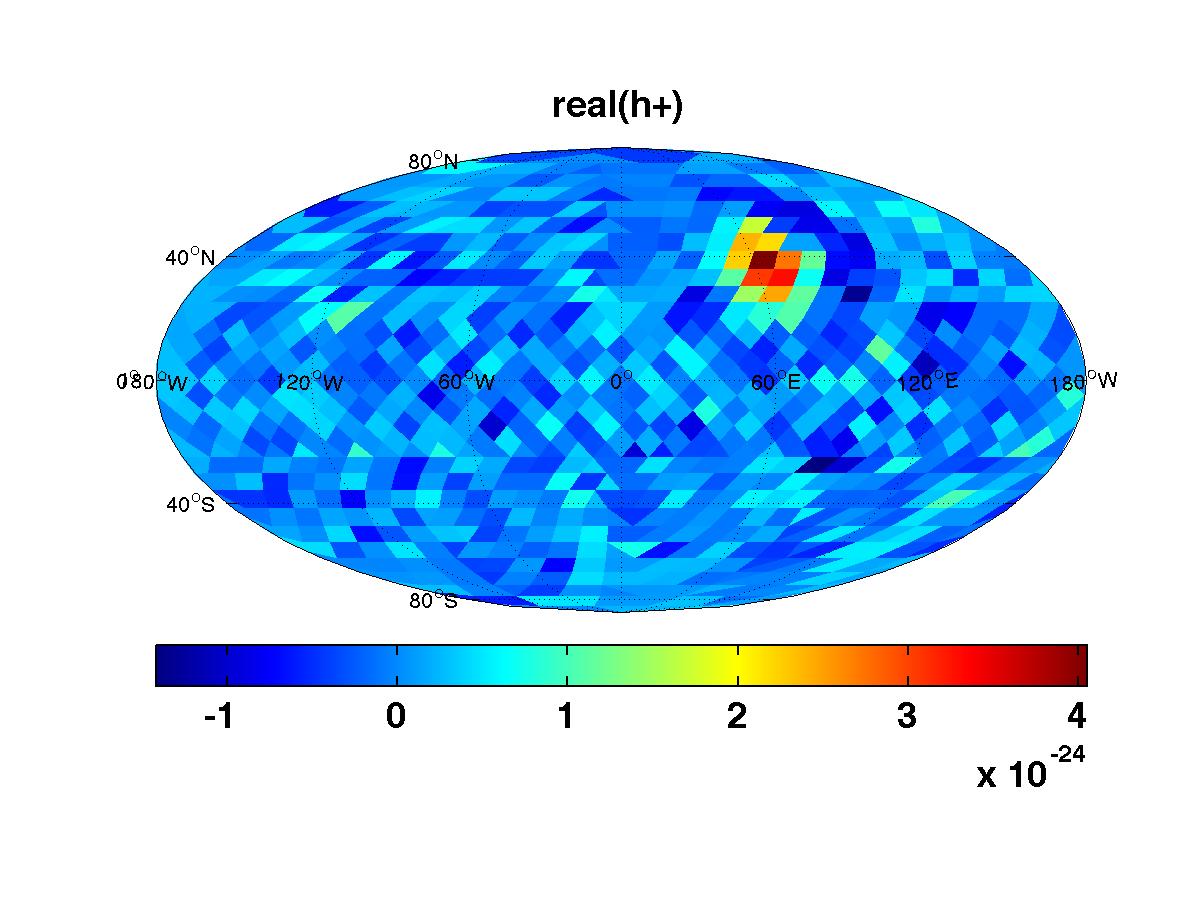}
\includegraphics[trim=3cm 4cm 3cm 2.5cm, clip=true, angle=0, width=0.24\textwidth]{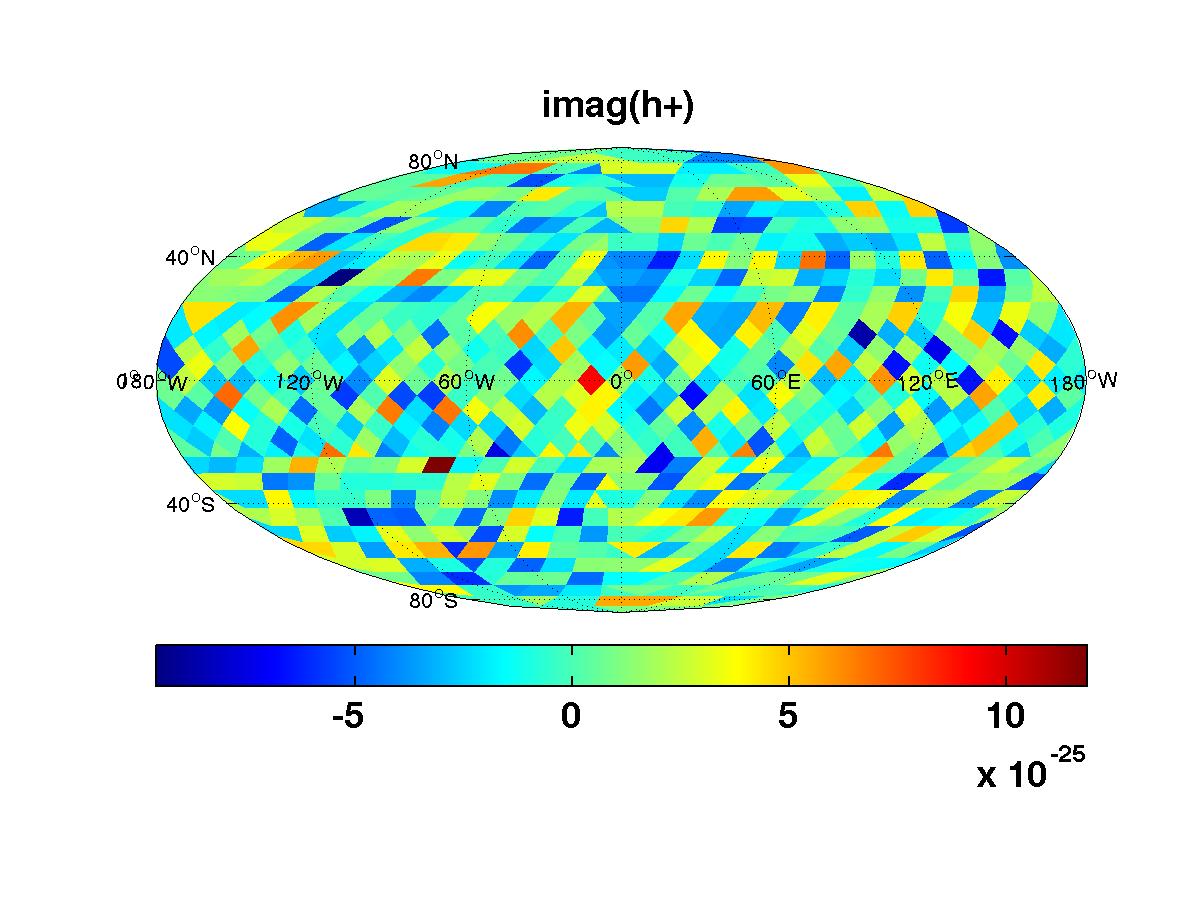}
\includegraphics[trim=3cm 4cm 3cm 2.5cm, clip=true, angle=0, width=0.24\textwidth]{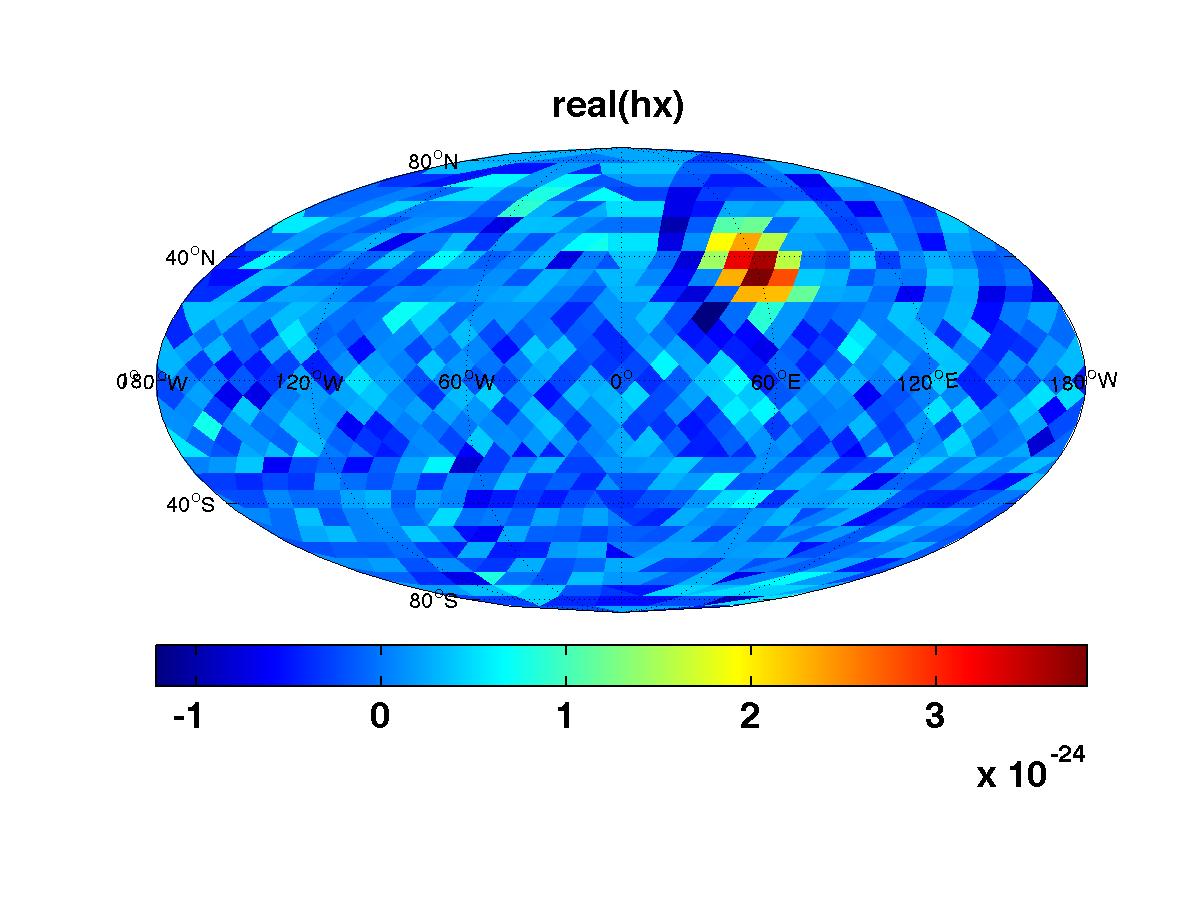}
\includegraphics[trim=3cm 4cm 3cm 2.5cm, clip=true, angle=0, width=0.24\textwidth]{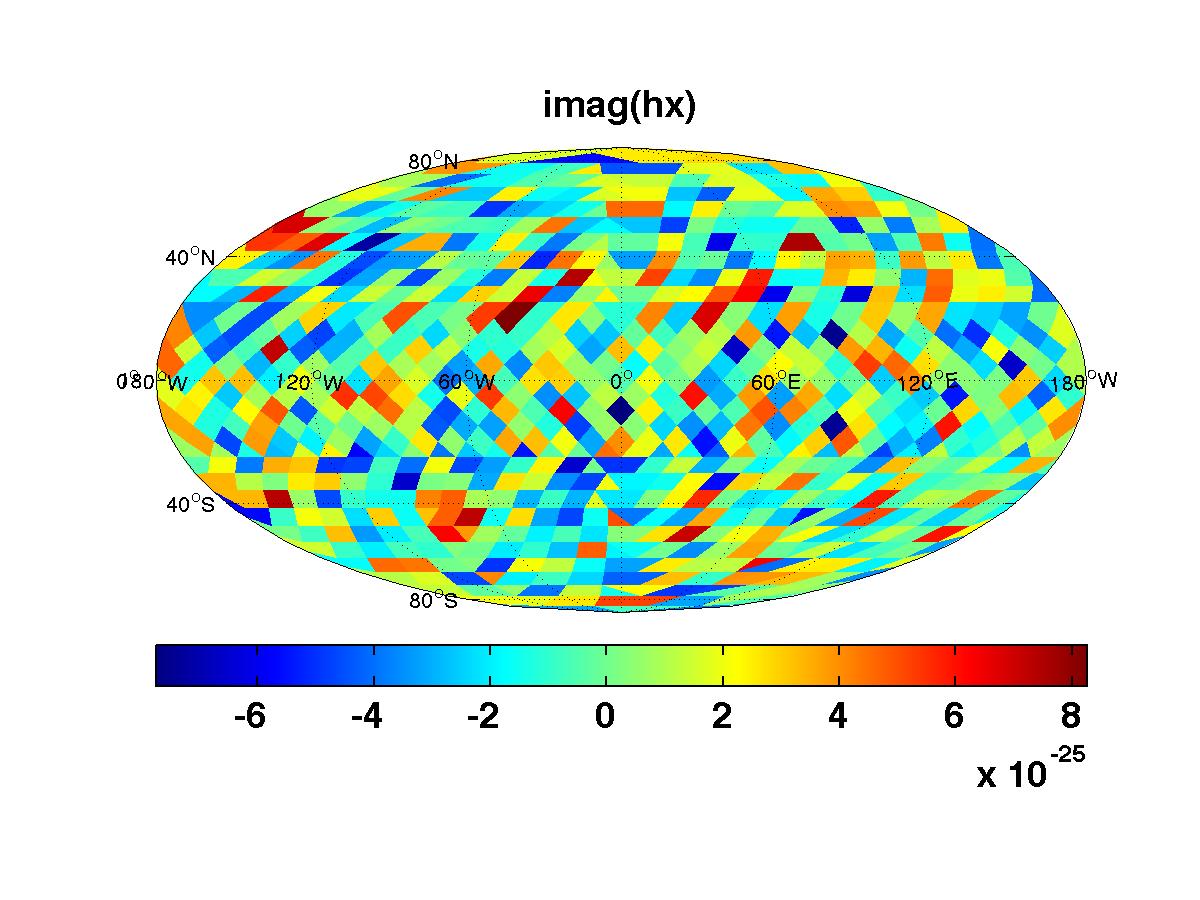}
\includegraphics[trim=3cm 4cm 3cm 2.5cm, clip=true, angle=0, width=0.24\textwidth]{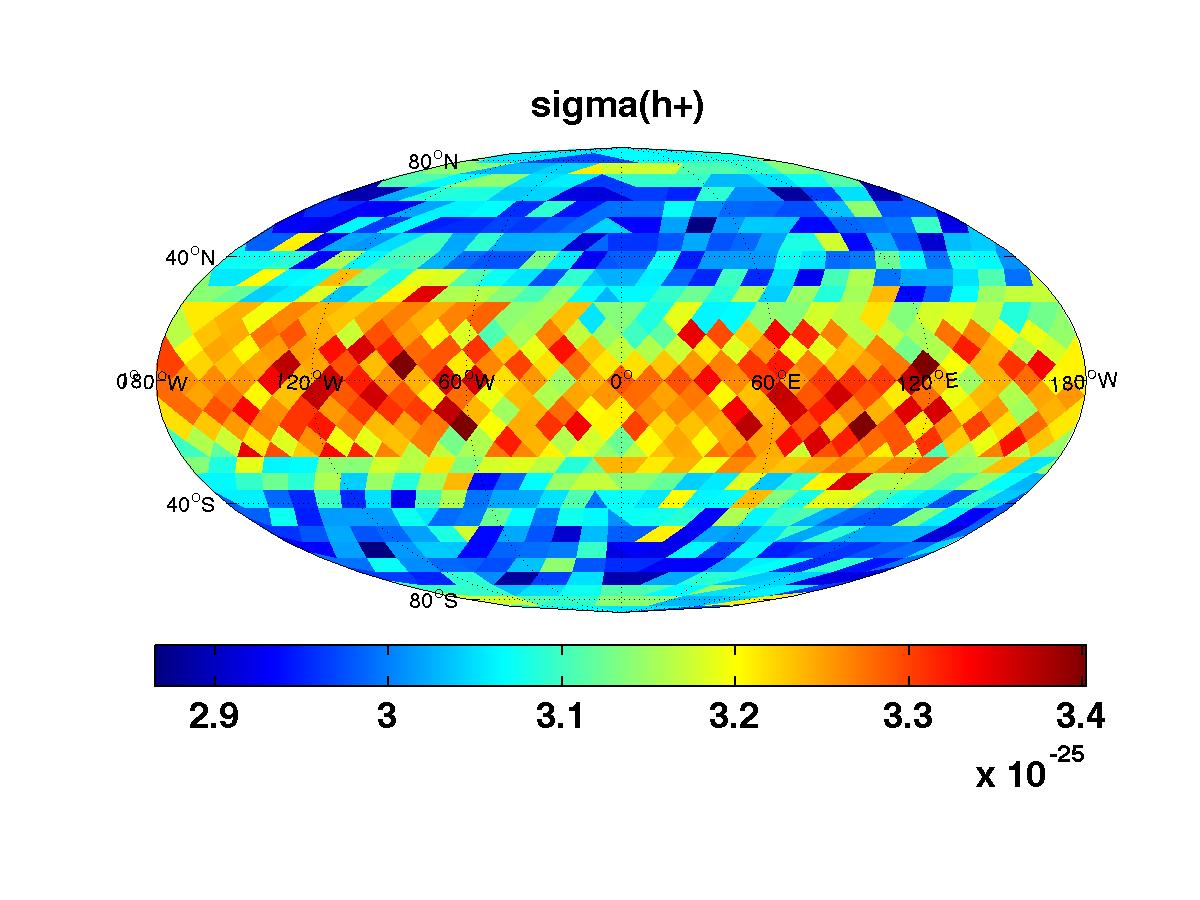}
\includegraphics[trim=3cm 4cm 3cm 2.5cm, clip=true, angle=0, width=0.24\textwidth]{sigma_hp_point}
\includegraphics[trim=3cm 4cm 3cm 2.5cm, clip=true, angle=0, width=0.24\textwidth]{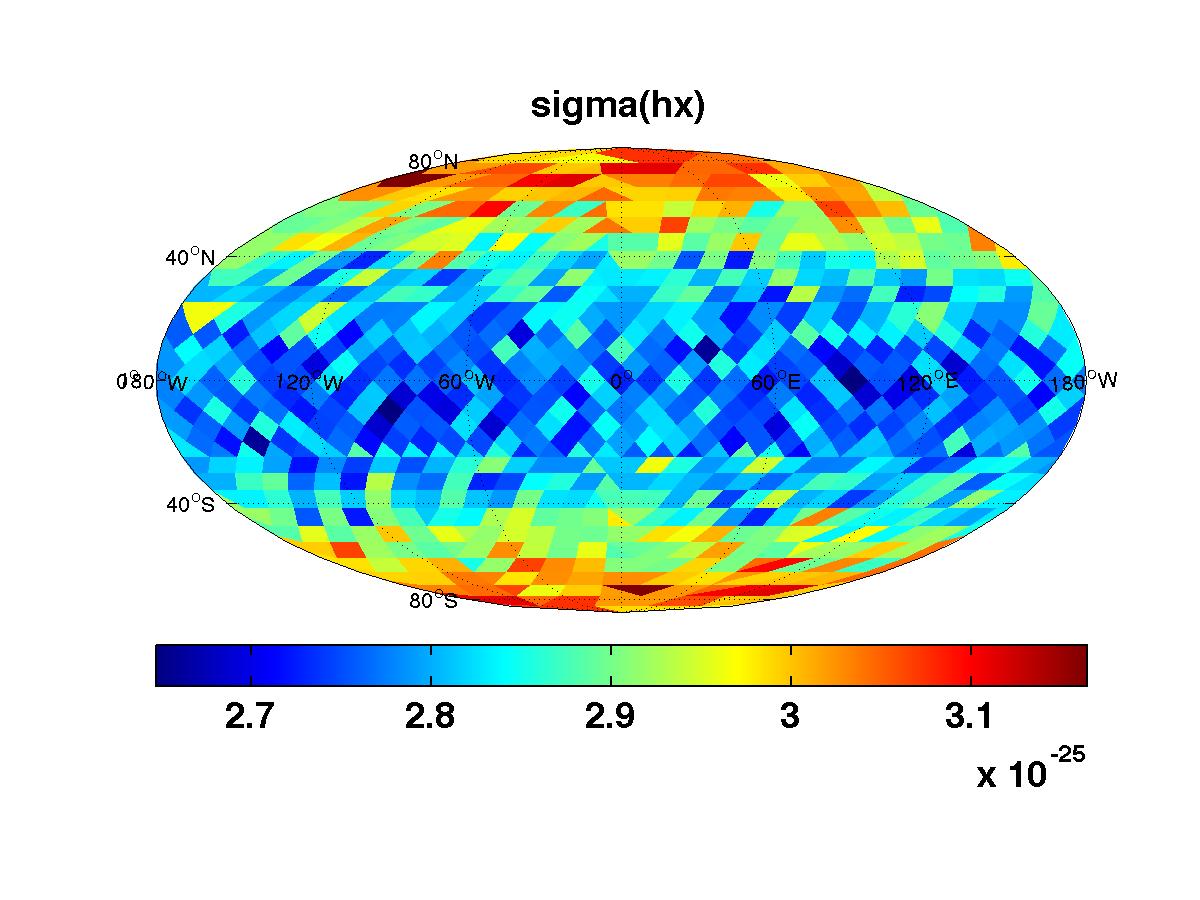}
\includegraphics[trim=3cm 4cm 3cm 2.5cm, clip=true, angle=0, width=0.24\textwidth]{sigma_hc_point}
\includegraphics[trim=3cm 4cm 3cm 2.5cm, clip=true, angle=0, width=0.24\textwidth]{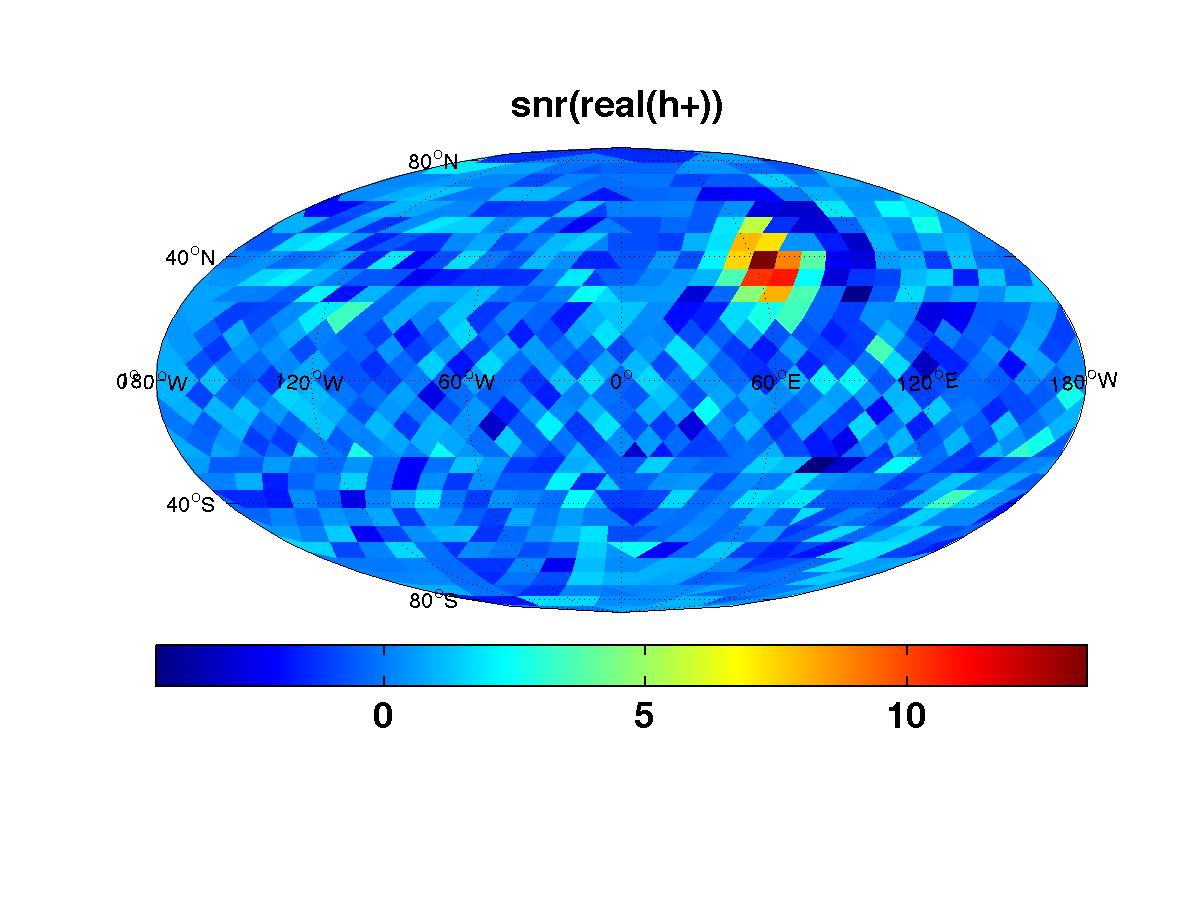}
\includegraphics[trim=3cm 4cm 3cm 2.5cm, clip=true, angle=0, width=0.24\textwidth]{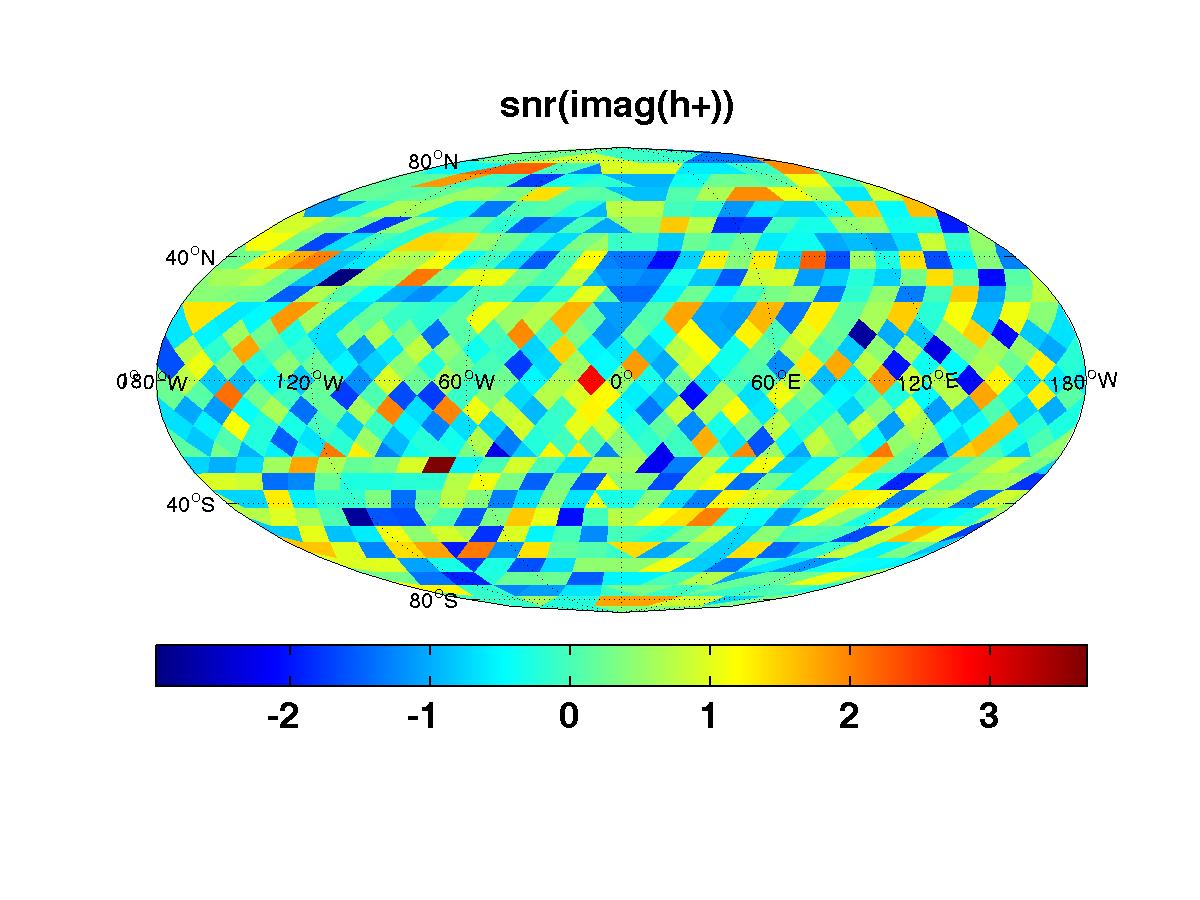}
\includegraphics[trim=3cm 4cm 3cm 2.5cm, clip=true, angle=0, width=0.24\textwidth]{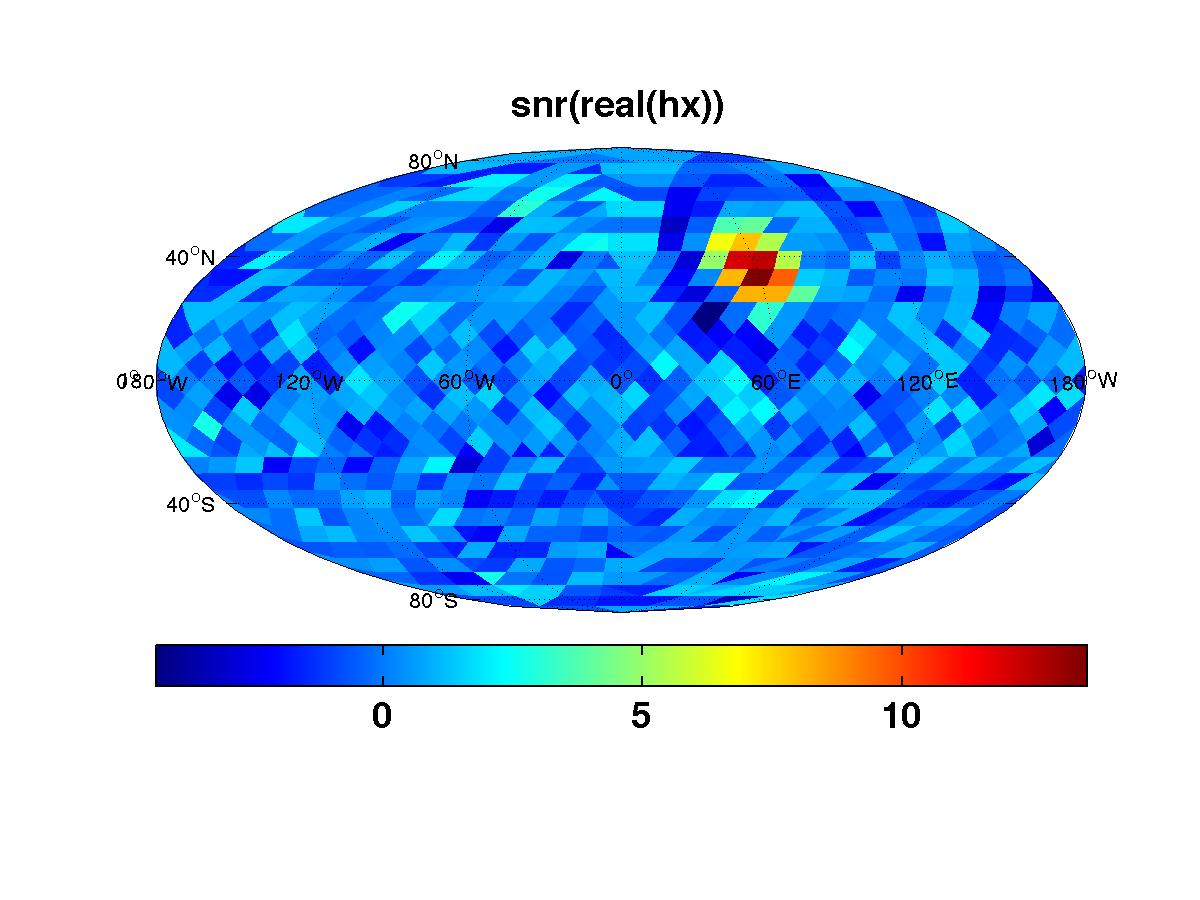}
\includegraphics[trim=3cm 4cm 3cm 2.5cm, clip=true, angle=0, width=0.24\textwidth]{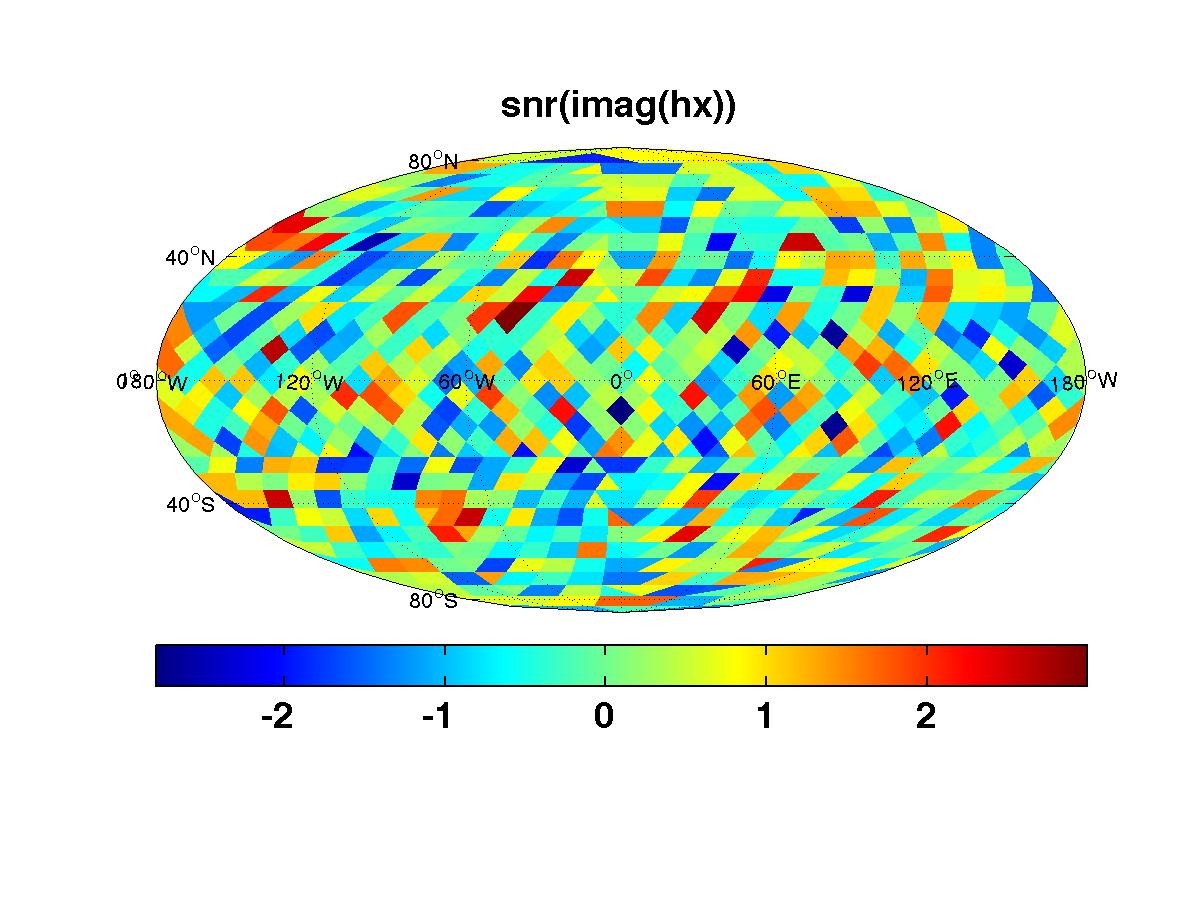}
\caption
{Recovery of the simulated point source in noise for the 6-detector network.
Injected maps (first row);
maximum-likehood recovered maps (second row);
uncertainty map (third row);
SNR map (last row).
Note that the uncertainties maps for the real and imaginary 
parts of $h_+$ (or $h_\times$) are the same.} 
\label{f:point_recovery}
\end{center}
\end{figure*}

Sky maps of the recovered point-source background 
injected into noisy data 
are shown in Fig.~\ref{f:point_recovery}.
These maps are for the 6-detector network 
with $N=2400$ total data points, 
corresponding to 6.67~days of total observation.
The first row shows the injected background.
The second row shows the maximum-likelihood sky map
estimates, which are the real and imaginary parts of
the $h_+$, $h_\times$ components of ${\mb h}_{\rm ML}$.
The third row shows the uncertainty maps, as specified
by ${\bs\sigma}_{\rm ML}$, and the fourth row shows the 
SNR maps.
The max SNR at the location of the point source is 
approximately~10.
The match is $\mu=0.64$ for this particular simulation.

Note that the uncertainty maps for the real and imaginary
parts of $h_+$ (or $h_\times$) are the same.
The uncertainty values are also fairly constant over the sky, 
with values around $3\times 10^{-25}$.
Thus, the SNR maps look very similar to the 
maximum-likehood maps but, of course, have different values
since they represent different quantities.
It is also the case that the uncertainty maps for the
other simulated backgrounds (grad-only and curl-only
background) will be identical to that for the 
point source background, since Eq.~(\ref{e:sigma_map})
for ${\bs\sigma}_{\rm ML}$ depends only
on the response matrix ${\mb R}$ and noise covariance
matrix ${\mb C}$ via (\ref{e:var2})---i.e.,
it is independent of the background that one is trying to
recover, at least in the weak-signal limit, which we
have assumed in our analyses.
So for the other two simulated backgrounds, we will 
show only the injected and 
maximum-likelihood recovered sky maps, and not the 
uncertainty and SNR maps.

\begin{figure*}[hbtp!]
\begin{center}
\includegraphics[trim=3cm 4cm 3cm 2.5cm, clip=true, angle=0, width=0.24\textwidth]{point_realhp}
\includegraphics[trim=3cm 4cm 3cm 2.5cm, clip=true, angle=0, width=0.24\textwidth]{point_imaghp}
\includegraphics[trim=3cm 4cm 3cm 2.5cm, clip=true, angle=0, width=0.24\textwidth]{point_realhc}
\includegraphics[trim=3cm 4cm 3cm 2.5cm, clip=true, angle=0, width=0.24\textwidth]{point_imaghc}
\includegraphics[trim=3cm 4cm 3cm 2.5cm, clip=true, angle=0, width=0.24\textwidth]{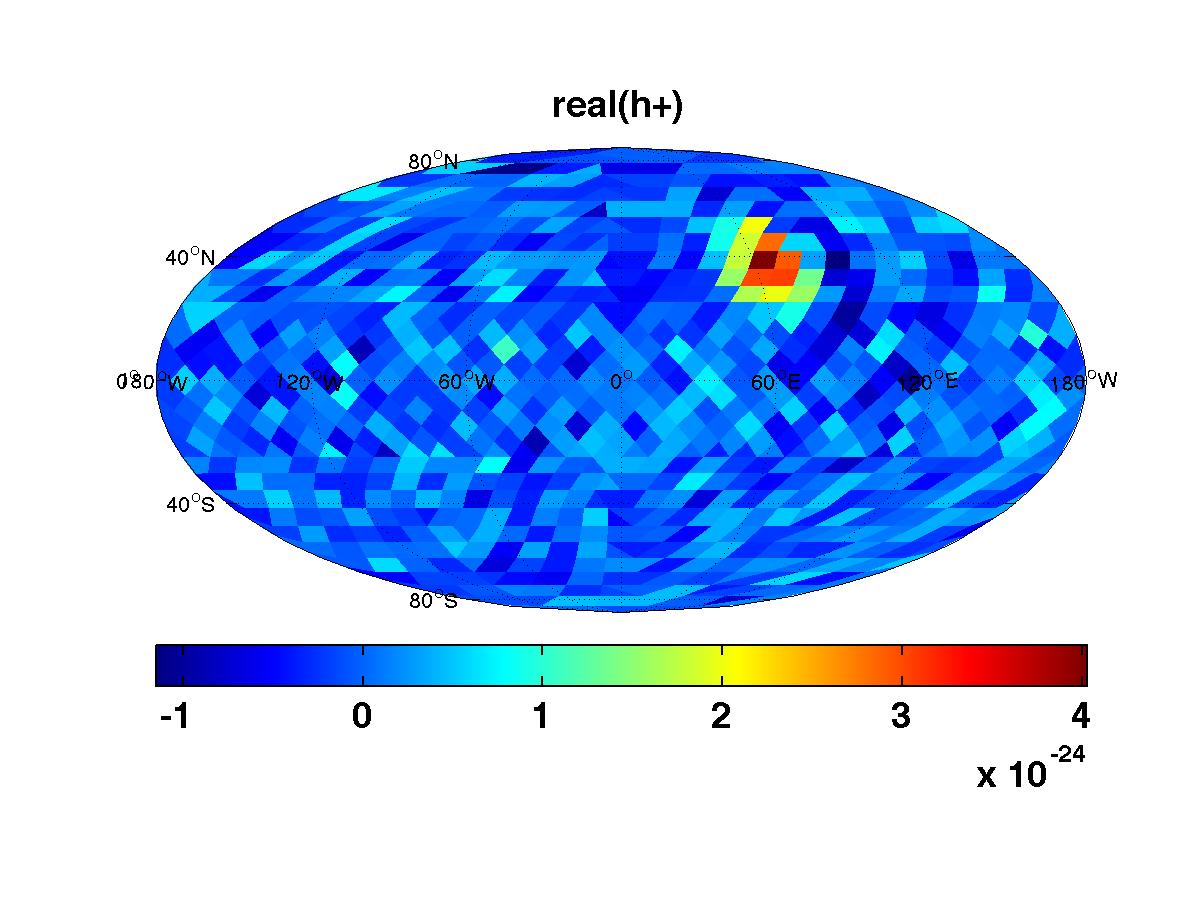}
\includegraphics[trim=3cm 4cm 3cm 2.5cm, clip=true, angle=0, width=0.24\textwidth]{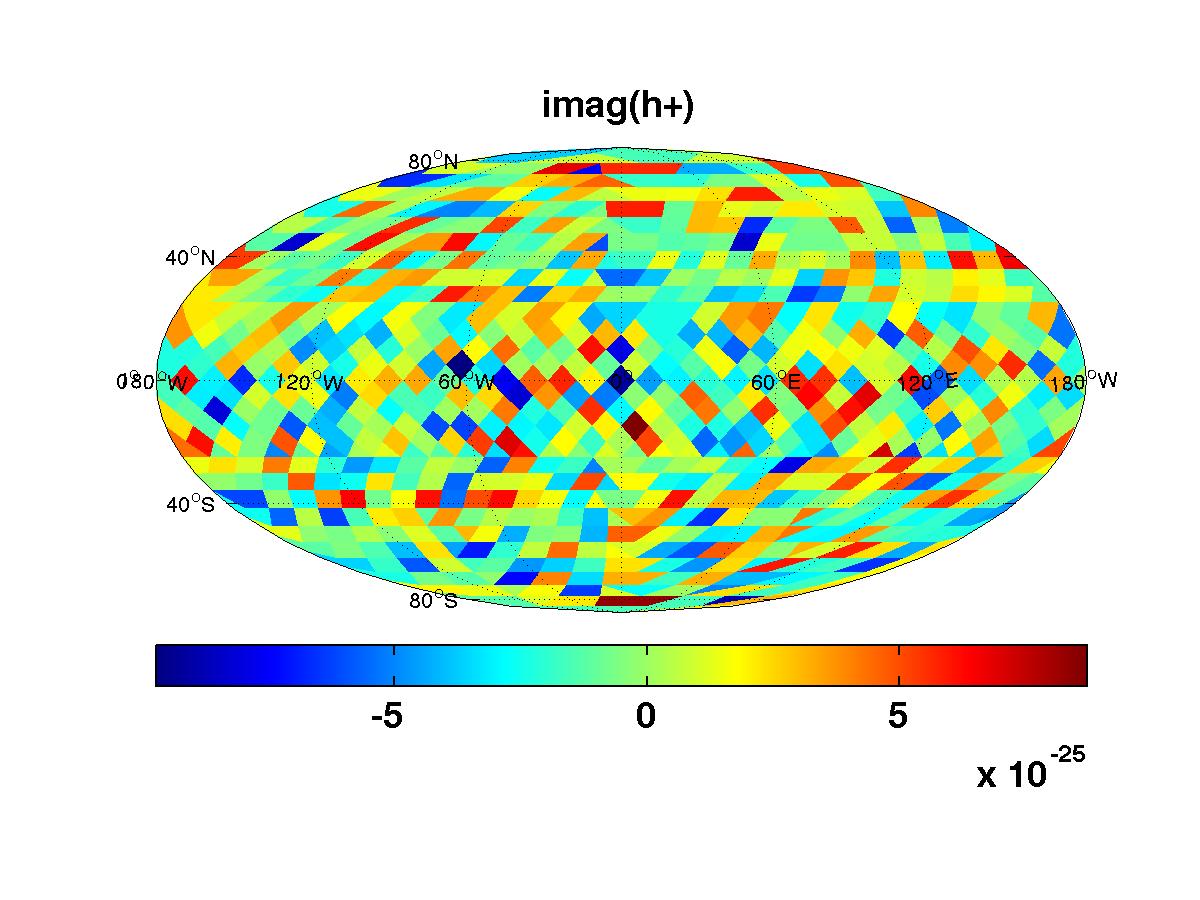}
\includegraphics[trim=3cm 4cm 3cm 2.5cm, clip=true, angle=0, width=0.24\textwidth]{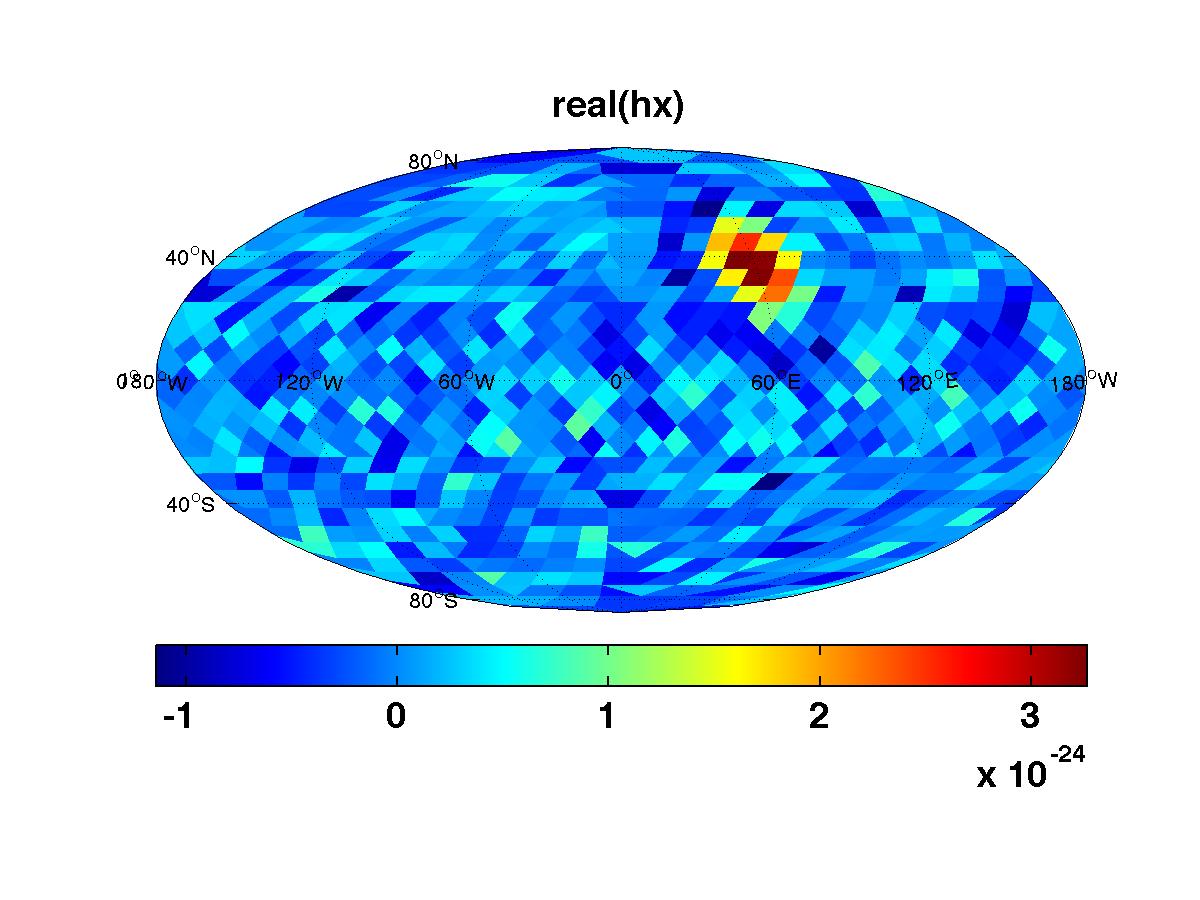}
\includegraphics[trim=3cm 4cm 3cm 2.5cm, clip=true, angle=0, width=0.24\textwidth]{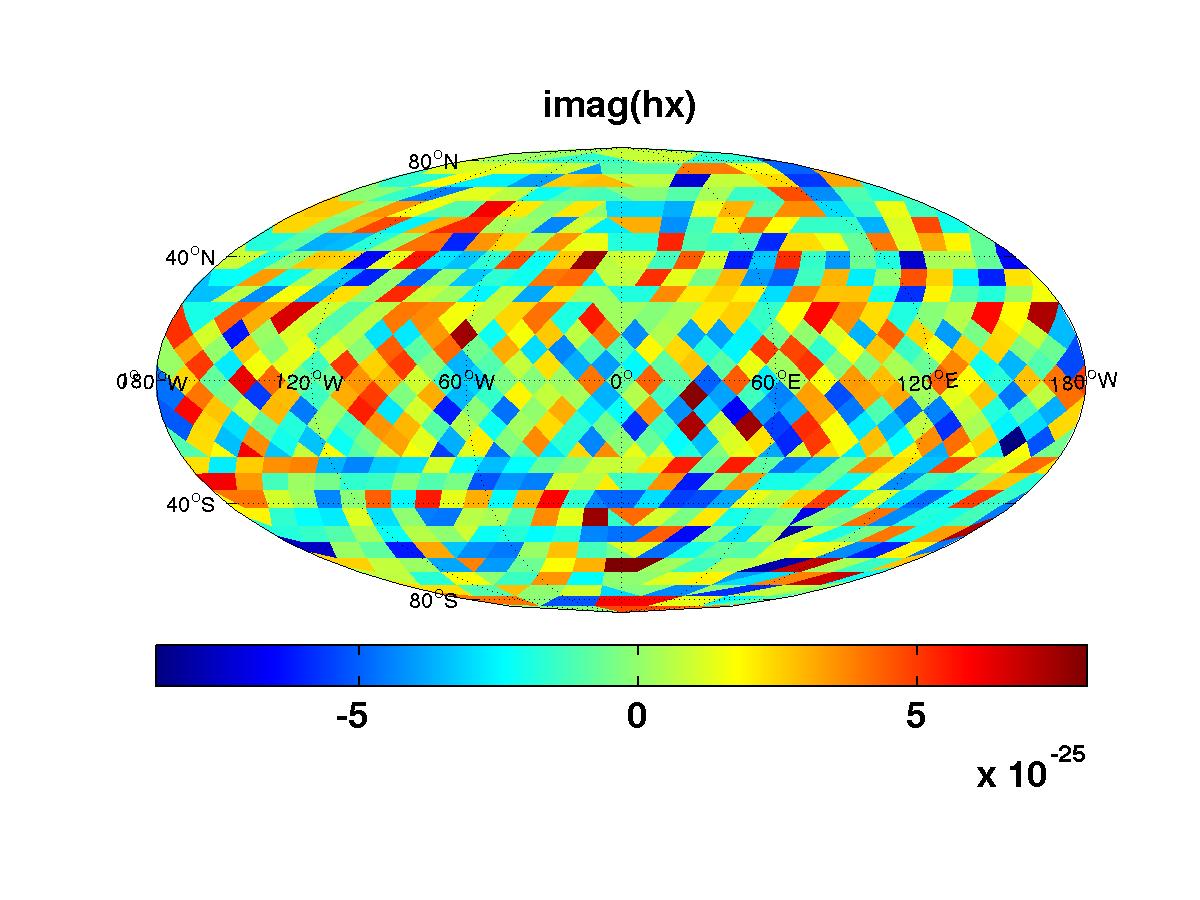}
\includegraphics[trim=3cm 4cm 3cm 2.5cm, clip=true, angle=0, width=0.24\textwidth]{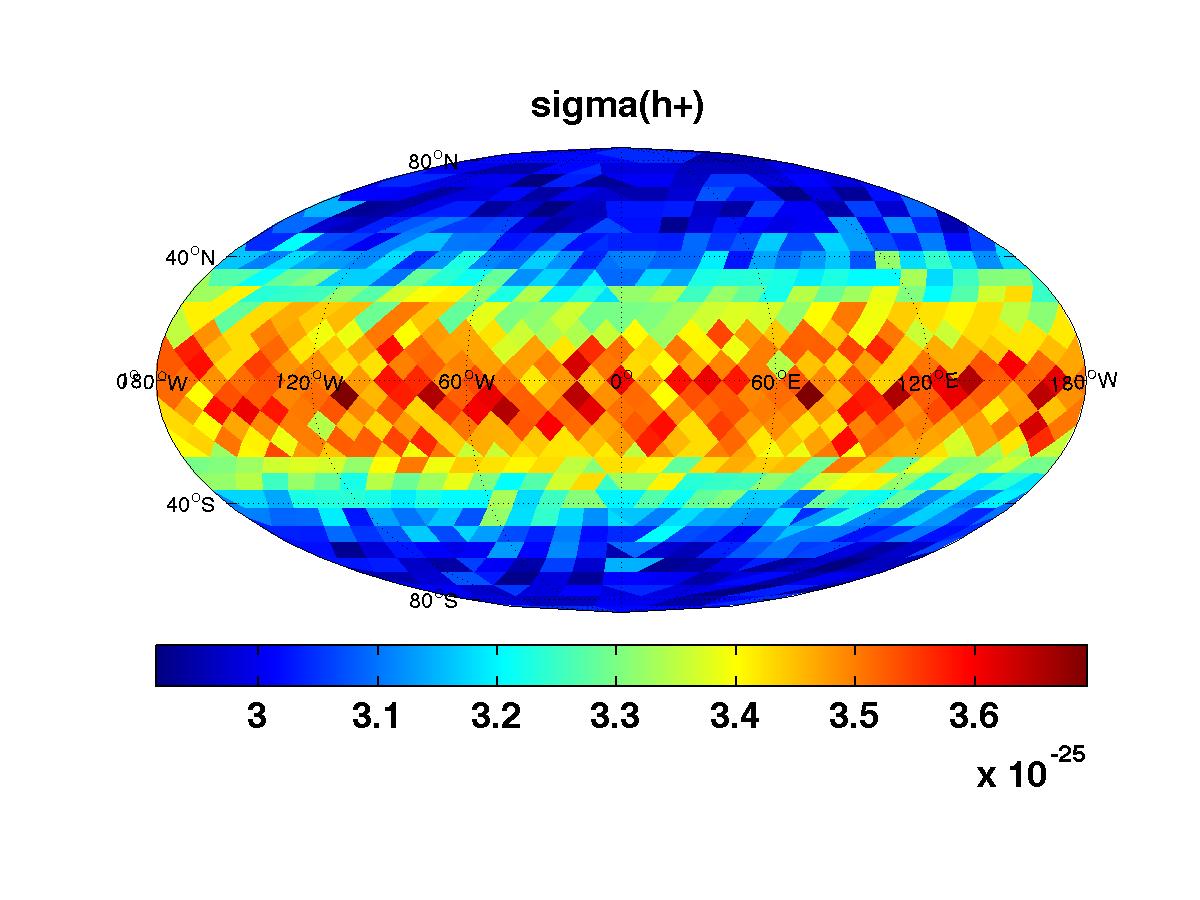}
\includegraphics[trim=3cm 4cm 3cm 2.5cm, clip=true, angle=0, width=0.24\textwidth]{sigma_hp_point_3_2}
\includegraphics[trim=3cm 4cm 3cm 2.5cm, clip=true, angle=0, width=0.24\textwidth]{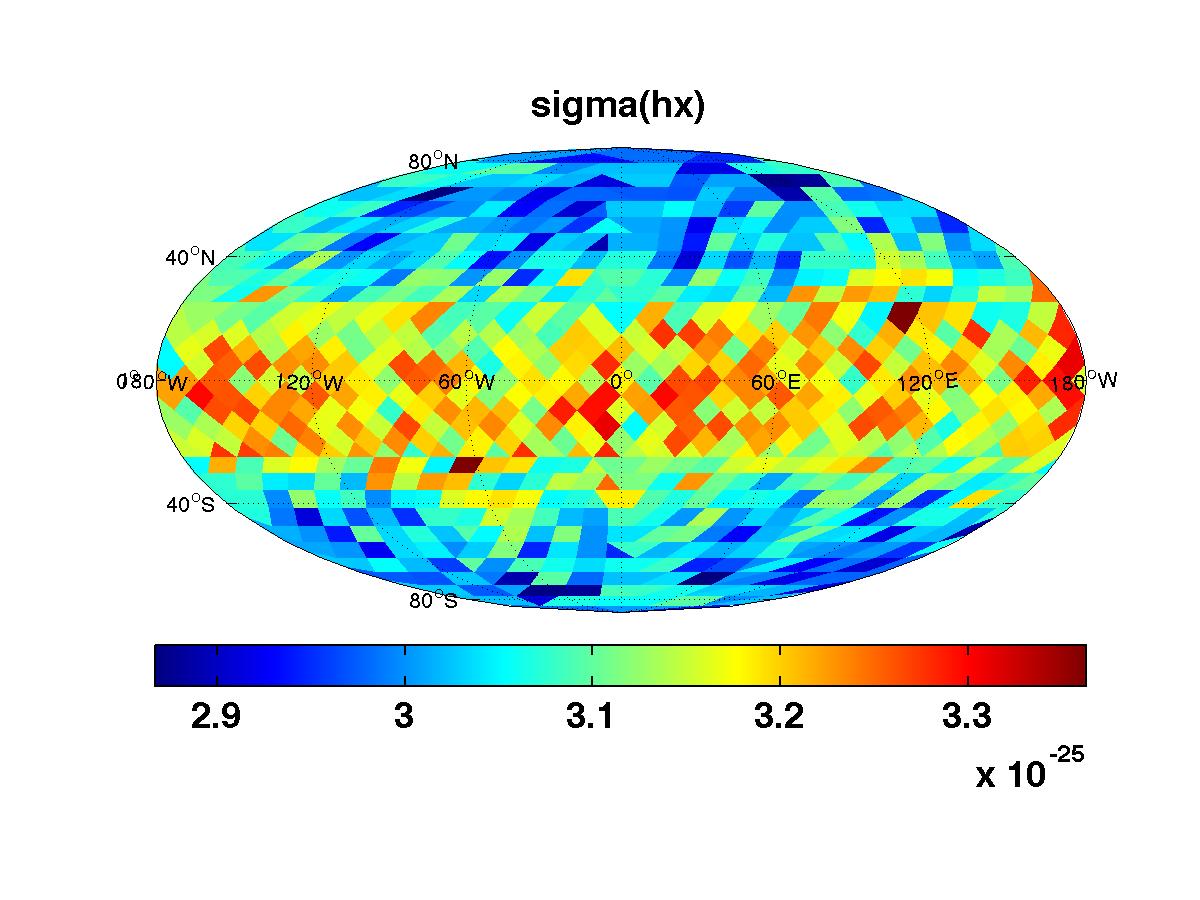}
\includegraphics[trim=3cm 4cm 3cm 2.5cm, clip=true, angle=0, width=0.24\textwidth]{sigma_hc_point_3_2}
\includegraphics[trim=3cm 4cm 3cm 2.5cm, clip=true, angle=0, width=0.24\textwidth]{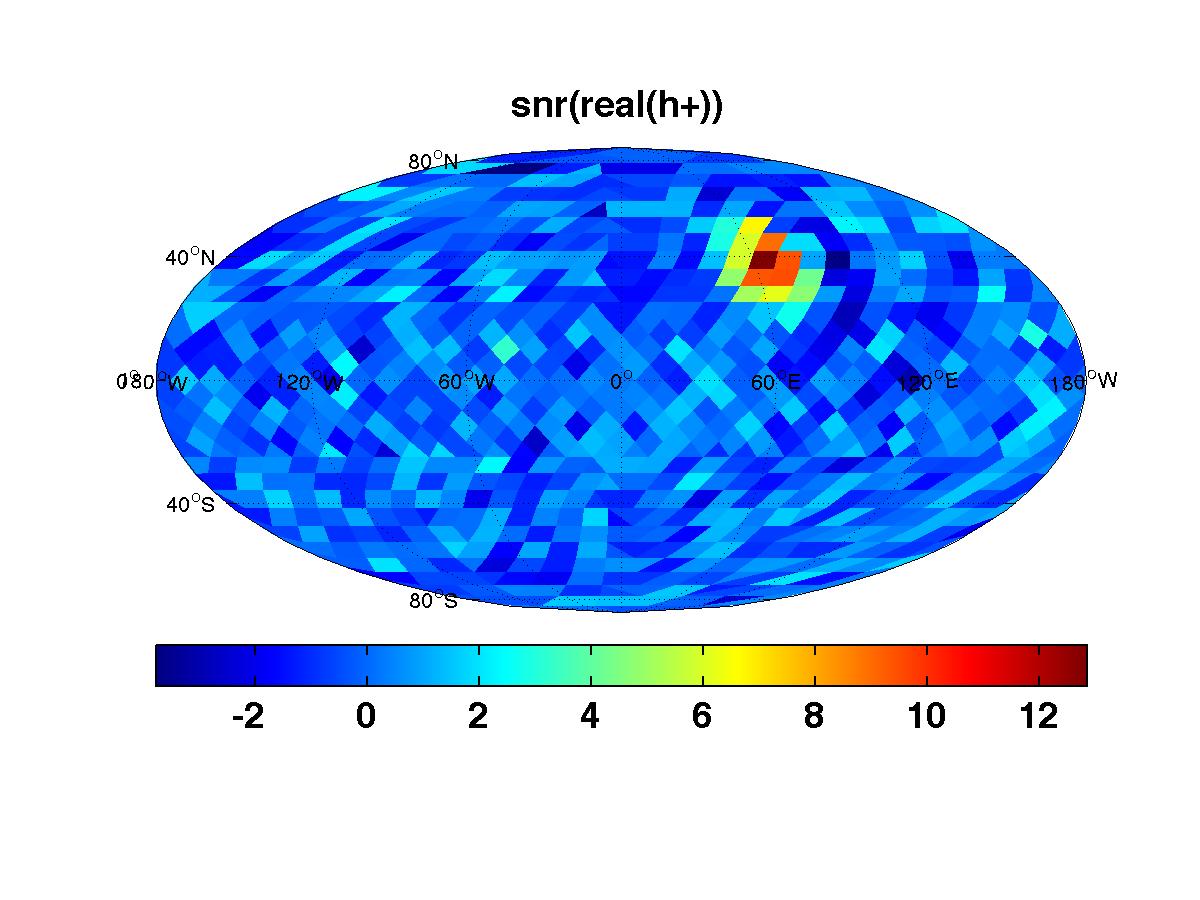}
\includegraphics[trim=3cm 4cm 3cm 2.5cm, clip=true, angle=0, width=0.24\textwidth]{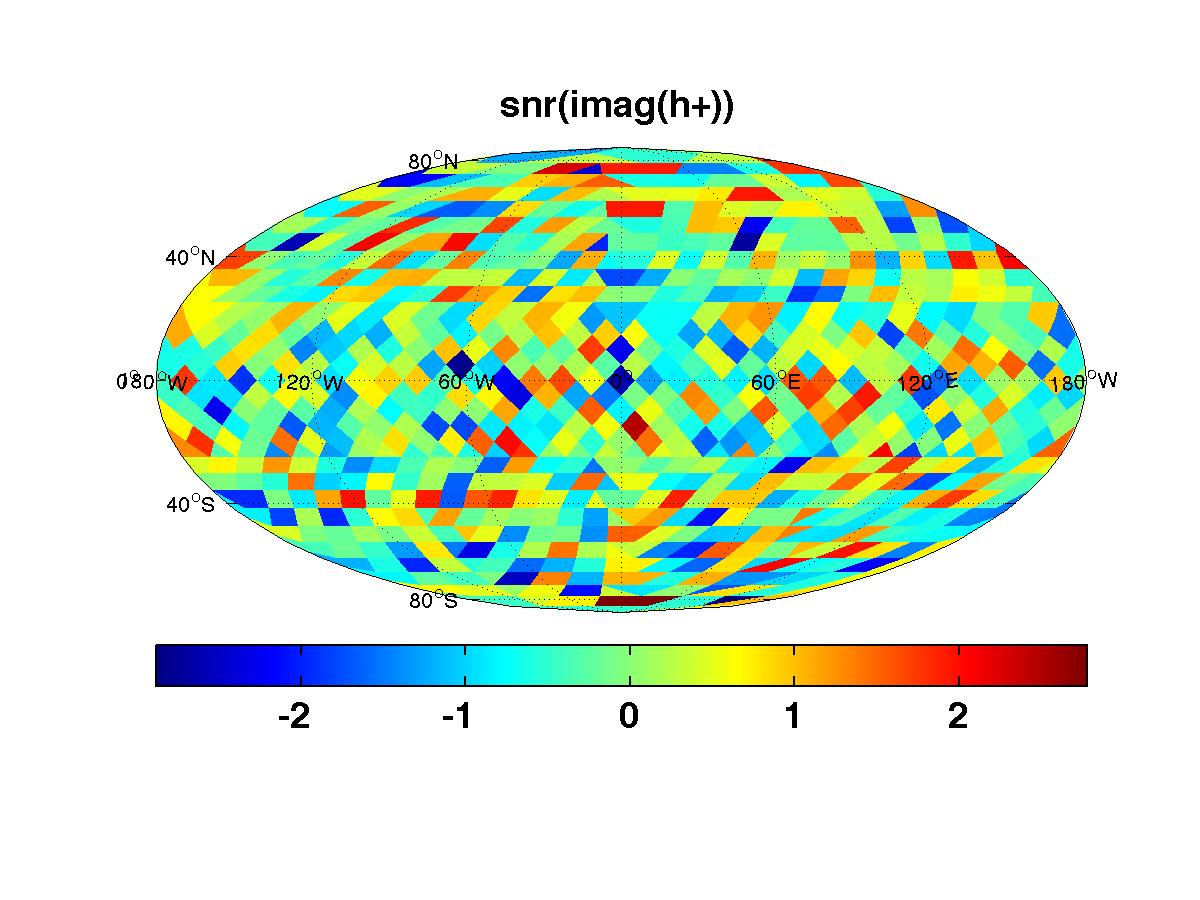}
\includegraphics[trim=3cm 4cm 3cm 2.5cm, clip=true, angle=0, width=0.24\textwidth]{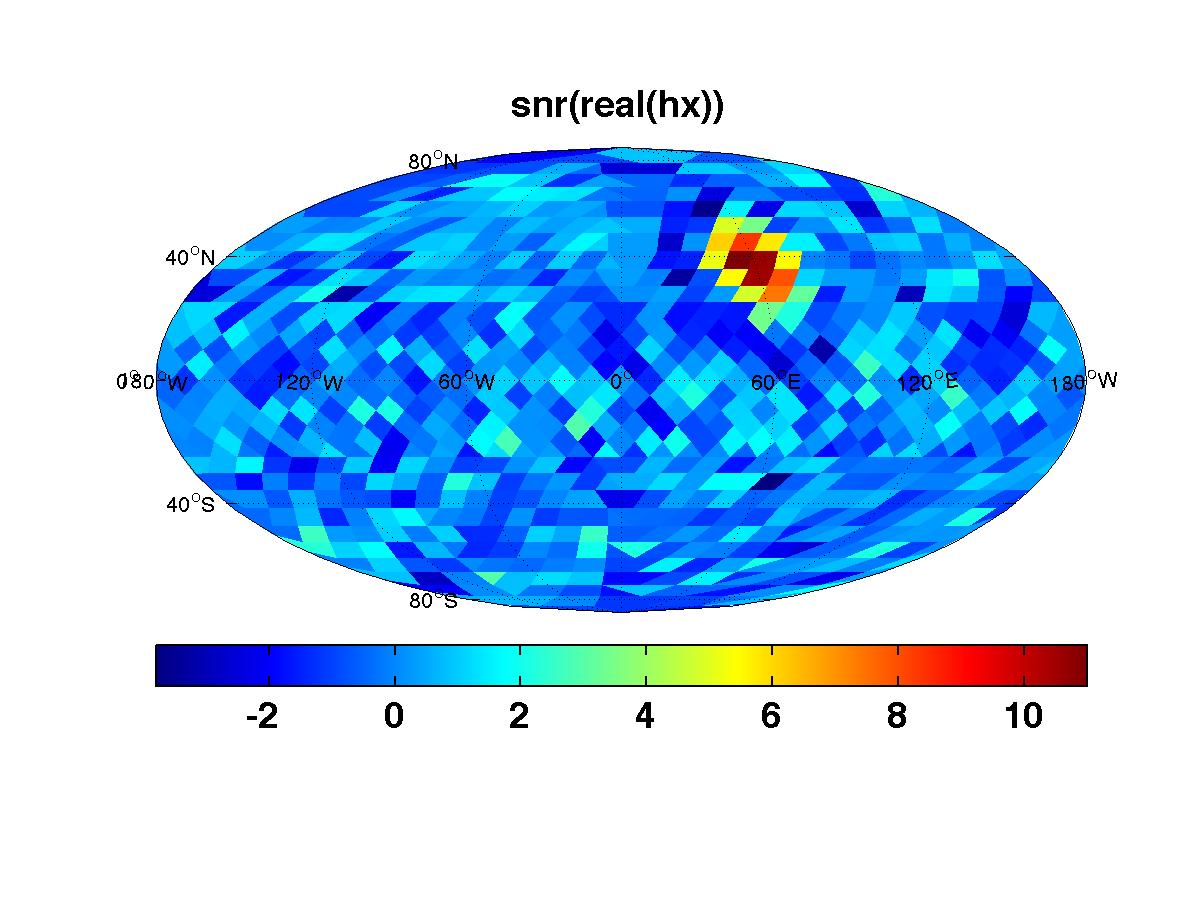}
\includegraphics[trim=3cm 4cm 3cm 2.5cm, clip=true, angle=0, width=0.24\textwidth]{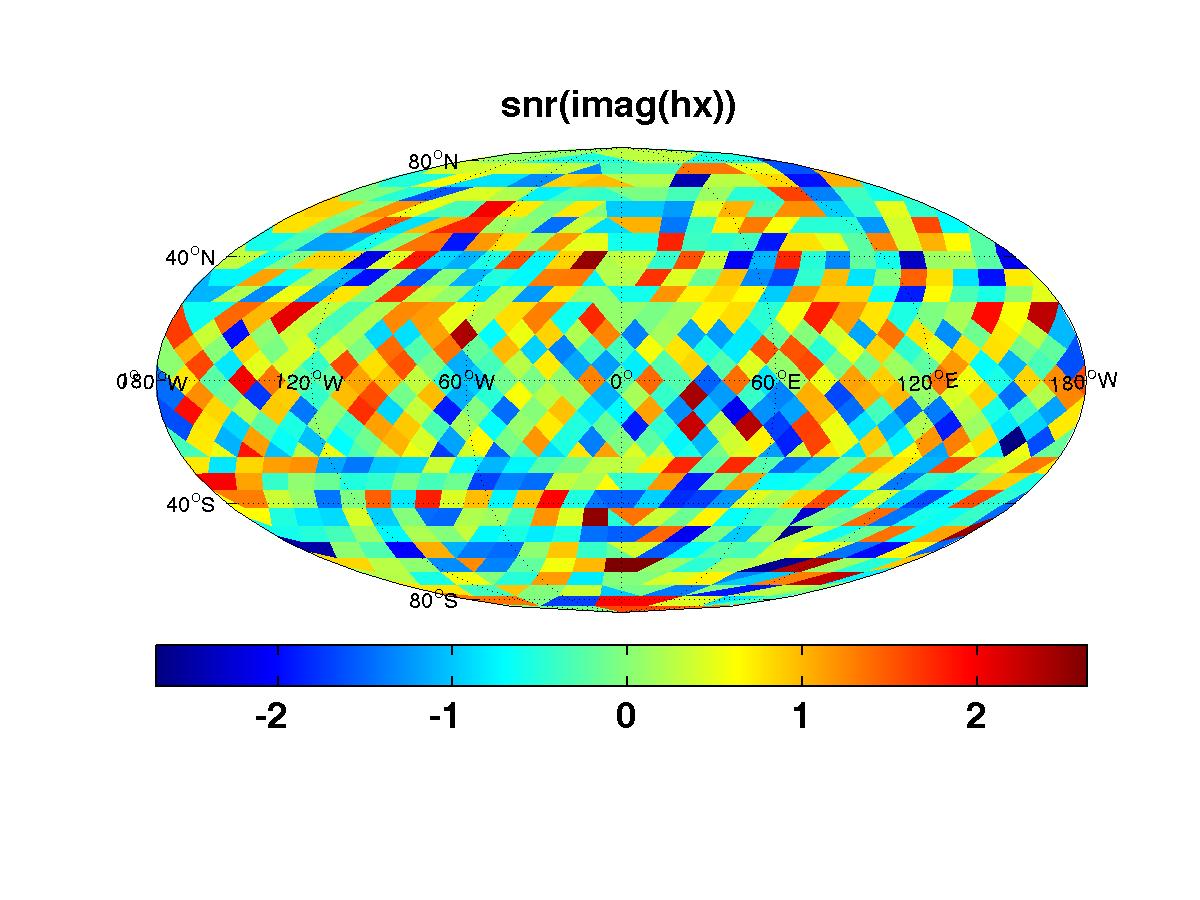}
\caption
{Same as Fig.~\ref{f:point_recovery}, but for the 3-detector network having the same 
number of total data points ($N=2400$) as the 6-detector network.
Injected maps (first row);
maximum-likehood recovered maps (second row);
uncertainty map (third row);
SNR map (last row).
The maps are more-or-less the same as for the 6-detector network shown in Fig.~\ref{f:point_recovery}.}
\label{f:point_recovery_3_2}
\end{center}
\end{figure*}

Figure~\ref{f:point_recovery_3_2} is identical to 
Fig.~\ref{f:point_recovery}, but for the 3-detector network
having the same number of total data points ($N=2400$)
as the 6-detector network.
The total observation time is thus twice as long, in order to 
compensate for the reduction in the number of interferometers.
Note that the uncertainty maps have values that 
are slightly larger than for the 6-detector network.
Also, the match is $\mu=0.59$,
which is slightly smaller than that for the 6-detector 
network.
Thus, we see from this simulation that the 3-detector 
network performs {\em nearly as well}
as the 6-detector network if we integrate long
enough to acquire the same number of total data points.
The performance of the 3-detector network is much
worse than the 6-detector network if we integrate 
for the same observation time, since then the total 
number of data points for the 3-detector network 
is half as large
(see the fourth rows of Figs.~\ref{f:grad_recovery}
and \ref{f:curl_recovery} below).

Maximum-likelihood recovered maps for the grad-only and 
curl-only backgrounds using the 6-detector and two 
3-detector networks mentioned
above (one having the same number of total data 
points as the 6-detector network; the other having half 
as many data points) are shown in 
Figs.~\ref{f:grad_recovery} and \ref{f:curl_recovery}.
The corresponding match values for the grad and curl 
recoveries are $\mu = 0.81$ and $0.85$ for the 
6-detector network.
For the two 3-detector networks, the match values 
are $\mu=0.80$ and $0.84$ when the total number of data
points is the same as for the 6-detector network,
and $\mu=0.55$ and $0.60$ when the total number of
data points is half as many.
The SNR values as a function of sky location for the
different recoveries range from about $-5$ to $5$ 
for the strong recoveries and $-4$ to $4$ for 
the weaker recoveries.
As can be seen from the fourth row of these two figures,
when the 3-detector network has only half as many total data 
points as the other two networks, the structure 
in the grad-only and curl-only sky maps is not nearly as clearly recovered
as for the other detector networks.

The most important take-home message is that 
the grad-only and curl-only backgrounds {\em can both 
be recovered} with a network of ground-based 
interferometers.
This is in contrast to the case for a pulsar timing array,
which is completely insensitive to a curl-only background
(see \cite{Gair-et-al:2014}, and in particular Fig.~11 from
that paper).
As mentioned earlier, the rotational and orbital motion
of the Earth synthesizes a set of virtual interferometers
that sample the gravitational-wave field from many 
different spatial locations.
This allows for the reconstruction of both grad and curl 
modes of the background, unlike the case for pulsar timing arrays.
\begin{figure*}[hbtp!]
\begin{center}
\includegraphics[trim=3cm 4cm 3cm 2.5cm, clip=true, angle=0, width=0.24\textwidth]{grad_realhp}
\includegraphics[trim=3cm 4cm 3cm 2.5cm, clip=true, angle=0, width=0.24\textwidth]{grad_imaghp}
\includegraphics[trim=3cm 4cm 3cm 2.5cm, clip=true, angle=0, width=0.24\textwidth]{grad_realhc}
\includegraphics[trim=3cm 4cm 3cm 2.5cm, clip=true, angle=0, width=0.24\textwidth]{grad_imaghc}
\includegraphics[trim=3cm 4cm 3cm 2.5cm, clip=true, angle=0, width=0.24\textwidth]{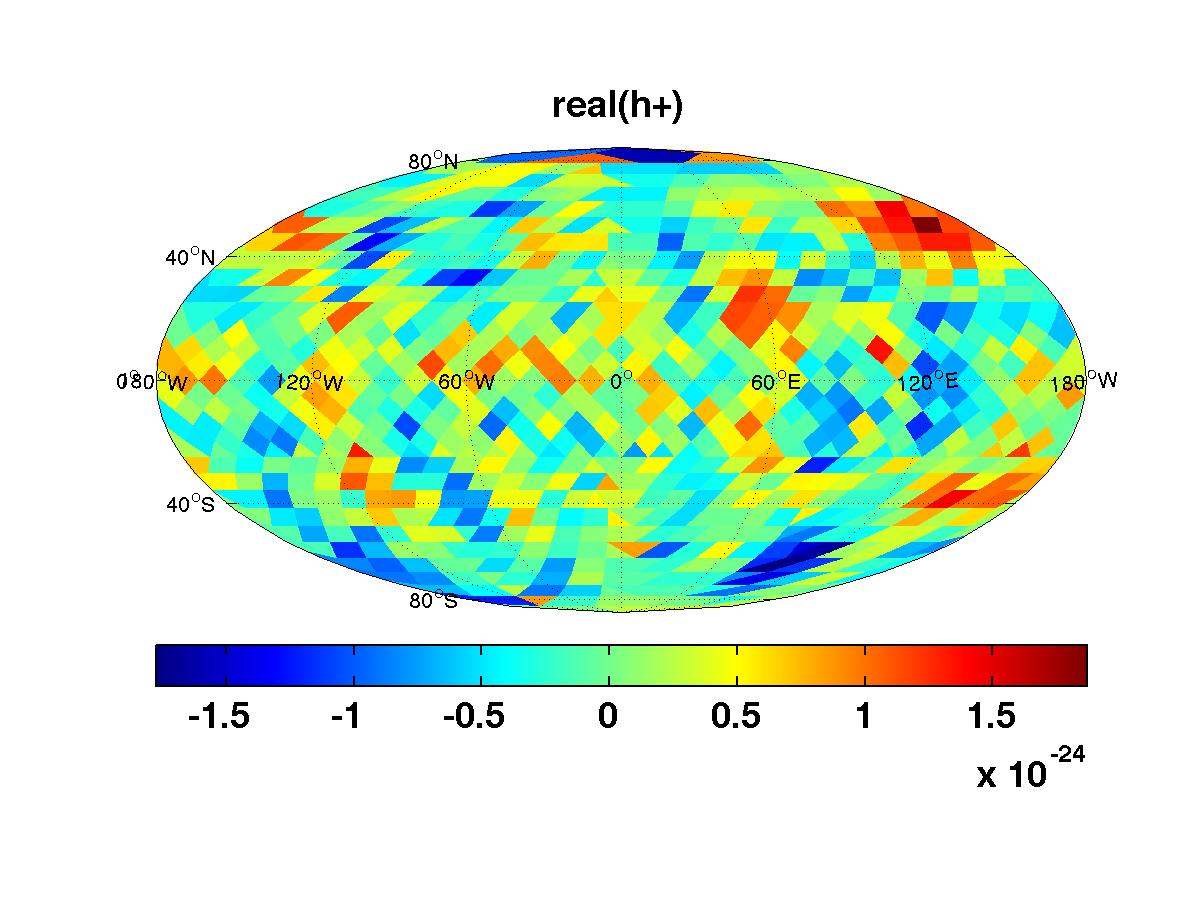}
\includegraphics[trim=3cm 4cm 3cm 2.5cm, clip=true, angle=0, width=0.24\textwidth]{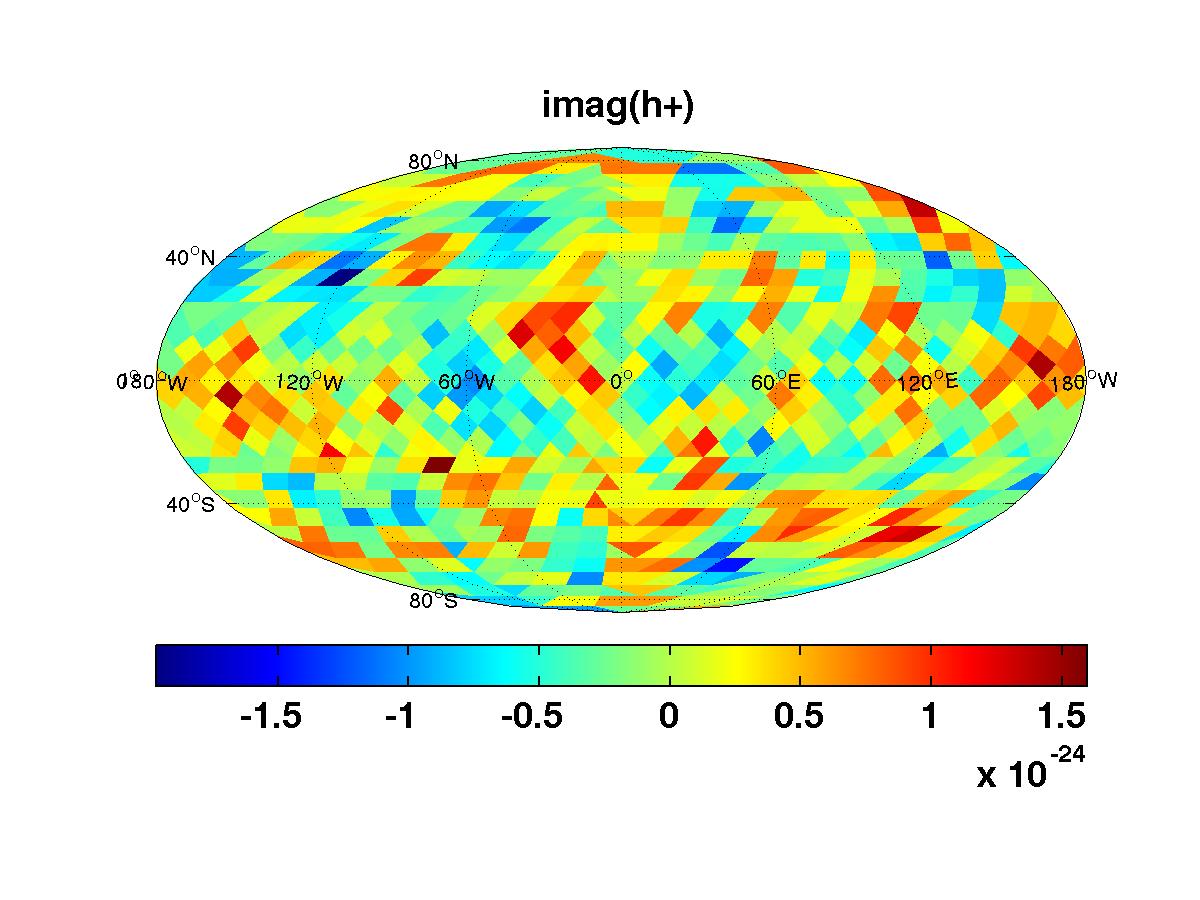}
\includegraphics[trim=3cm 4cm 3cm 2.5cm, clip=true, angle=0, width=0.24\textwidth]{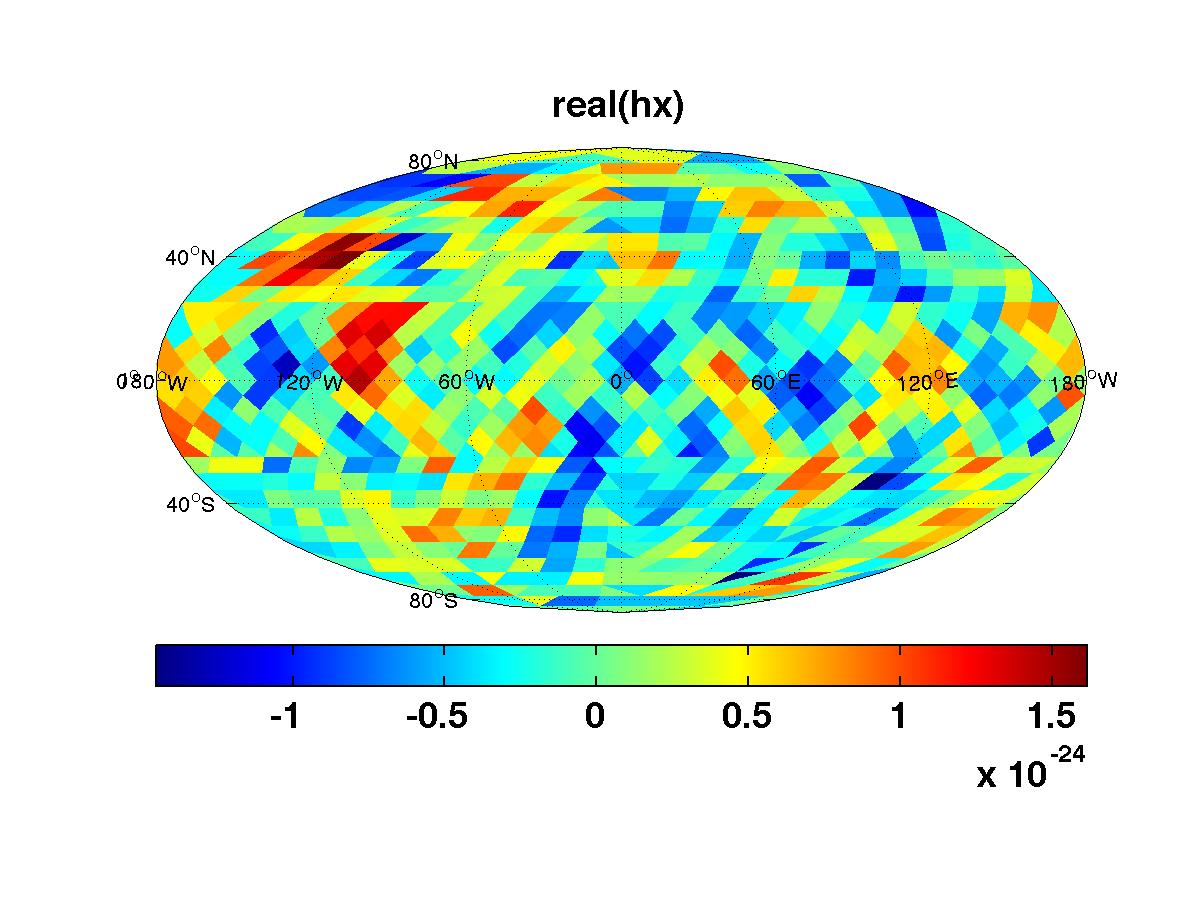}
\includegraphics[trim=3cm 4cm 3cm 2.5cm, clip=true, angle=0, width=0.24\textwidth]{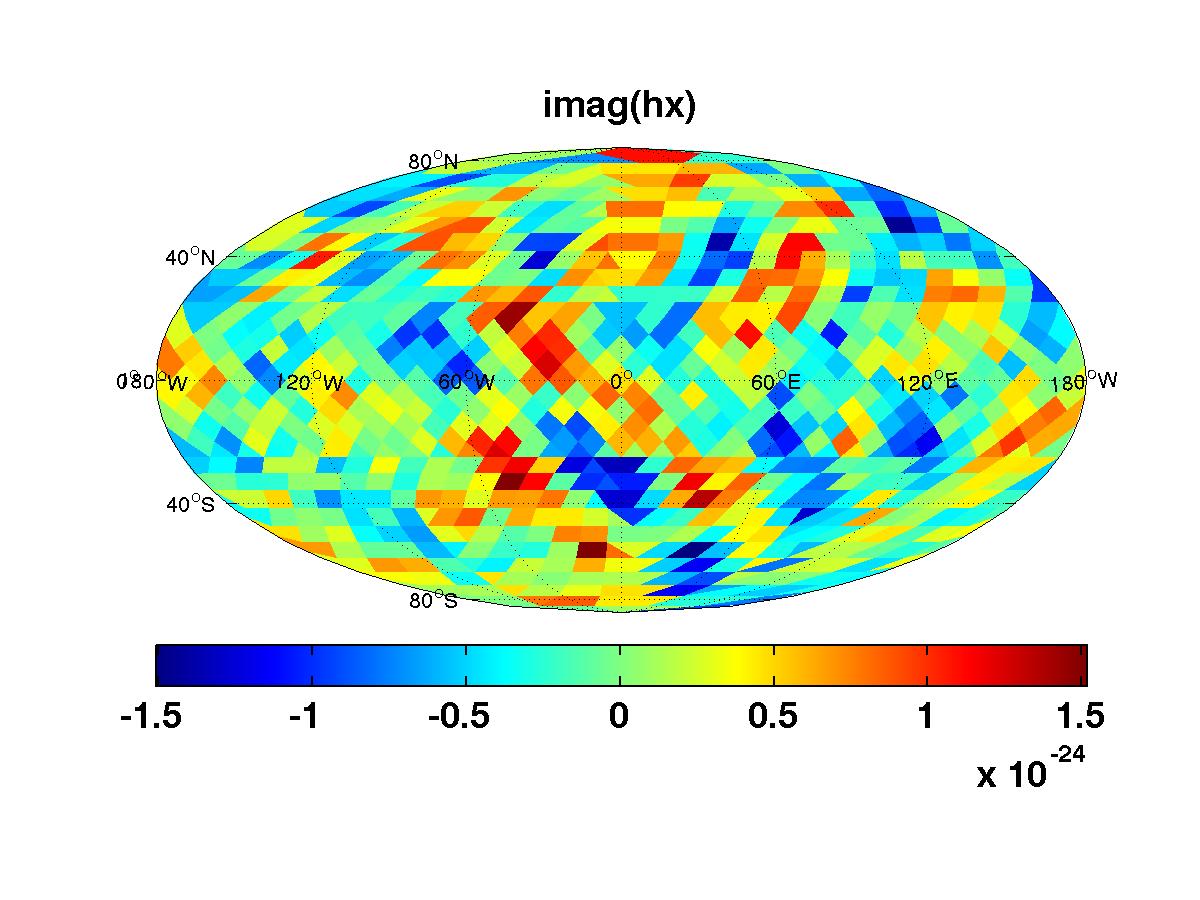}
\includegraphics[trim=3cm 4cm 3cm 2.5cm, clip=true, angle=0, width=0.24\textwidth]{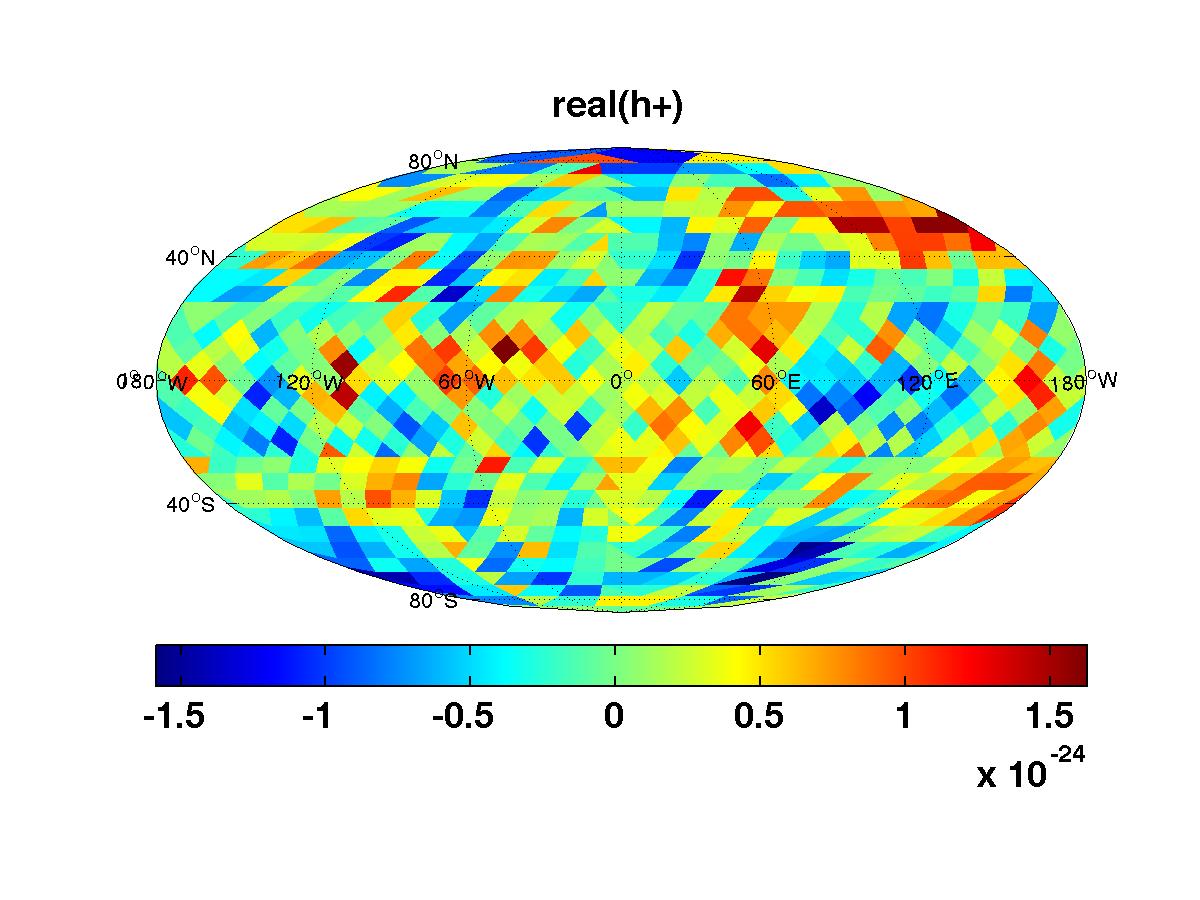}
\includegraphics[trim=3cm 4cm 3cm 2.5cm, clip=true, angle=0, width=0.24\textwidth]{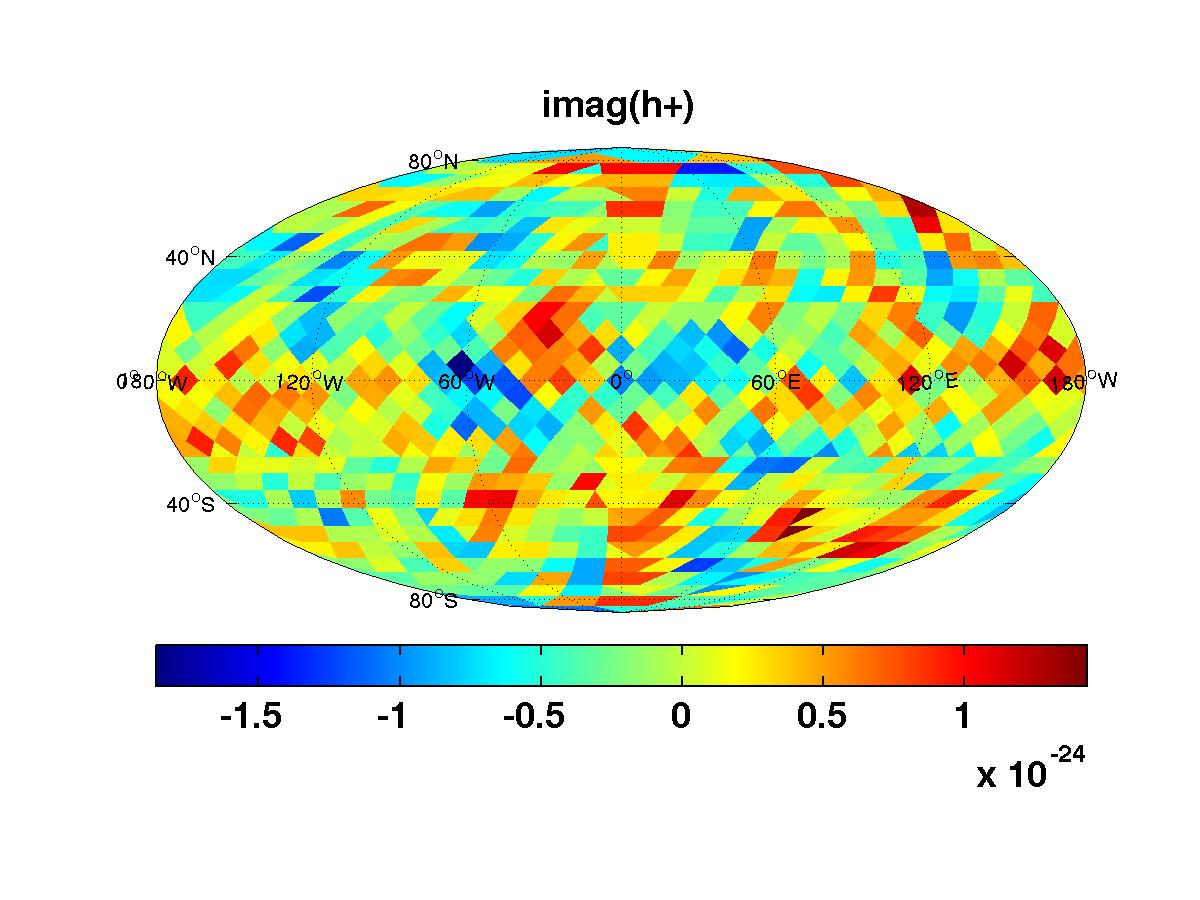}
\includegraphics[trim=3cm 4cm 3cm 2.5cm, clip=true, angle=0, width=0.24\textwidth]{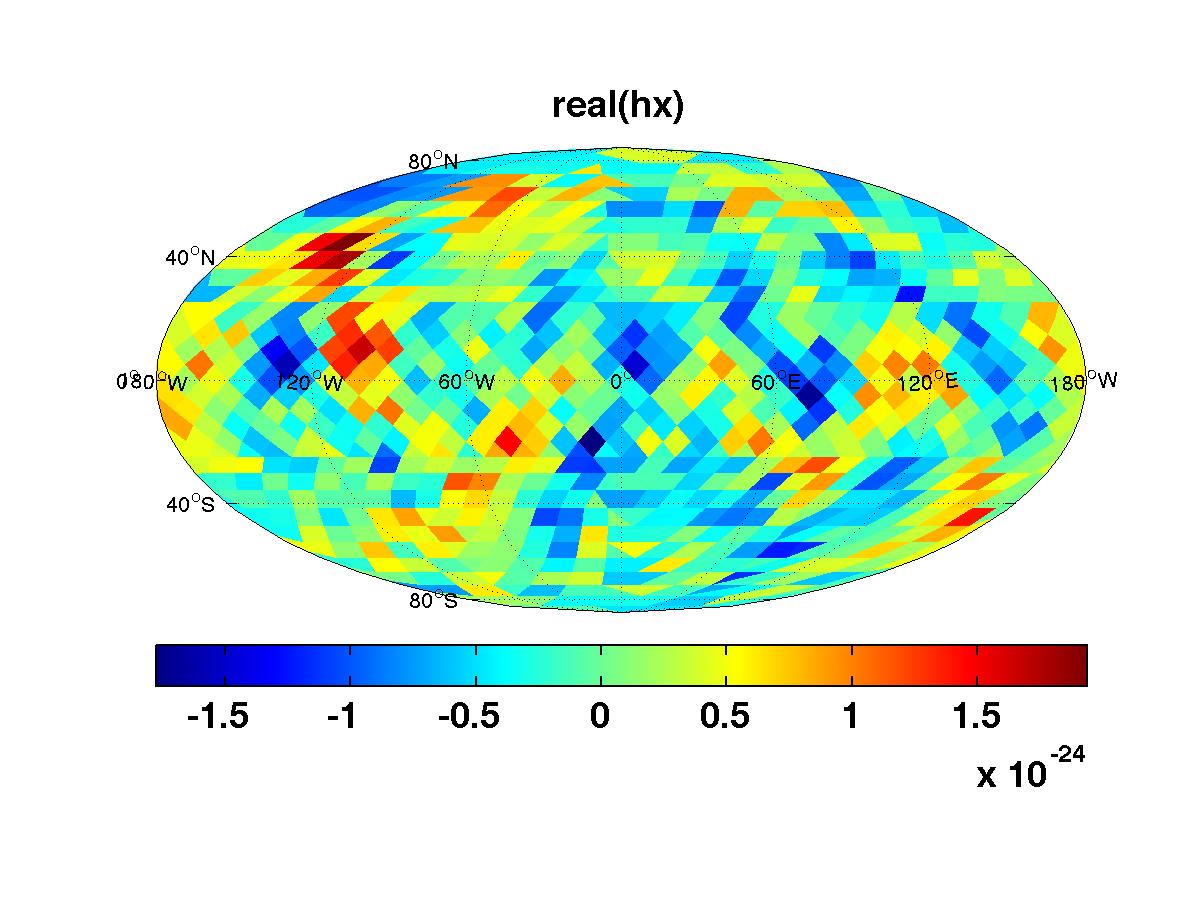}
\includegraphics[trim=3cm 4cm 3cm 2.5cm, clip=true, angle=0, width=0.24\textwidth]{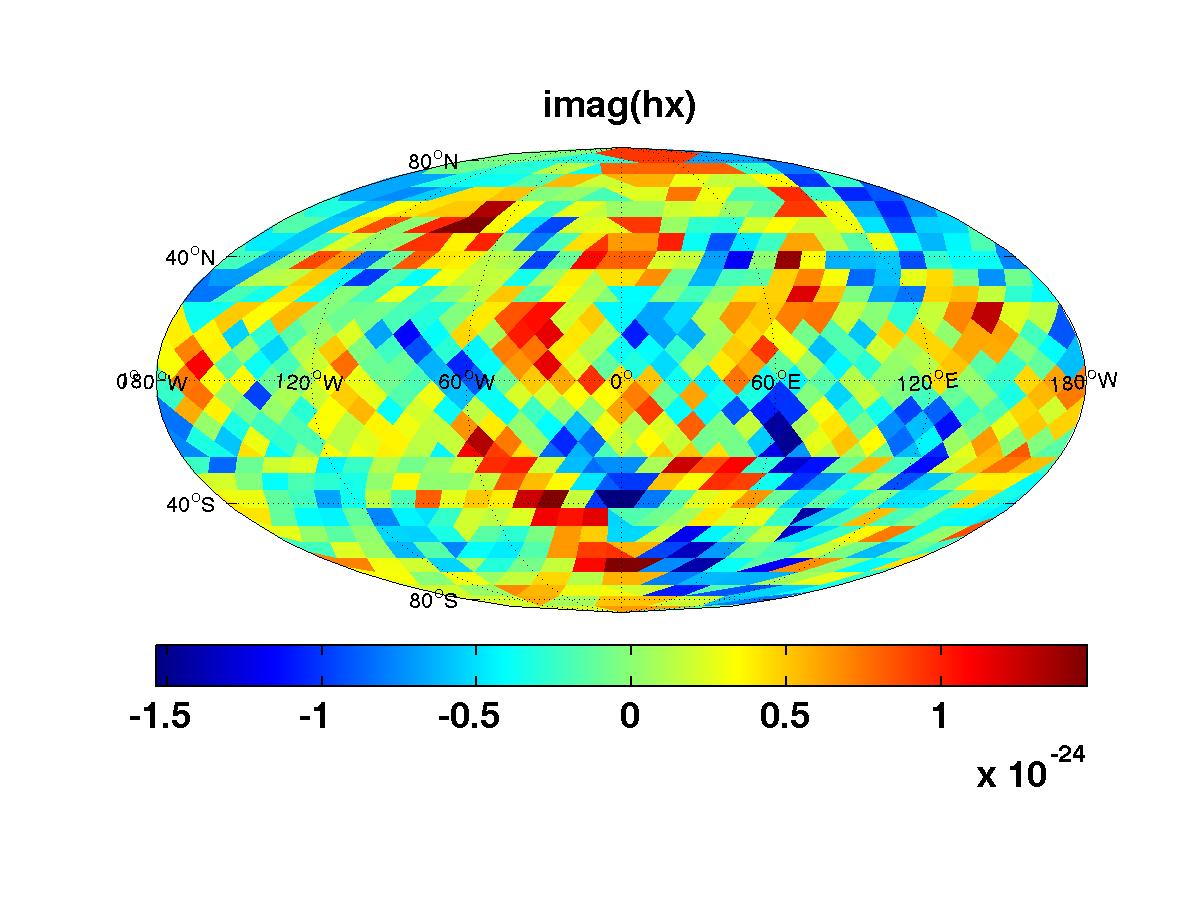}
\includegraphics[trim=3cm 4cm 3cm 2.5cm, clip=true, angle=0, width=0.24\textwidth]{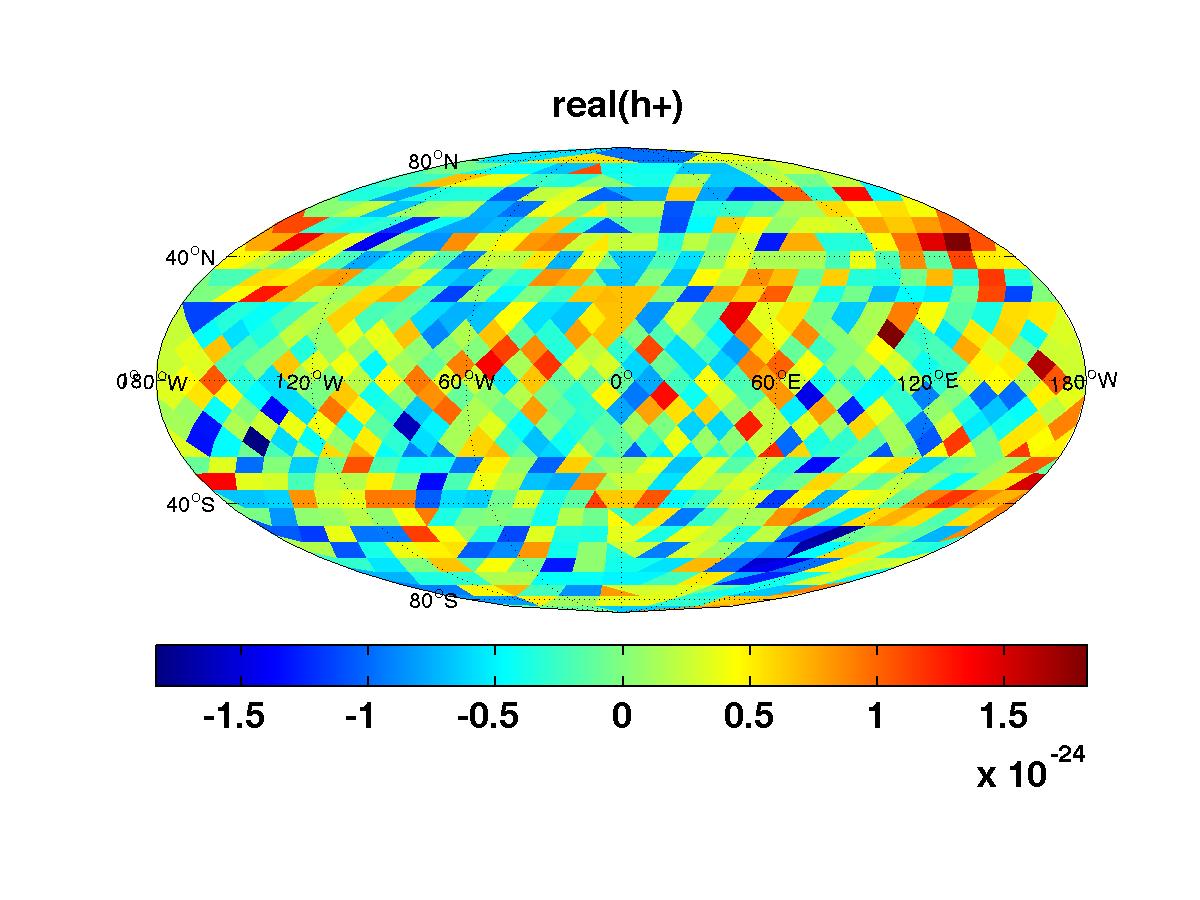}
\includegraphics[trim=3cm 4cm 3cm 2.5cm, clip=true, angle=0, width=0.24\textwidth]{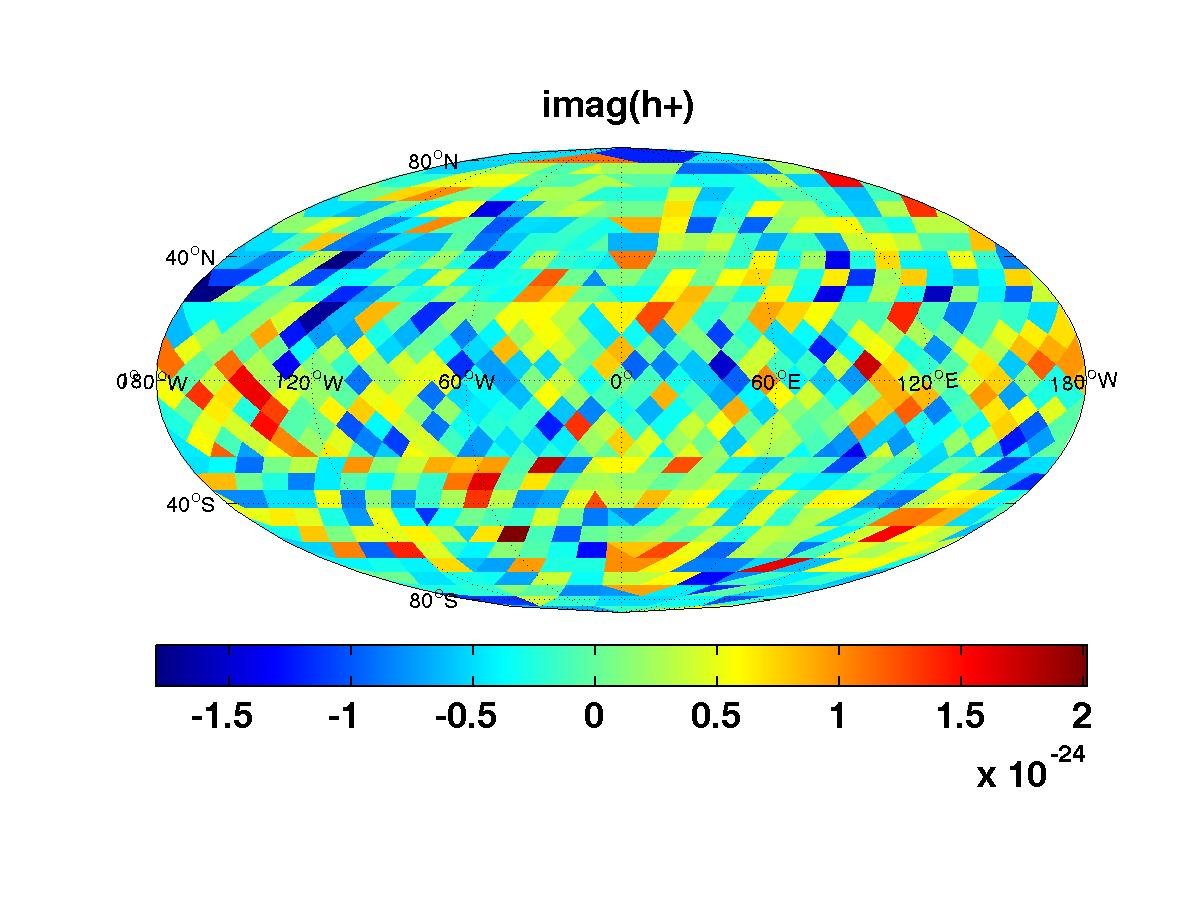}
\includegraphics[trim=3cm 4cm 3cm 2.5cm, clip=true, angle=0, width=0.24\textwidth]{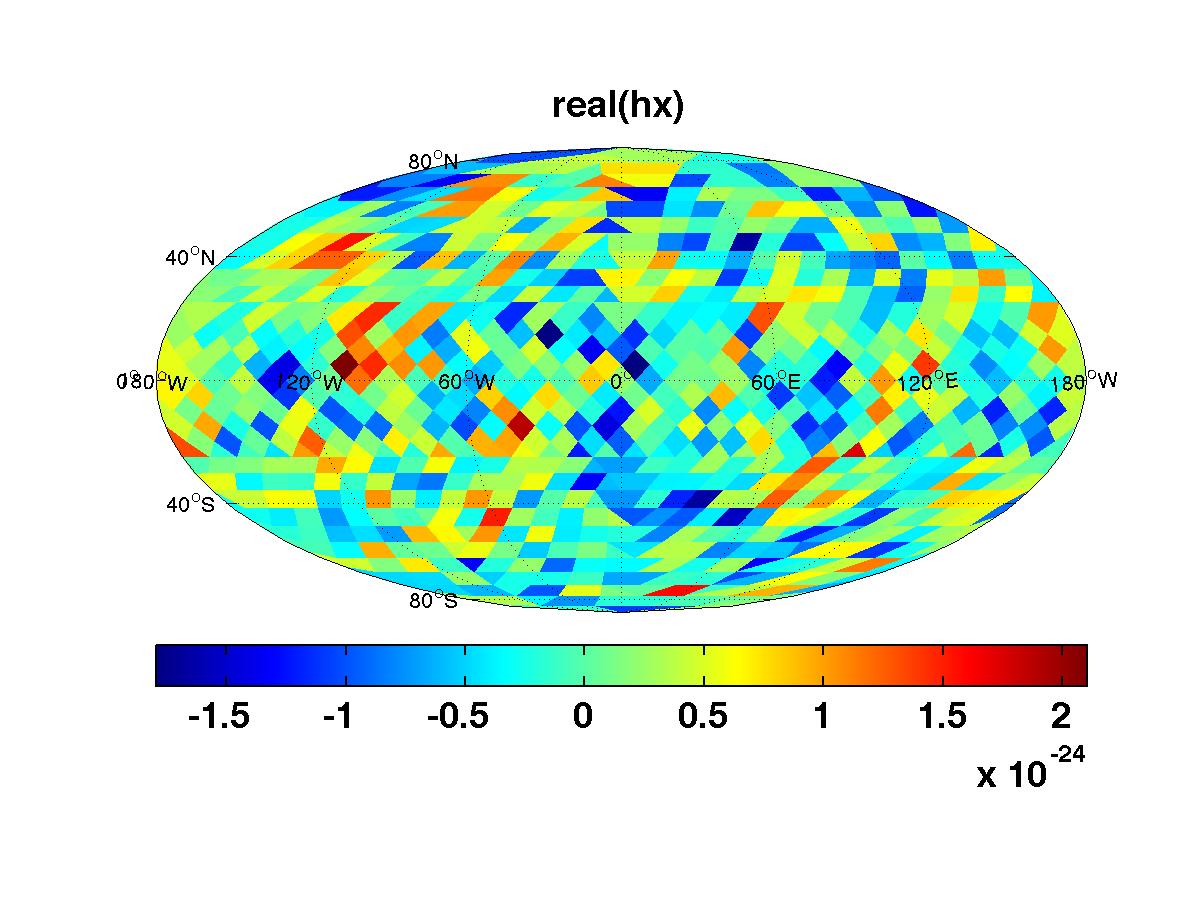}
\includegraphics[trim=3cm 4cm 3cm 2.5cm, clip=true, angle=0, width=0.24\textwidth]{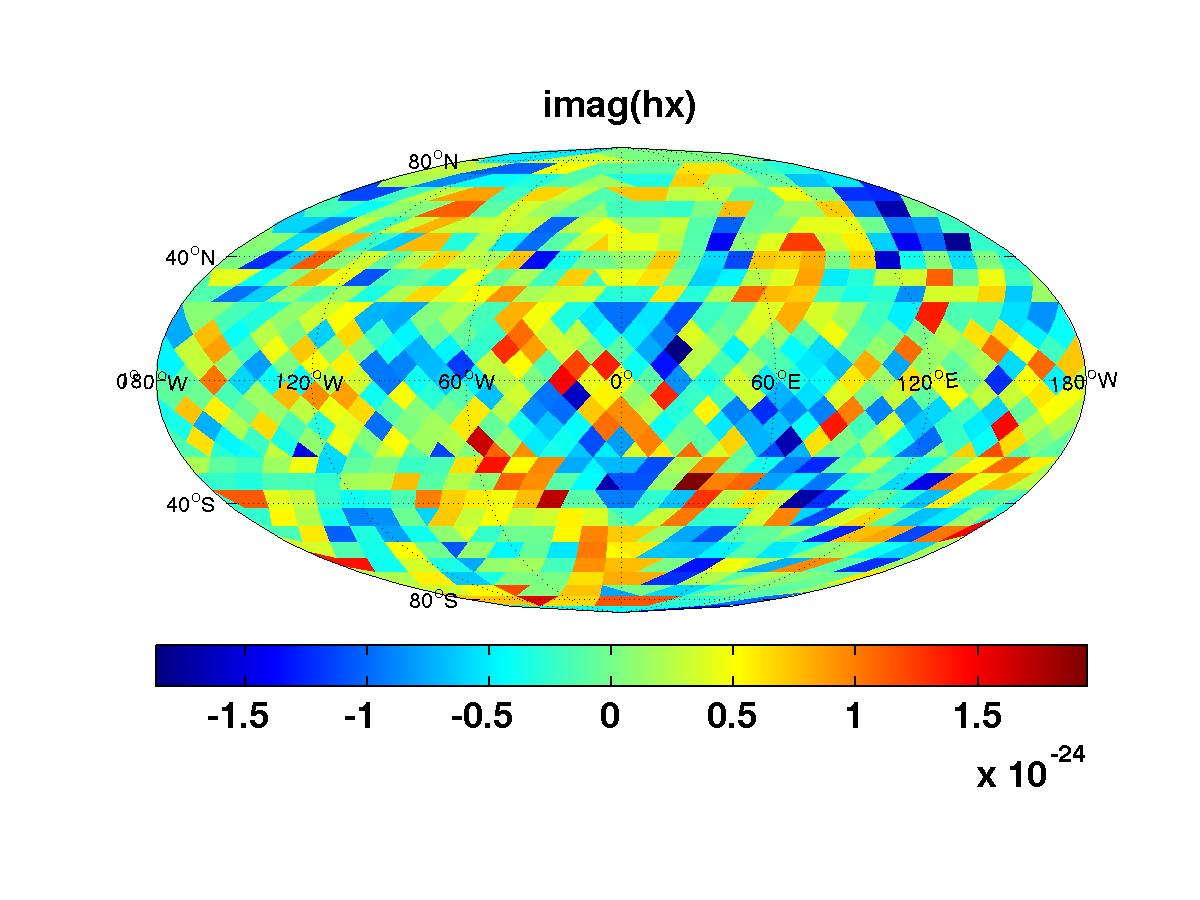}
\caption
{Recovery of the grad-only background in noise.
Injected maps (first row);
recovered maps for the 6-detector network (second row);
recovered maps for the 3-detector network having the same number of total data points 
($N=2400$) as the 6-detector network (third row);
recovered maps for the 3-detector network having half as many total data points 
($N=1200$) as the 6-detector network (fourth row).}
\label{f:grad_recovery}
\end{center}
\end{figure*}
\begin{figure*}[hbtp!]
\begin{center}
\includegraphics[trim=3cm 4cm 3cm 2.5cm, clip=true, angle=0, width=0.24\textwidth]{curl_realhp}
\includegraphics[trim=3cm 4cm 3cm 2.5cm, clip=true, angle=0, width=0.24\textwidth]{curl_imaghp}
\includegraphics[trim=3cm 4cm 3cm 2.5cm, clip=true, angle=0, width=0.24\textwidth]{curl_realhc}
\includegraphics[trim=3cm 4cm 3cm 2.5cm, clip=true, angle=0, width=0.24\textwidth]{curl_imaghc}
\includegraphics[trim=3cm 4cm 3cm 2.5cm, clip=true, angle=0, width=0.24\textwidth]{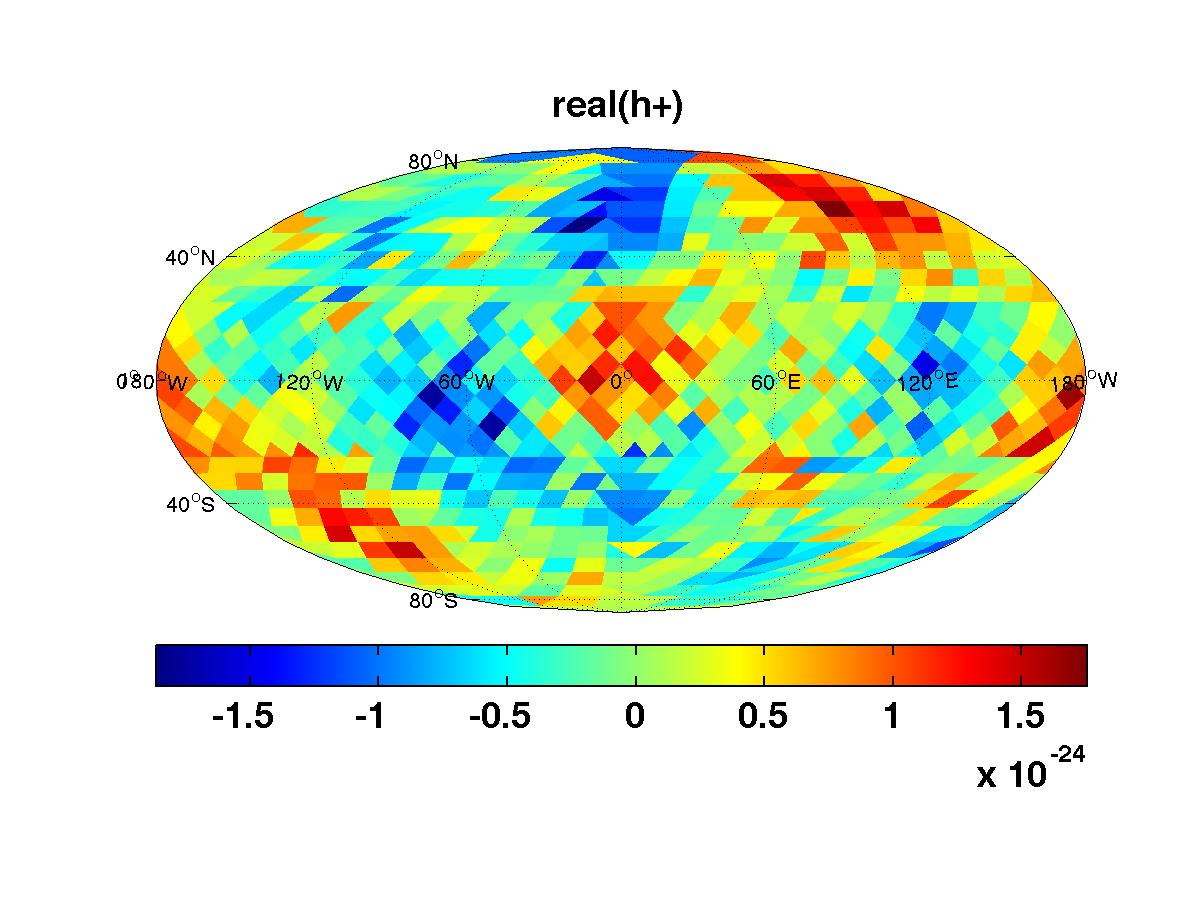}
\includegraphics[trim=3cm 4cm 3cm 2.5cm, clip=true, angle=0, width=0.24\textwidth]{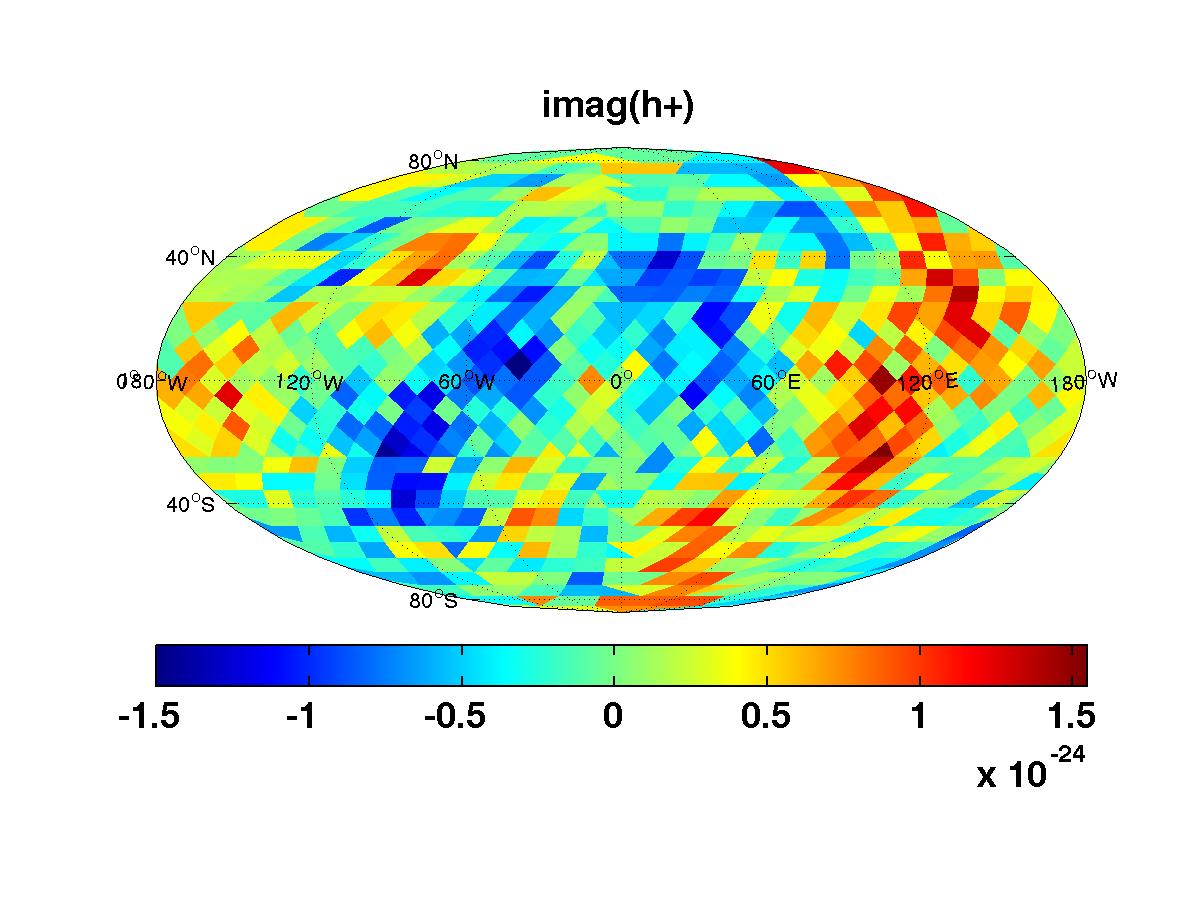}
\includegraphics[trim=3cm 4cm 3cm 2.5cm, clip=true, angle=0, width=0.24\textwidth]{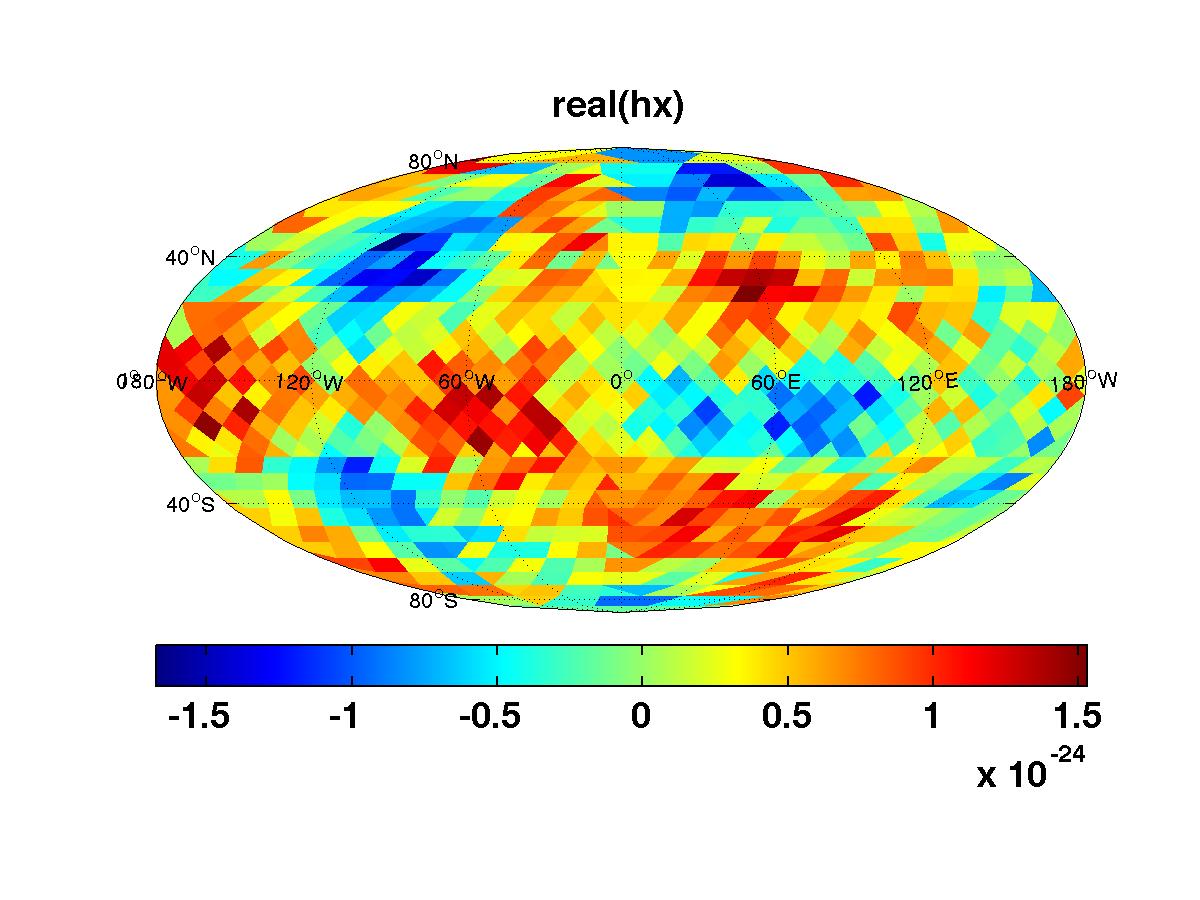}
\includegraphics[trim=3cm 4cm 3cm 2.5cm, clip=true, angle=0, width=0.24\textwidth]{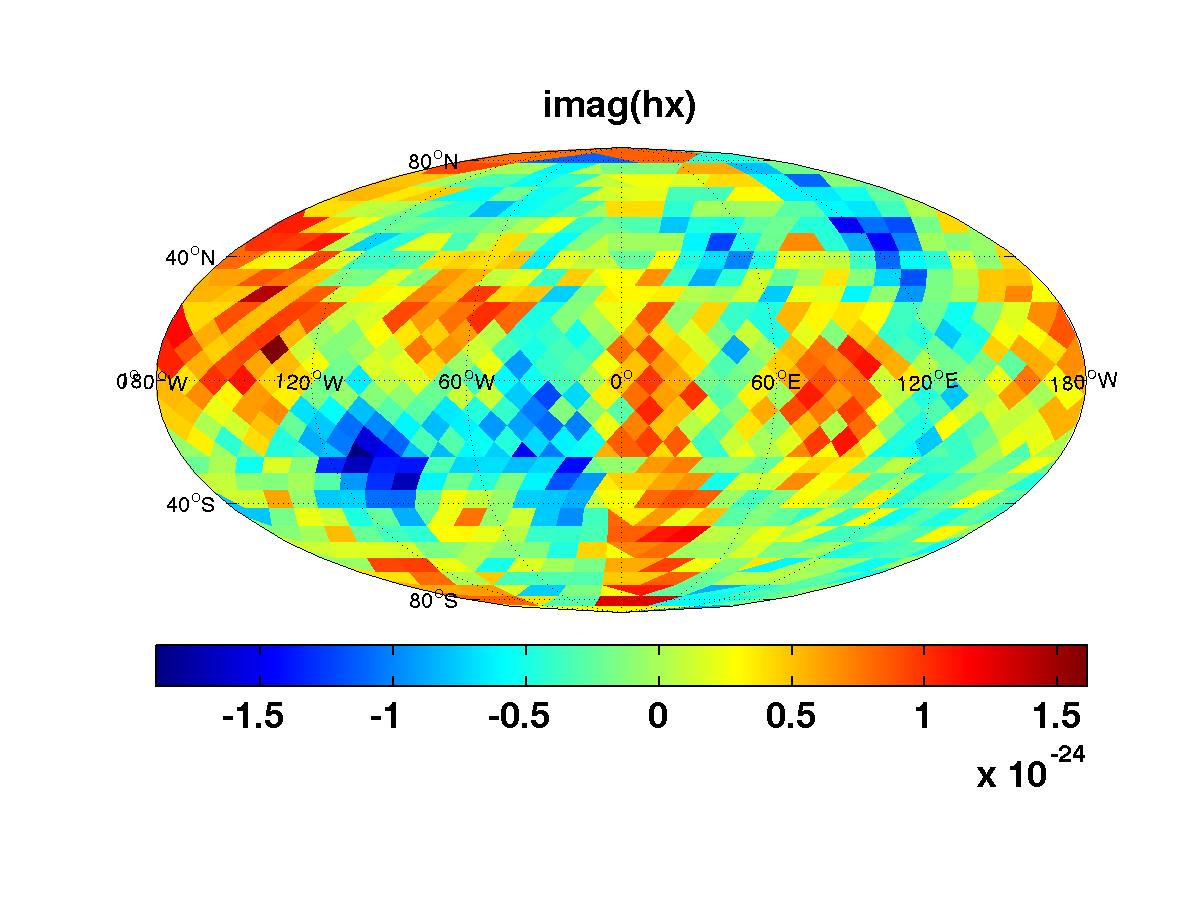}
\includegraphics[trim=3cm 4cm 3cm 2.5cm, clip=true, angle=0, width=0.24\textwidth]{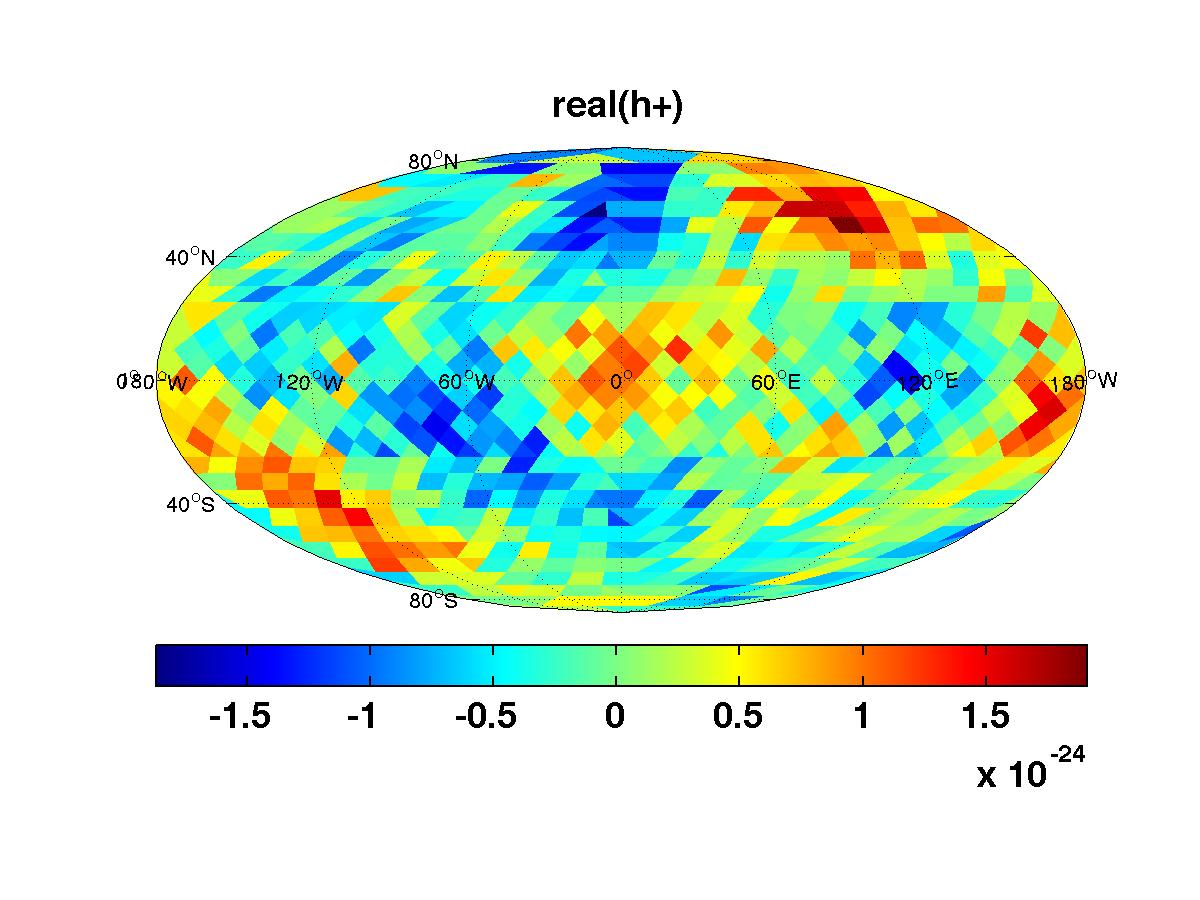}
\includegraphics[trim=3cm 4cm 3cm 2.5cm, clip=true, angle=0, width=0.24\textwidth]{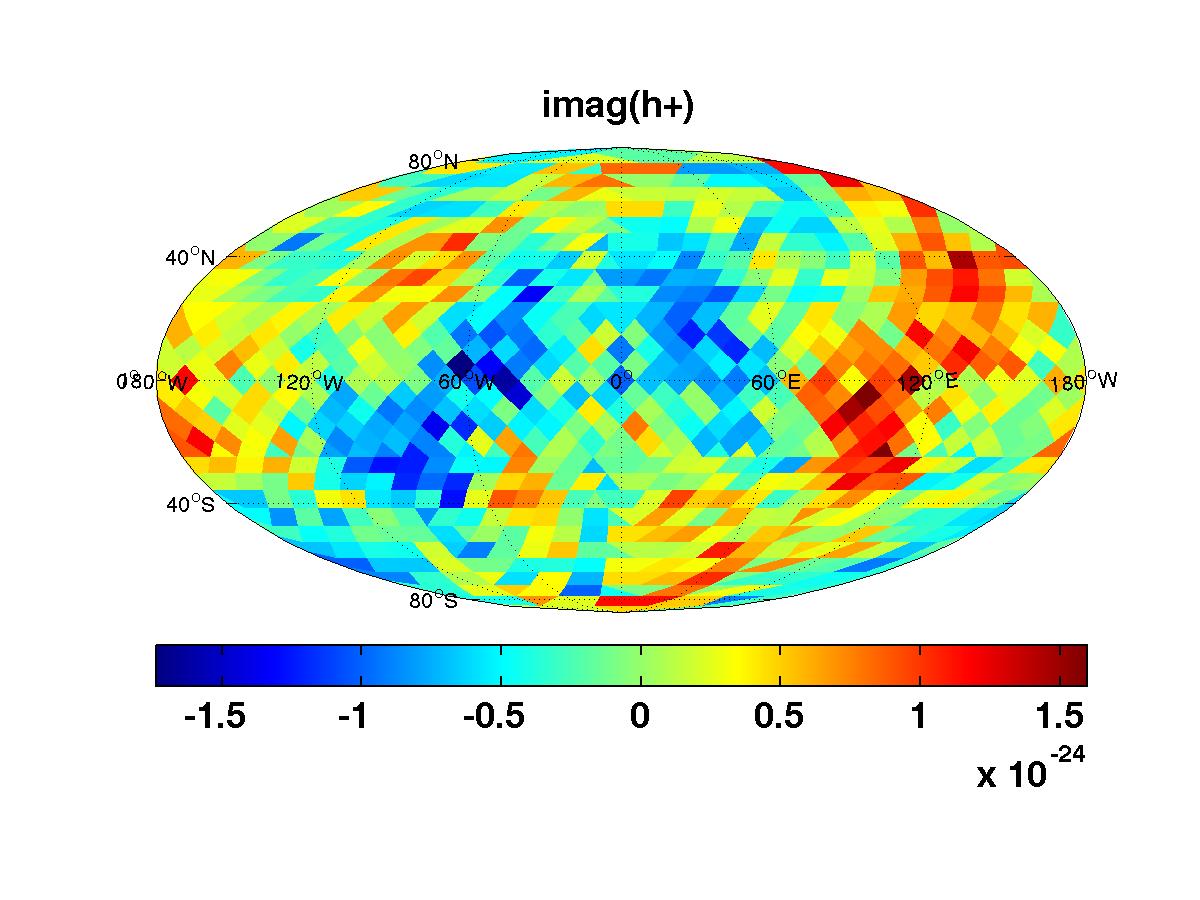}
\includegraphics[trim=3cm 4cm 3cm 2.5cm, clip=true, angle=0, width=0.24\textwidth]{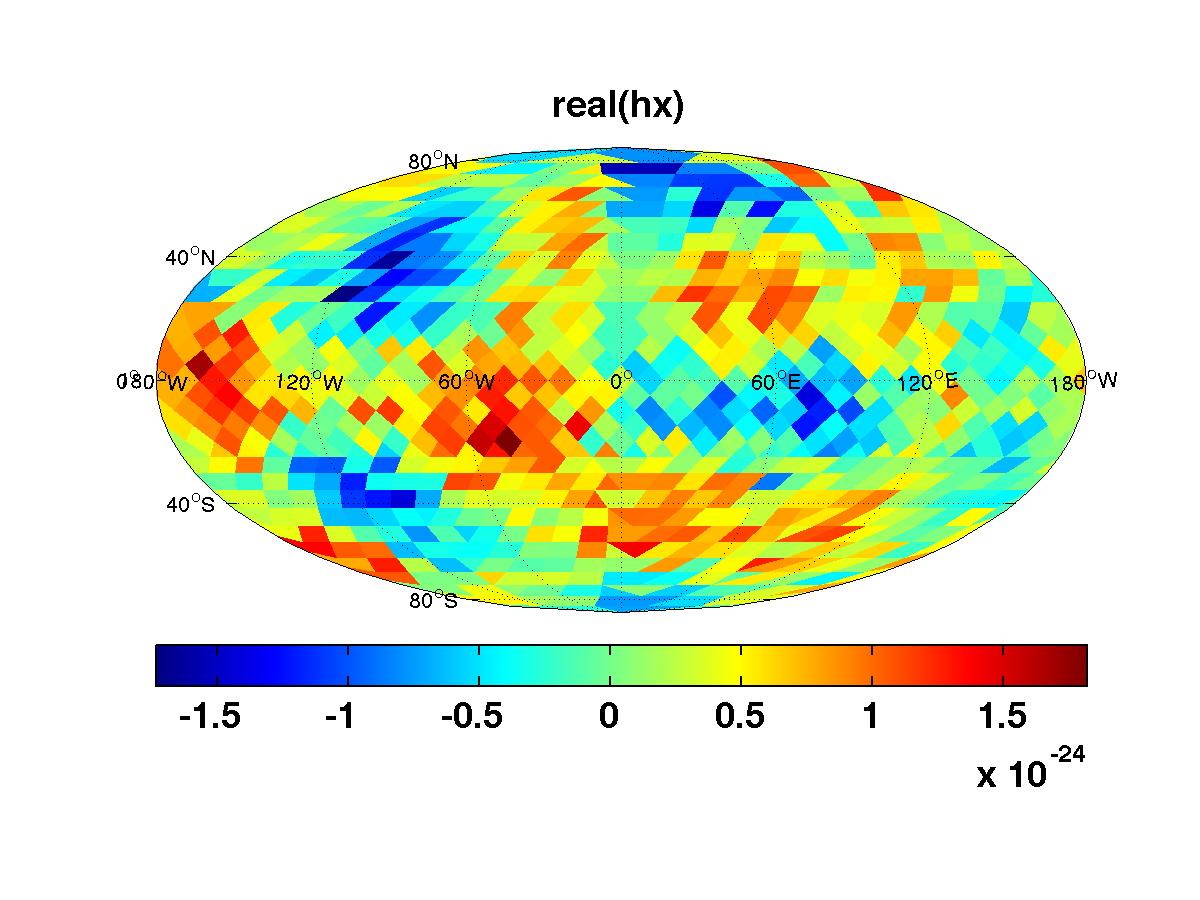}
\includegraphics[trim=3cm 4cm 3cm 2.5cm, clip=true, angle=0, width=0.24\textwidth]{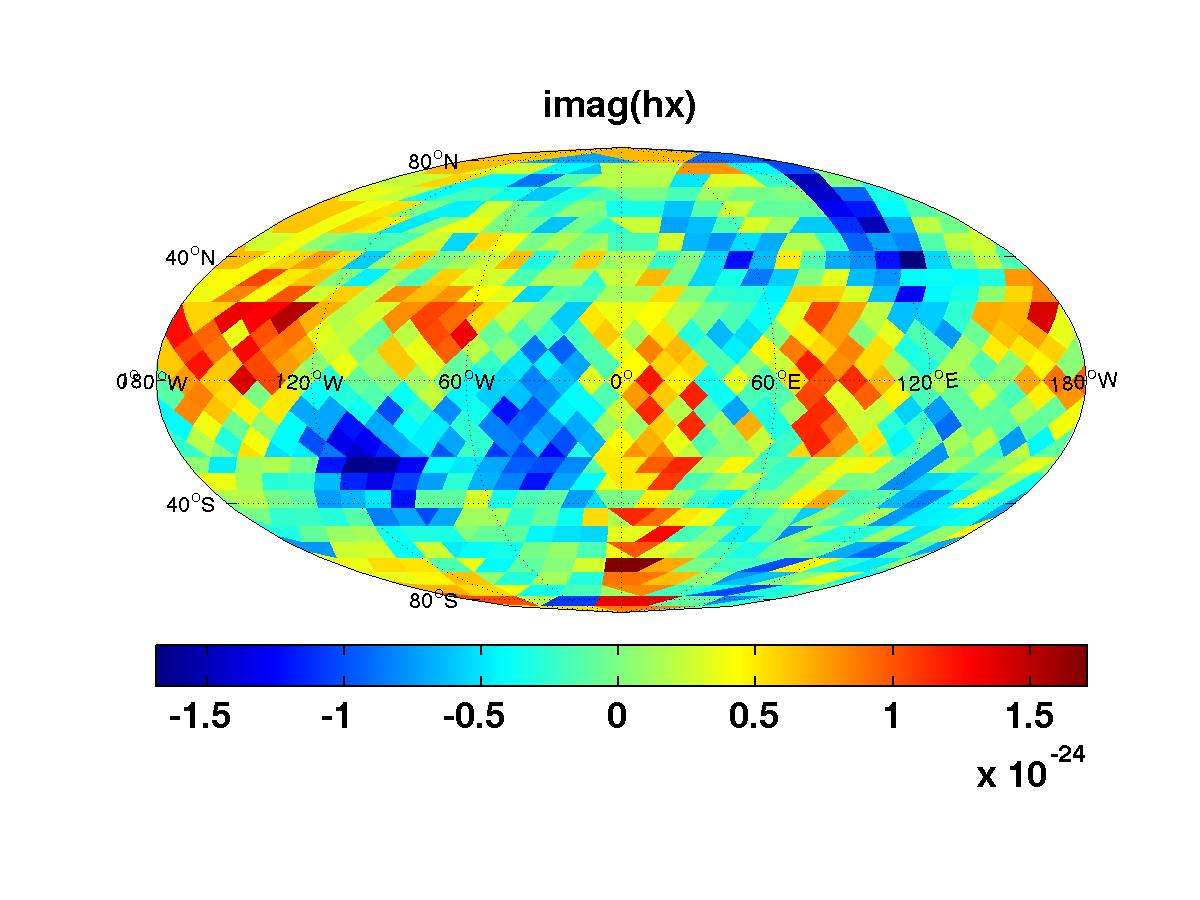}
\includegraphics[trim=3cm 4cm 3cm 2.5cm, clip=true, angle=0, width=0.24\textwidth]{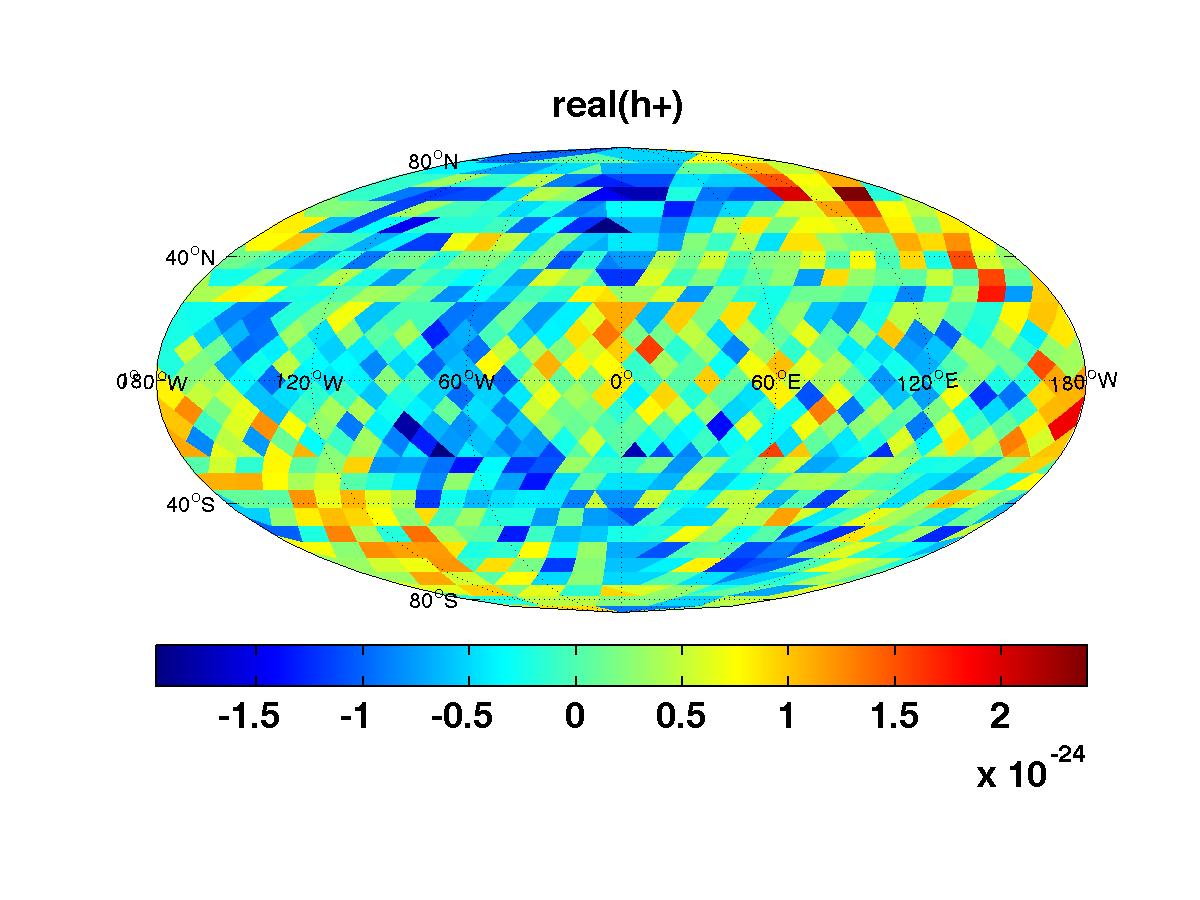}
\includegraphics[trim=3cm 4cm 3cm 2.5cm, clip=true, angle=0, width=0.24\textwidth]{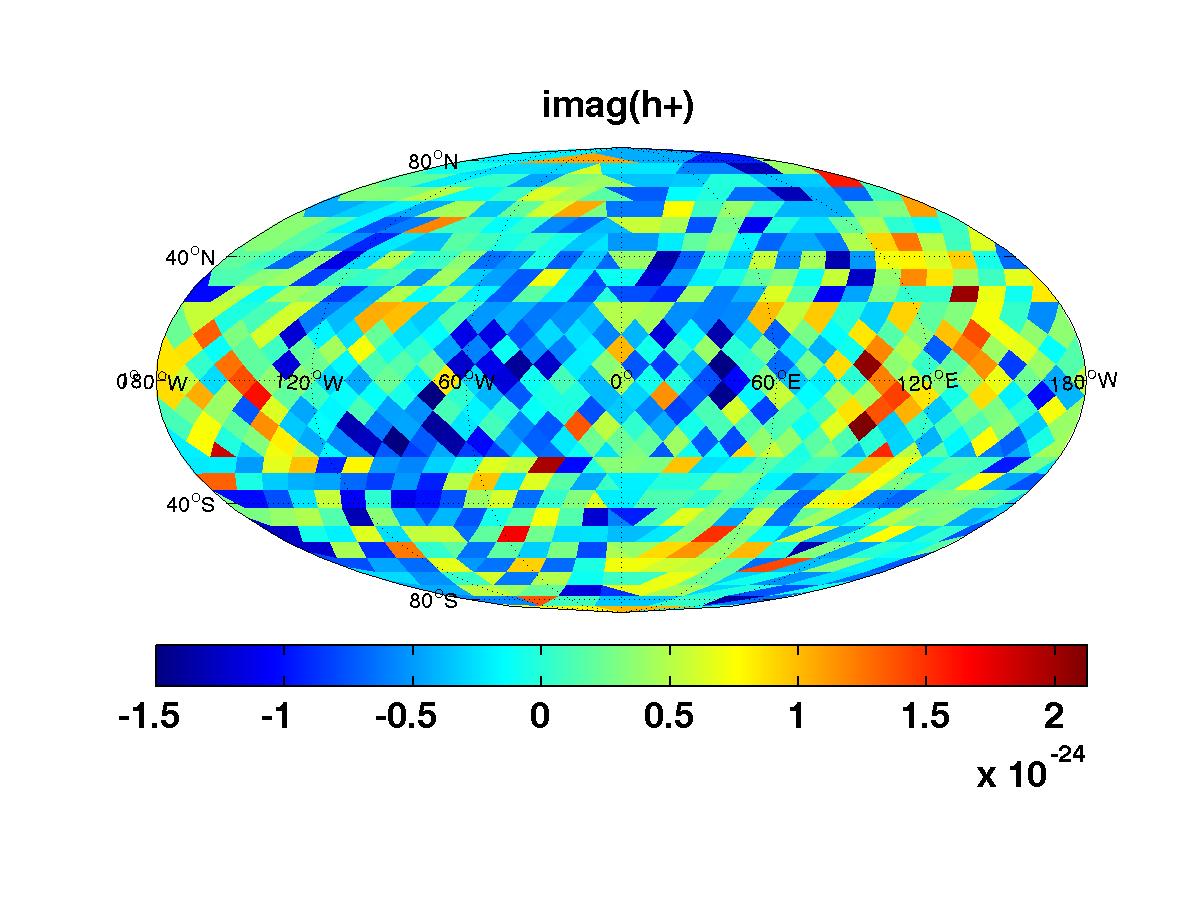}
\includegraphics[trim=3cm 4cm 3cm 2.5cm, clip=true, angle=0, width=0.24\textwidth]{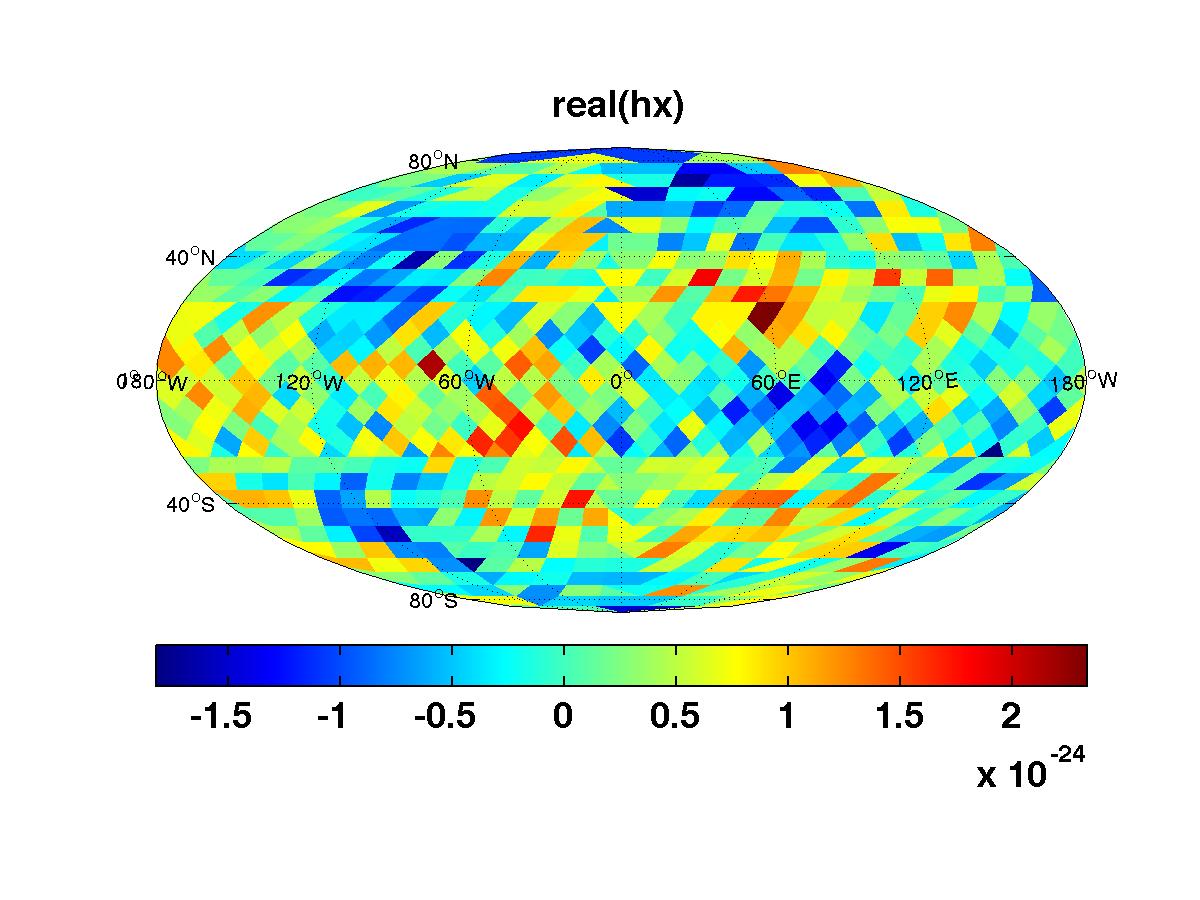}
\includegraphics[trim=3cm 4cm 3cm 2.5cm, clip=true, angle=0, width=0.24\textwidth]{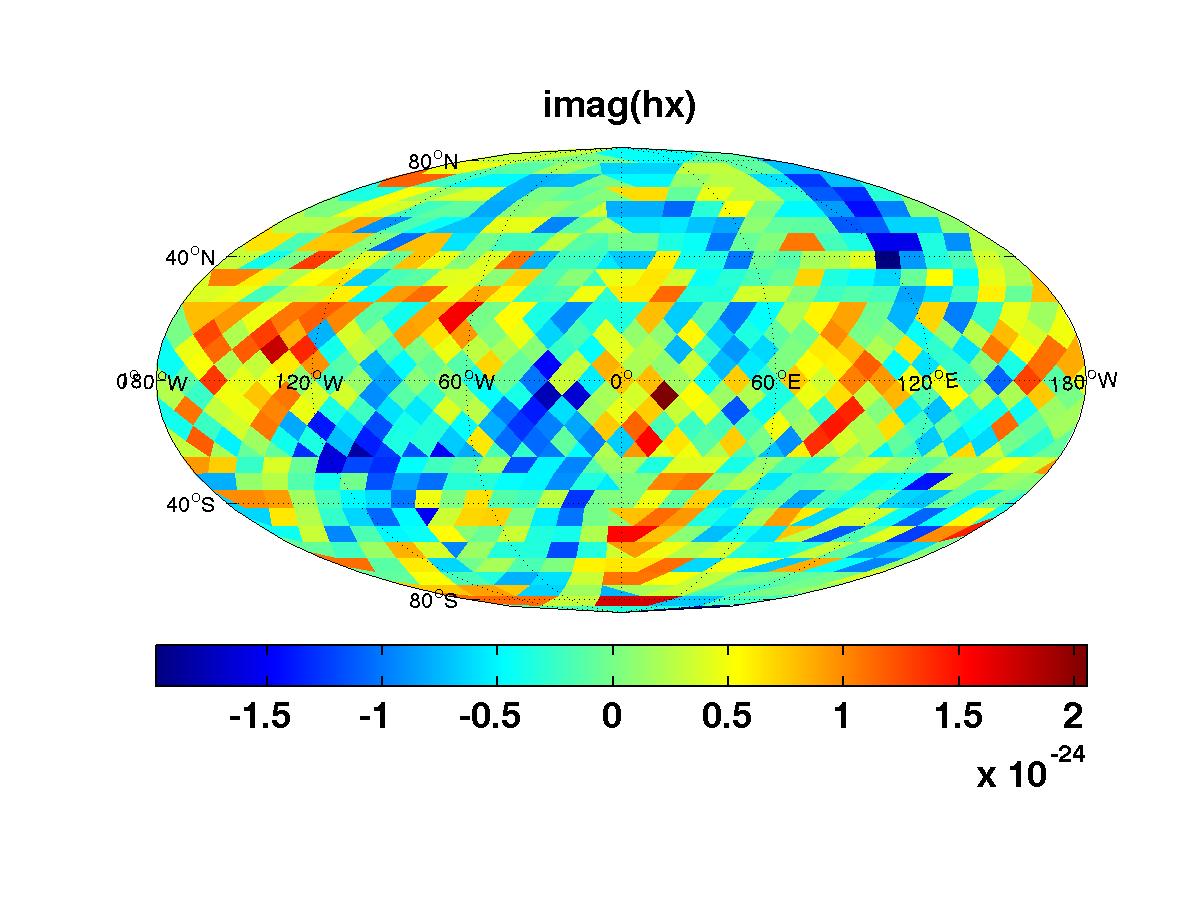}
\caption
{Recovery of the curl-only background in noise.
Injected maps (first row);
recovered maps for the 6-detector network (second row);
recovered maps for the 3-detector network having the same number of total data points 
($N=2400$) as the 6-detector network (third row);
recovered maps for the 3-detector network having half as many total data points 
($N=1200$) as the 6-detector network (fourth row).}
\label{f:curl_recovery}
\end{center}
\end{figure*}
%
%

\subsection{Minimum duration between data segments for 
independent measurements}
\label{s:min_duration}

Having shown our map recovery techniques to be successful in the
context of noisy simulated injections, we now return to an issue
discussed at the end of Sec.~\ref{s:rotation-orbit}---namely, the
minimum time increment between observations required to
synthesize a network of {\it independent} virtual interferometers, and
thus avoid degeneracies in the information content of our measured
strain signal. We consider two cases: (i) a single AdvLIGO Livingston
detector, and (ii) the full $6$-detector advanced network previously
discussed. In both cases we assume a total of $1200$ strain
measurements of the gravitational-wave sky have been recorded. 
As previously, 
the $h_+$, $h_\times$ components of the sky are decomposed into $768$
pixels for a total of $1536$ unknown parameters to be determined by
our search. We compute maximum likelihood maps from the
$1200$ observations, which are carried out over various total timespans to
investigate how the match of the recovered map with the injected map
scales with $\Delta t$, the time between observations.

\begin{figure*}[hbtp!]
\begin{center}
\includegraphics[angle=0,width=0.6\textwidth]{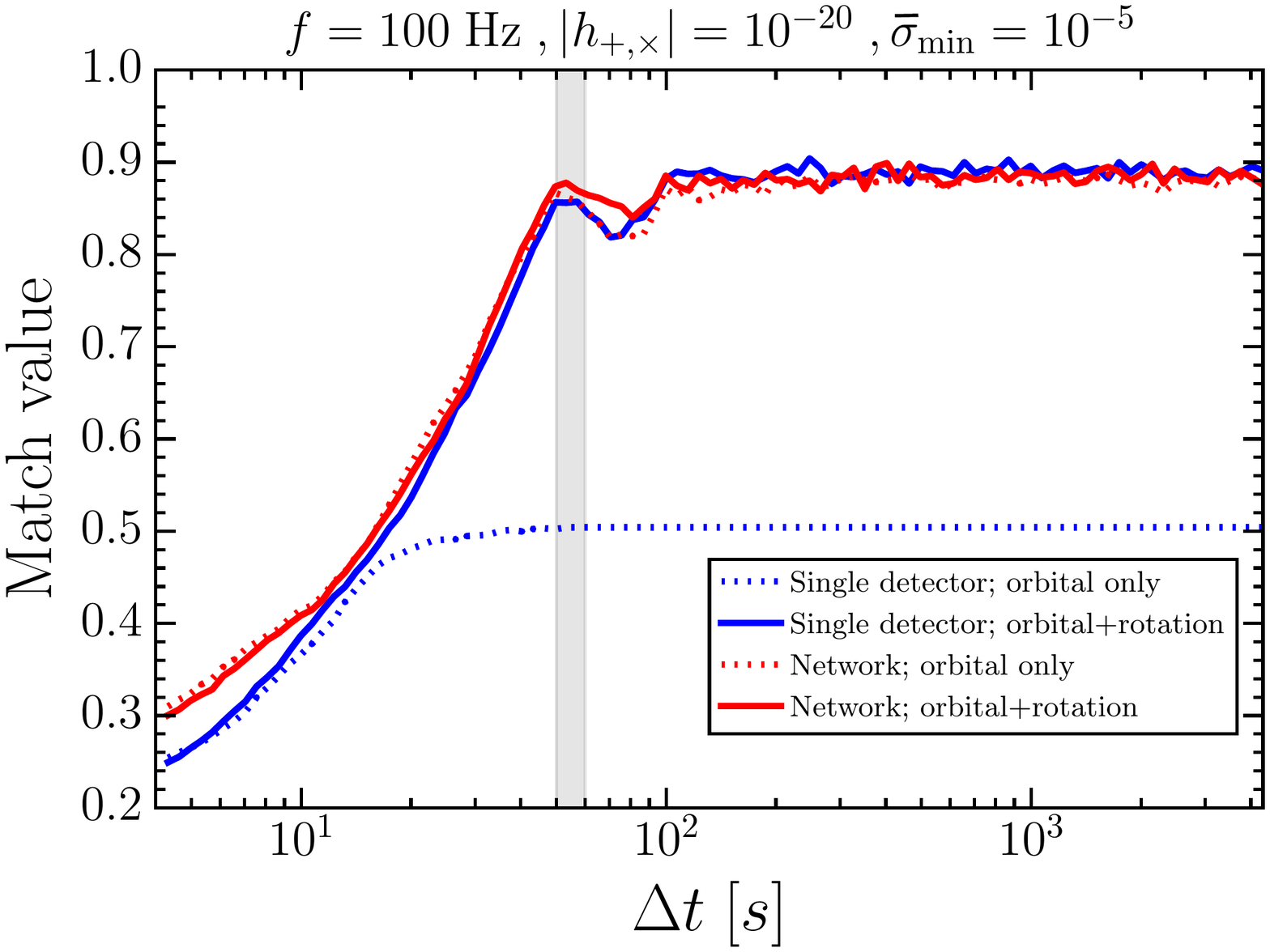}
\caption{Match values for various noisy map injections (averaged
  over $500$ noise realizations) as a
  function of the time increment between $1200$ observations, for a
  single detector and a $6$-detector network. For a single detector,
  orbital-only motion is only able to achieve a match value of $\sim
  0.5$ at large $\Delta t$. Daily
rotation provides additional information by sweeping the antenna beam
pattern across the sky, thus giving excellent plateau match
values of $\sim 0.9$. 
The global placement of a
network of detectors (and their differing antenna beam pattern
orientations) gives excellent match values even for orbital-only
motion, and daily rotation does not significantly improve this match. 
We see that the first peak in the match
value occurs at $\sim 50-60$ seconds (shown as a grey strip), 
when the detectors first
decorrelate from themselves and are no longer driven in coincidence by
a passing gravitational wave. The small dip after
the first peak is due to the virtual detectors being driven in
anti-coincidence by the gravitational wave, thereby
losing some of their independence and diminishing the match. However at
larger $\Delta t$ this behavior is averaged out over several
gravitational-wave wavelengths, allowing the match value to recover.}
\label{f:match-versus-deltaT}
\end{center}
\end{figure*}

\begin{figure*}[hbtp!]
\begin{center}
\includegraphics[angle=0, width=.49\textwidth]{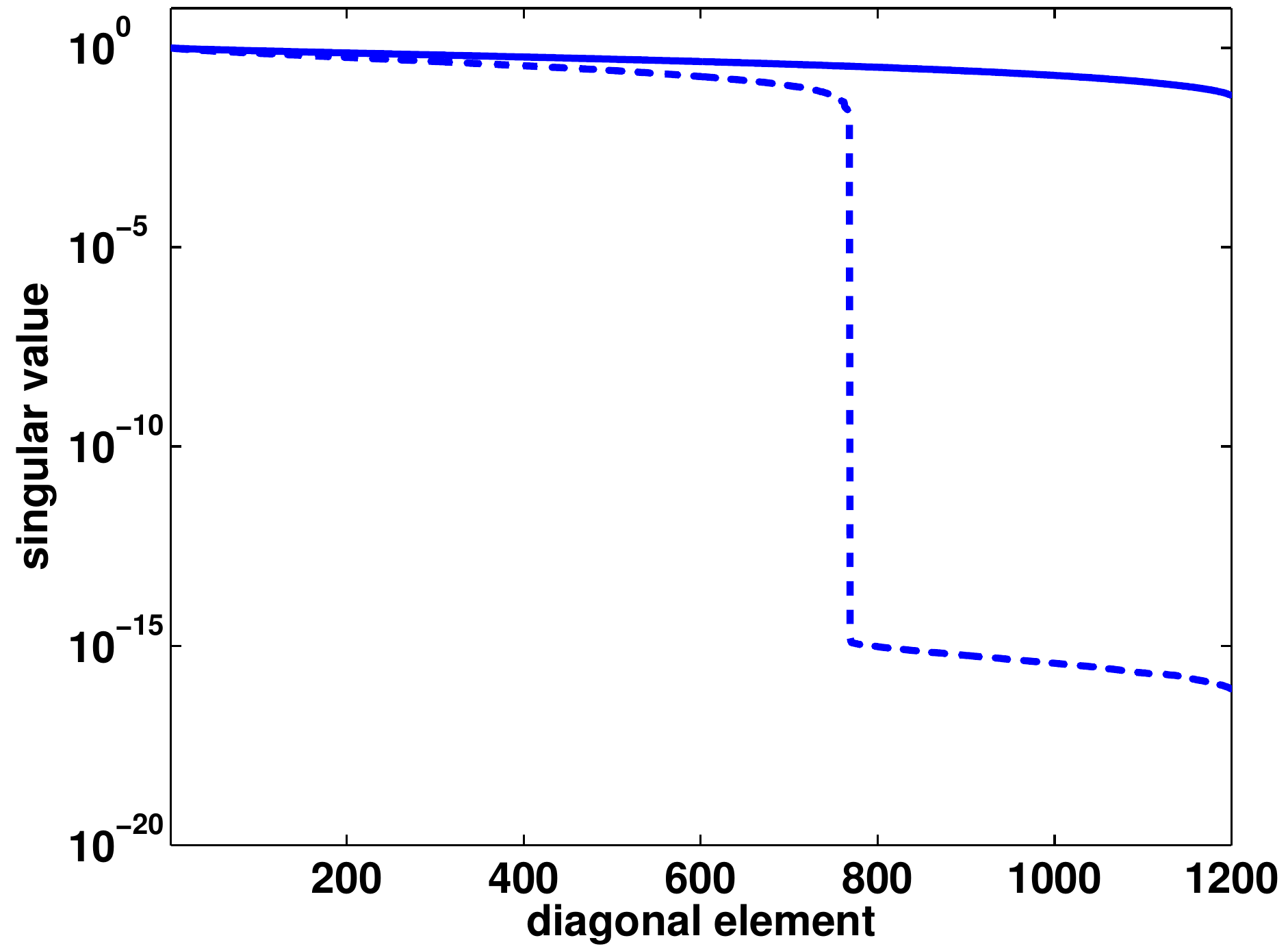}
\caption{Singular values of the response matrix for a single 
  LIGO-Livingston interferometer, for $1200$ total observations
  with $\Delta t = (1~{\rm sidereal\ day})/60\approx 1436~{\rm s}$. 
  (The singular values are normalized by the largest singular
  value, corresponding to the first diagonal element.)
  The dashed blue line shows the singular values when the
  detector is affixed to an Earth undergoing {\it orbital motion}
  only, whilst the solid blue line shows the singular values when the
  Earth is both {\it rotating and orbiting}. This indicates that
in the orbiting-only case, regularization of the response matrix at
machine-level precision ($\epsilon\sim 10^{-16}$) will not remove 
the very small singular values after diagonal element 768, 
requiring a more stringent cutoff level. 
In contrast, when the extra influence of the Earth's rotation is introduced,
the additional information provided by the changing detector arm
orientations (which sweep the antenna beam pattern across the sky)
acts to drastically improve the conditioning of the response matrix.}
\label{f:singular-values}
\end{center}
\end{figure*}

Our results are summarized in Fig.~\ref{f:match-versus-deltaT}. For all
cases, we find that the match of the recovered maps with the injected
map is poor for small time increments between observations, since the
detector(s) will not have moved far enough to establish independence
from its previous position. With
only orbital motion of a single detector, the match
values are only able to plateau at $\sim\!0.5$. 
Adding in the influence of daily Earth
rotation seems to ameliorate this poor match behaviour. The origin of
this effect can be deduced from the singular values of the response
matrix in both cases. This is shown in Fig.~\ref{f:singular-values}
for a single LIGO Livingston detector, where the addition of Earth
rotation acts to break degeneracies in the response matrix and
conditions it to have a much smaller dynamic range of singular values.
Rotating the Earth acts to sweep the antenna
beam pattern of the detector across the sky and provides additional
information with which to measure the gravitational-wave background. 
With only
orbital motion the arms of the detector remain in fixed orientations,
and hence so does the detector's antenna beam pattern. 
We also show in Fig.~\ref{f:match-versus-deltaT}
the match behaviour for a full $6$-detector advanced
network. In this case we
already have information from multiple orientations of the antenna
beam patterns by virtue of the different global placements of the
detectors. Hence, the inclusion of daily Earth rotation makes little
impact on the match value, which plateaus at $\sim\!0.9$ even with
orbital-only motion.

The key lessons here are that Earth's daily rotation is an important influence
on top of the orbital motion of the Earth around the Sun, since it sweeps
the detector antenna beam patterns across the sky to gather additional
information about any gravitational-wave signal of interest. 
Furthermore, from Fig.~\ref{f:match-versus-deltaT} we can clearly see 
that the first peak of the match value occurs
at $\sim 50-60$ s, when the detectors decorrelate from themselves for the first time
(see Fig.\ \ref{f:decorr-times}) and are no longer driven in
coincidence by a passing gravitational wave. With this time increment the detector's strain
measurements are effectively independent from their preceding or
subsequent measurements, thereby allowing us to synthesize a large network
of virtual interferometers from the daily and orbital motion of the Earth. The small dip after
the first peak may be due to the detectors being driven in
anti-coincidence, thereby
losing some of their independence. However, the match value recovers in the limit of
large $\Delta t$, since the detectors are then separated by several 
gravitational-wave wavelengths and this behaviour is averaged out.

\section{Discussion}
\label{s:discussion}

We have presented a new method for mapping the 
gravitational-wave sky using a network of ground-based laser 
interferometers.
This method extends the formalisms developed in 
\cite{Gair-et-al:2014, Cornish-vanHaasteren:2014}, which 
were originally applied to the case of pulsar timing arrays.
We have shown that we can recover both the gradient and curl 
components of a gravitational-wave background, as a 
consequence of the spatial separation of the individual
interferometers in the network, or of a single interferometer
at different times during its rotational and orbital motion
around the Sun.
This is in contrast to the case for a pulsar timing array, 
which is completely insensitive to the curl modes.
Also, by mapping both the amplitude and phase of 
$h_+(f,\hat k)$ and $h_\times(f,\hat k)$ 
as functions of direction on the sky 
(as referenced from the SSB), our method extends previous 
approaches~\cite{allen-ottewill, Cornish:2001hg, mitra-et-al, 
Thrane-et-al:2009, S5anisotropic:2011, Mingarelli:2013, TaylorGair:2013} for 
anisotropic backgrounds, which map the distribution of 
gravitational-wave {\em power}, 
$|h_+|^2+|h_\times|^2$.
Our formalism can be cast in terms of either the traditional
$+$ and $\times$ polarization modes of the background
$\{h_+(f,\hat k), h_\times(f,\hat k)\}$, or the gradient 
and curl modes $\{a^G_{(lm)}(f), a^C_{(lm)}(f)\}$, 
with respect to a decomposition of the metric perturbations
in terms of spin-weighted or tensor (gradient and curl) 
spherical harmonics~\cite{Gair-et-al:2014}.

The results of the simulations presented in Sec.~\ref{s:simulations} 
can be thought as a {\em proof-of-principle demonstration} of the 
general map-making formalism described in the rest of the paper.
The actual analysis of {\em real} data from a network of advanced 
interferometers will most likely differ from this simplified scenario
in several ways:

(i) The amplitudes of the simulated backgrounds were chosen to 
be sufficently large, so as to allow for fairly decent recovery 
after only a few days of observation.
Much weaker backgrounds will require an increased observation 
time, of order months or years,
noting that the (power) signal-to-noise ratio scales like
$A^2\sqrt{T}$, where $A$ is the amplitude of the background
and $T$ is the total observation time.

(ii) For initial analyses, it might be easier to work in the 
tensor spherical harmonic basis, and estimate the grad and curl 
components of the background 
$\{a^G_{(lm)}(f)$, $a^C_{(lm)}(f)\}$
out to some relatively small value of 
$l_{\rm max}$, e.g., $l_{\rm max}=10$.  
This would reduce the number of modes that we would need to recover 
from $2 N_{\rm pix}$ ($=1536$ for example) to 
$2 N_{\rm modes} = 234$ at each discrete frequency.
The estimates of the grad and curl components can then be converted 
to sky maps of 
$h_+(f,\hat k)$, $h_\times(f,\hat k)$ 
using (\ref{e:h+hx}).

(iii) Varying noise levels in the detectors (on a time scale
$\gtrsim$ the segment duration $\tau$ of our short-term
Fourier transforms) will complicate somewhat the expression for 
the noise covaraince matrix ${\mb C}$.
The $N_d$ block matrices that enter the expression for ${\mb C}$ 
will no longer be proportional to the unit matrix 
$\mathds{1}_{N_t\times N_t}$, but rather will have diagonal 
elements proportional to 
$S_I(f;t_{Ii})$, $i=1,2,\cdots,N_t$, reflecting the 
time-varying noise levels in detector $I$.

(iv) As ground-based interferometers are broad-band detectors, 
we will have measurements at a set of discrete 
frequencies $f_j$, $j=1,2,\cdots,N_f$, where 
$N_f\sim$~several hundred to a few thousand 
depending on the frequency bin size $\delta f$.
For initial analyses, it will probably
be simplest to average the estimates
of $h_+(f_j,\hat k)$, $h_\times(f_j,\hat k)$ over the 
different frequency components.

(v) If one would like to compare the consistency of 
different models of a stochastic background with the 
measured data---e.g., is the measured data consistent with 
an unpolarized, isotropic background or with a background
having a non-zero dipole component or with correlated 
emission on the sky, etc.---a Bayesian 
formulation of the problem would be more appropriate.
The different models would be defined by the appropriate
choice of variables for the stochastic background and prior 
probably distributions for these variables.
Bayesian model selection would then be used to select between
the competing models.

Perhaps the most compelling reason for using the formalism 
presented here is that it provides a completely 
generic approach to mapping the gravitational-wave sky.
It allows us to construct a map of the background that 
extracts all of the information that is possible to extract 
from the measured data.
With the advanced ground-based interferometers coming on-line 
at the end of this year, and with the first detection of 
gravitational waves expected to follow shortly thereafter, 
it seems appropriate to utilize approaches such as this that 
attempt to maximize the science return of the data.

\acknowledgments 
JDR acknowledges support from NSF Awards PHY-1205585 and CREST
HRD-1242090. 
This research
was in part supported by ST's appointment to the NASA Postdoctoral
Program at the Jet Propulsion Laboratory, administered by Oak Ridge
Associated Universities through a contract with NASA.
NJC acknowledges support from NSF PHY-1306702 and 
the NANOGrav Physics Frontier Center, NSF PFC-1430284.
JG's work is supported by the Royal Society. 
CMFM's work is supported by a Marie Curie International Outgoing Fellowship
within the 7th European Community Framework Programme.
RvH acknowledges support by NASA through Einstein Fellowship grants PF3-140116.
JDR thanks Malik Rakhmanov for useful discussions regarding 
pseudo-inverse calculations when the system of equations is
under-determined.
This research has made use of Python and its standard libraries: 
numpy and matplotlib.  
We have also made use of MEALPix (a Matlab implementation of 
HEALPix~\cite{HEALPix}), developed by the GWAstro Research Group 
and available from {\tt http://gwastro.psu.edu}.
This work was performed using the Darwin Supercomputer of the 
University of Cambridge High Performance Computing Service 
(http://www.hpc.cam.ac.uk/), provided by Dell Inc.~using Strategic
Research Infrastructure Funding from the Higher Education Funding Council 
for England and funding from the Science and Technology Facilities Council.
This paper has been assigned LIGO DCC number LIGO-P1500065.

\begin{appendix}
\section{Derivation of gradient and curl response functions}
\label{s:derivation}

Here we derive the gradient and curl response functions for an
interferometer in the small-antenna limit, allowing for a non-zero
displacement $\vec x_0$ of the vertex of the interferometer
from the origin of coordinates.

Expressions for the response functions evaluated in a reference
frame whose origin is located at the vertex of the interferometer
were derived in Appendix~D of \cite{Gair-et-al:2014}:
\be
\begin{aligned}
\bar R^G_{(lm)}(f) &= \delta_{l2}\,\frac{4\pi}{5}\sqrt{\frac{1}{3}}
\left[Y_{2m}(\hat u) - Y_{2m}(\hat v)\right]\,,
\\
\bar R^C_{(lm)}(f) &= 0\,,
\end{aligned}
\ee
where $\hat u$, $\hat v$ are unit vectors in the directions of the 
two arms of the interferometer.
We have put bars on the above expressions to distinguish them from 
similar unbarred quantities that we will calculate in a reference frame 
whose origin is at the solar system barycentre (SSB).
Note that $\bar R^G_{(lm)}$ is independent of frequency and is
non-zero only for the quadrupole modes, $l=2$.

Under a translation of reference frames from the SSB to the vertex 
of the interferometer located at $\vec x_0$, the Fourier components 
$h_{ab}(f,\hat k)$ of the metric perturbations $h_{ab}(t,\vec x)$ 
in the ``cosmic" (or SSB) frame transform to
\begin{equation}
\bar h_{ab}(f,\hat k) = h_{ab}(f,\hat k) e^{-i2\pi f\hat k\cdot \vec x_0/c}
\label{e:barh_ab}
\end{equation}
in the detector frame.
The correponding mode expansions in the two frames are given by
\be
\begin{aligned}
h_{ab}(f,\hat k) 
&= \sum_{(lm)}\sum_P a^P_{(lm)}(f) Y_{(lm)ab}^P(\hat k)\,,
\\
\bar h_{ab}(f,\hat k) 
&= \sum_{(lm)}\sum_P \bar a^P_{(lm)}(f) Y_{(lm)ab}^P(\hat k)\,.
\end{aligned}
\ee
This last equation for $\bar h_{ab}(f,\hat k)$ can be inverted
to find $\bar a^P_{(lm)}(f)$ in terms of $a^P_{(lm)}(f)$
using (\ref{e:barh_ab}) and the orthogonality of the gradient 
and curl spherical harmonics:
\begin{widetext}
\be
\begin{aligned}
\bar a^P_{(lm)}(f) 
&= \int_{S^2}{\rm d}^2\Omega_{\hat k}\>
\bar h_{ab}(f,\hat k) Y^P_{(lm)}{}^{ab\,*}(\hat k)
\\
&= \int_{S^2}{\rm d}^2\Omega_{\hat k}\>
h_{ab}(f,\hat k) 
e^{-i2\pi f \hat k\cdot \vec x_0/c}\,
Y^P_{(lm)}{}^{ab\,*}(\hat k)
\\
&= \int_{S^2}{\rm d}^2\Omega_{\hat k}\
\sum_{(l'm')}\sum_{P'} a^{P'}_{(l'm')}(f) Y^{P'}_{(l'm')ab}(\hat k)
e^{-i2\pi f \hat k\cdot \vec x_0/c}\,
Y^P_{(lm)}{}^{ab\,*}(\hat k)\,.
\end{aligned}
\ee
Using the identity:
\be
e^{-i2\pi f\hat k\cdot\vec x_0/c}
= 4\pi \sum_{L=0}^\infty (-i)^L j_L(\alpha) 
\sum_{M=-L}^L Y^*_{LM}(\hat x_0)Y_{LM}(\hat k)\,,
\qquad
\alpha \equiv 2\pi f|\vec x_0|/c\,,
\ee
we obtain
\begin{align}
\bar a^P_{(lm)}(f) 
= 
\sum_{(l'm')}\sum_{P'} 
a^{P'}_{(l'm')}(f) 
\sum_{L=0}^\infty\sum_{M=-L}^L
4\pi(-i)^L j_L(\alpha) Y^*_{LM}(\hat x_0)
\int_{S^2}{\rm d}^2\Omega_{\hat k}\
Y^{P'}_{(l'm')ab}(\hat k)
Y^P_{(lm)}{}^{ab\,*}(\hat k)
Y_{LM}(\hat k)
\end{align}
relating the mode coefficients in the two frames.

To make the connection between the mode coefficients 
and the corresponding response functions, we note that the 
detector response $\tilde r(f)$ (or $r(t)$) 
to the gravitational-wave background will have the 
same value regardless of which frame we choose to evaluate 
it in.
Thus,
\be
\begin{aligned}
\tilde r(f)
&=\sum_{(lm)}\sum_P \bar R^P_{(lm)}\bar a^P_{(lm)}(f)
\\
&=\sum_{m=-2}^2 \bar R^G_{(2m)}\bar a^G_{(2m)}(f)
\\
&=\sum_{m=-2}^2 \bar R^G_{(2m)}
\sum_{(l'm')}\sum_{P'} 
a^{P'}_{(l'm')}(f) 
\sum_{L=0}^\infty\sum_{M=-L}^L
4\pi(-i)^L j_L(\alpha) Y^*_{LM}(\hat x_0)
\int_{S^2}{\rm d}^2\Omega_{\hat k}\
Y^{P'}_{(l'm')ab}(\hat k)
Y^G_{(2m)}{}^{ab\,*}(\hat k)
Y_{LM}(\hat k)
\\
&=
\sum_{(l'm')}\sum_{P'} 
R^{P'}_{(l'm')}(f)
a^{P'}_{(l'm')}(f)\,, 
\end{aligned}
\ee
where
\be
R^P_{(lm)}(f) =
\sum_{m'=-2}^2 
\sum_{L=0}^\infty
\sum_{M=-L}^L 
\bar R^G_{(2m')}
4\pi(-i)^L j_L(\alpha)Y^*_{LM}(\hat x_0)
\int_{S^2}{\rm d}^2\Omega_{\hat k}\
Y^{P}_{(lm)ab}(\hat k)
Y^G_{(2m')}{}^{ab\,*}(\hat k)
Y_{LM}(\hat k)\,.
\ee
We can write this last expression explicitly in 
terms of Wigner-3$j$ symbols if we replace the grad and 
curl spherical harmonics in the integral by spin-2 spherical 
harmonics using:
\be
\begin{aligned}
Y^{G}_{(lm)ab}(\hat k) Y^G_{(2m')}{}^{ab\,*}(\hat k)
&=\frac{1}{2}\left[
{}_{-2}Y_{lm}(\hat k) {}_{-2}Y_{2m'}^*(\hat k) 
+
{}_{2}Y_{lm}(\hat k) {}_{2}Y_{2m'}^*(\hat k) 
\right]\,,
\\
Y^{C}_{(lm)ab}(\hat k) Y^G_{(2m')}{}^{ab\,*}(\hat k)
&=\frac{1}{2i}\left[
{}_{-2}Y_{lm}(\hat k) {}_{-2}Y_{2m'}^*(\hat k) 
-
{}_{2}Y_{lm}(\hat k) {}_{2}Y_{2m'}^*(\hat k) 
\right]\,.
\end{aligned}
\ee
This leads to 
\be
\begin{aligned}
R^G_{(lm)}(f)
=&\sum_{m'=-2}^2 
\sum_{L=0}^\infty
\sum_{M=-L}^L 
\bar R^G_{(2m')}
4\pi (-i)^L j_L(\alpha)Y^*_{LM}(\hat x_0)
\\
&\frac{(-1)^{m'}}{2}
\sqrt{\frac{(2\cdot 2+1)(2l+1)(2L+1)}{4\pi}}
\left(\begin{array}{ccc}
2 & l & L\\
-m' & m & M 
\end{array}\right)
\left[\left(\begin{array}{ccc}
2 & l & L\\
-2 & 2 & 0 
\end{array}\right)
+
\left(\begin{array}{ccc}
2 & l & L\\
2 & -2 & 0 
\end{array}\right)\right]\,,
\\
R^C_{(lm)}(f)
=&\sum_{m'=-2}^2 
\sum_{L=0}^\infty
\sum_{M=-L}^L 
\bar R^G_{(2m')}
4\pi (-i)^L j_L(\alpha)Y^*_{LM}(\hat x_0)
\\
&\frac{(-1)^{m'}}{2i}
\sqrt{\frac{(2\cdot 2+1)(2l+1)(2L+1)}{4\pi}}
\left(\begin{array}{ccc}
2 & l & L\\
-m' & m & M 
\end{array}\right)
\left[\left(\begin{array}{ccc}
2 & l & L\\
-2 & 2 & 0 
\end{array}\right)
-
\left(\begin{array}{ccc}
2 & l & L\\
2 & -2 & 0 
\end{array}\right)\right]\,.
\end{aligned}
\ee
These expressions can be further simplified 
using the symmetry property
\be
\left(\begin{array}{ccc}
l_1 & l_2 & l\\
m_1 & m_2 & m
\end{array}\right)
= (-1)^{l_1+l_2 + l}
\left(\begin{array}{ccc}
l_1 & l_2 & l\\
-m_1 & -m_2 & -m 
\end{array}\right)
\ee
to eliminate one the Wigner 3-$j$ symbols in terms of
the other, and the triangle inequality
\be
|l_1-l_2|\le l\le l_1+l_2
\quad\Rightarrow\quad
l-2 \le L \le l+2
\ee
to collapse the infinite sums over $L$ to sums over just 5 terms.
The final expressions are
\be
\begin{aligned}
R^G_{(lm)}(f)
=&\sum_{m'=-2}^2 
\sum_{L=l-2}^{l+2}
\sum_{M=-L}^L 
\bar R^G_{(2m')}
4\pi (-i)^L j_L(\alpha)Y^*_{LM}(\hat x_0)
\\
&\frac{(-1)^{m'}}{2}
\sqrt{\frac{(2\cdot 2+1)(2l+1)(2L+1)}{4\pi}}
\left(\begin{array}{ccc}
2 & l & L\\
-m' & m & M 
\end{array}\right)
\left(\begin{array}{ccc}
2 & l & L\\
2 & -2 & 0 
\end{array}\right)
\left[(-1)^{l+L}+1\right]\,,
\\
R^C_{(lm)}(f)
=&\sum_{m'=-2}^2 
\sum_{L=l-2}^{l+2}
\sum_{M=-L}^L 
\bar R^G_{(2m')}
4\pi (-i)^L j_L(\alpha)Y^*_{LM}(\hat x_0)
\\
&\frac{(-1)^{m'}}{2i}
\sqrt{\frac{(2\cdot 2+1)(2l+1)(2L+1)}{4\pi}}
\left(\begin{array}{ccc}
2 & l & L\\
-m' & m & M 
\end{array}\right)
\left(\begin{array}{ccc}
2 & l & L\\
2 & -2 & 0 
\end{array}\right)
\left[(-1)^{l+L}-1\right]\,.
\end{aligned}
\ee
The Wigner 3-$j$ symbol selection rule
\be
-m'+m+M=0 
\ee
implies that the sum over $M$ 
collapses to only those values 
satisfying $M=m'-m$ and $|M|<L$.
Note also that in a reference frame with the 
$z$-axis chosen along $\hat x_0$, 
\be
Y^*_{LM}(\hat x_0) = Y^*_{LM}(0,\phi) 
= \delta_{M0}\,\sqrt{\frac{2L+1}{4\pi}}\,,
\ee
so for this case the sum over $M$ reduces to just the $M=0$ value.
The selection rule $-m'+m=0$ and $|m'|\le 2$ then imply
non-zero values for only $|m|\le 2$ in this special frame.
\\

\end{widetext}

\section{Equivalence of whitened and non-whitened analyses}
\label{s:whitening}

As discussed in Sec.~\ref{s:ML-estimation},
we are interested in finding the maximum-likelihood value
${\mb h}_{\rm ML}$ of the likelihood function
\be
p({\mb d}|{\mb C},{\mb h}) 
\propto
\exp\left[-
({\mb d}-{\mb R}{\mb h})^\dagger
{\mb C}^{-1}
({\mb d}-{\mb R}{\mb h})\right]\,,
\ee
when the Fisher matrix 
${\mb F}\equiv {\mb R}^\dagger{\mb C}^{-1}{\mb R}$ is not
invertible.
One approach, described in \cite{Gair-et-al:2014}, is to
work with the SVD of the response matrix ${\mb R}$:
\be
{\mb R} = {\mb U}{\bs\Sigma}{\mb V}^\dagger\,.
\label{e:svd(R)}
\ee
In general we can write ${\mb U} = [{\mb U}_r {\mb U}_n]$, 
where ${\mb U}_r$ is an $N\times r$ matrix denoting the 
range of the response matrix ${\mb R}$, 
where $r$ equals 
the number of non-zero singular values in ${\bs\Sigma}$. 
We then replace ${\mb R}{\mb h}$ in the likelihood by 
${\mb U}_r{\mb b}$, where ${\mb b}$ is a vector of dimension 
$r$, and then proceed as for a non-singular response. 
The maximum-likelihood value for ${\mb b}$ is then
\be
{\mb b}_{\rm ML}
=({\mb U}_r^\dagger {\mb C}^{-1} {\mb U}_r)^{-1}
{\mb U}_r^\dagger {\mb C}^{-1}{\mb d}\,,
\label{e:bML}
\ee
and the corresponding maximum-likelihood estimate of the
gravitational-wave sky is
\be
{\mb h}_{\rm ML}
={\mb V}{\bs\Sigma}_r^{+}{\mb b}_{\rm ML}\,,
\label{e:bML_svd}
\ee
where ${\bs\Sigma}_r$ is the $r\times M$ dimensional
matrix obtained by crossing out the 
last $N-r$ rows of ${\bs\Sigma}$, and ${\bs\Sigma}_r^{+}$
is the pseudo-inverse of ${\bs\Sigma}_r$,
obtained by taking the reciprocal of each non-zero
singular value of ${\bs\Sigma}_r$, 
and then transposing the resulting matrix.

An alternative approach, which we described in 
Sec.~\ref{s:ML-estimation}, is to work with the 
the whitened data $\bar{\mb d}\equiv {\mb L}^\dagger{\mb d}$ 
and whitened response matrix
$\bar{\mb R}\equiv {\mb L}^\dagger{\mb R}$,
where ${\mb L}$ is a lower triangular matrix defined by the
Cholesky decomposition of the inverse covariance matrix,
${\mb C}^{-1} = {\mb L}{\mb L}^\dagger$.
Working with the SVD of $\bar{\mb R}$:
\be
\bar{\mb R} = \bar{\mb U}\bar{\bs\Sigma}\bar{\mb V}^\dagger\,,
\ee
we have
\be
{\mb h}_{\rm ML} 
=\bar{\mb V}\bar{\bs\Sigma}^+\bar{\mb U}^\dagger
\bar{\mb d}\,.
\ee
Now the SVD of a product cannot be simply written in terms
of the SVDs of the individual matrices.
However, we can show the equivalence of these two approaches 
for maximizing the likelihood, 
by working with the equivalent likelihood that was 
introduced in the first approach, i.e.,
\be
p({\mb d}|{\mb C},{\mb h}) 
\propto
\exp\left[-
({\mb d}-{\mb U}_r{\mb b})^\dagger
{\mb C}^{-1}
({\mb d}-{\mb U}_r{\mb b})\right]\,,
\ee
and obtaining ${\mb h}_{\rm ML}$ from ${\mb b}_{\rm ML}$ using
Eq.~(\ref{e:bML_svd}).
The first approach requires no modification, but for the 
second approach we now need the SVD of the whitened 
${\mb U}_r$ matrix, ${\mb L}^\dagger{\mb U}_r$:
\be
{\mb L}^\dagger{\mb U}_r = \tilde{\mb U}\tilde{\bs\Sigma}
\tilde{\mb V}^\dagger\,,
\label{e:svd(LdaggerUr}
\ee
for which
\be
{\mb b}_{\rm ML}
=\tilde{\mb V}\tilde{\bs\Sigma}^+\tilde{\mb U}^\dagger
\bar{\mb d}\,.
\ee
Since we are now working only with the range of ${\mb R}$ and 
the noise covariance matrix ${\mb C}$ is positive definite, 
the rank of ${\mb L}^\dagger{\mb U}_r$ must equal the rank 
of ${\mb U}_r$.
As before, we can write 
$\tilde{\mb U}=[\tilde{\mb U}_r \tilde{\mb U}_n]$,
where $\tilde{\mb U}_r$ is an $N\times r$ matrix,
which gives the range of ${\mb L}^\dagger {\mb U}_r$.
Thus, we can equivalently write Eq.~(\ref{e:svd(LdaggerUr}) 
as
\be
{\mb L}^\dagger{\mb U}_r = \tilde{\mb U}_r\tilde{\bs\Sigma}_r
\tilde{\mb V}^\dagger\,,
\label{e:svd(LdaggerUr)alt}
\ee
where $\tilde{\bs\Sigma}_r$ is an invertible, 
square $r\times r$ matrix
obtained, as before, by crossing out the last $N-r$ rows
of $\tilde{\bs\Sigma}$.
From this last equation we now see that
\be
\tilde{\mb U}_r = {\mb L}^\dagger {\mb U}_r\tilde{\mb V}
\tilde{\bs\Sigma}_r^{-1}\,,
\ee
and
\be
\begin{aligned}
{\mb b}_{\rm ML}
&=\tilde{\mb V}\tilde{\bs\Sigma}_r^{-1}\tilde{\mb U}_r^\dagger\bar{\mb d}
\\
&=\tilde{\mb V}\tilde{\bs\Sigma}_r^{-1}
\tilde{\bs\Sigma}_r^{-1}\tilde{\mb V}^\dagger
{\mb U}_r^\dagger{\mb L}\bar{\mb d}
\\
&=({\mb U}_r^\dagger{\mb C}^{-1}{\mb U}_r)^{-1}
{\mb U}_r^\dagger{\mb C}^{-1}{\mb d}\,,
\end{aligned}
\ee
where the final equality follows from the observation that
${\mb U}_r^\dagger {\mb C}^{-1}{\mb U}_r
={\mb U}_r^\dagger {\mb L}{\mb L}^\dagger{\mb U}_r
=\tilde{\mb V}\tilde{\bs\Sigma}_r^{2}\tilde{\mb V}^\dagger$
(which is a consequence of Eq.~(\ref{e:svd(LdaggerUr)alt})).
We have thus recovered the result given in Eq.~(\ref{e:bML}),
which was obtained without whitening the data.

\end{appendix}

\bibliography{manuscript}

\end{document}